\documentstyle[12pt,epsf]{article}

\advance \topmargin by -\headheight
\advance \topmargin by -\headsep

\evensidemargin \oddsidemargin
\newcommand{\beq}{\begin{equation}}
\newcommand{\eeq}{\end{equation}}
\newcommand{\rc}{\nonumber\\}
\newcommand{\bear}{\begin{eqnarray}}
\newcommand{\eear}{\end{eqnarray}}

\def\tr{{\hbox{\rm tr}}}

\def\ie{{\em i.e.}}
\def\ie{\hbox{\it i.e.}}

\def\CC{{\mathchoice
{\rm C\mkern-8mu\vrule height1.45ex depth-.05ex
width.05em\mkern9mu\kern-.05em}
{\rm C\mkern-8mu\vrule height1.45ex depth-.05ex
width.05em\mkern9mu\kern-.05em}
{\rm C\mkern-8mu\vrule height1ex depth-.07ex
width.035em\mkern9mu\kern-.035em}
{\rm C\mkern-8mu\vrule height.65ex depth-.1ex
width.025em\mkern8mu\kern-.025em}}}

\def\RR{{\rm I\kern-1.6pt {\rm R}}}

\def\ZZ{{\rm Z}\kern-3.8pt {\rm Z} \kern2pt}
\def\IB{\relax{\rm I\kern-.18em B}}
\def\ID{\relax{\rm I\kern-.18em D}}
\def\II{\relax{\rm I\kern-.18em I}}
\def\IP{\relax{\rm I\kern-.18em P}}

\def\np{Nucl. Phys.}
\def\pl{Phys. Lett.}
\def\prl{Phys. Rev. Lett.}
\def\pr{Phys. Rev.}

\def\mpl{Mod. Phys. Lett.}

\def\atmp{Adv. Theor. Math. Phys. }
\def\jhep{J. High Energy Phys.}
\def\ptp{Prog. Theor. Phys.}
\def\jgp{J. Geom. Phys.}
\def\atmp{Adv. Theor. Math. Phys.}

\renewcommand{\theequation}{{\rm\thesection.\arabic{equation}}}

\begin{document}

\begin{titlepage}

\begin{center} \Large \bf Holographic flavor on the Higgs branch

\end{center}

\vskip 0.3truein
\begin{center}
Daniel Are\'an${}^{\,*}$\footnote{arean@fpaxp1.usc.es},
Alfonso V. Ramallo${}^{\,*}$\footnote{alfonso@fpaxp1.usc.es} and
Diego Rodr\'\i guez-G\'omez${}^{\,\dagger}
$\footnote{drodrigu@princeton.edu}

\vspace{0.3in}

${}^{\,*}$Departamento de F\'\i sica de Part\'\i culas, Universidade de
Santiago de Compostela \\and\\
Instituto Galego de F\'\i sica de Altas Enerx\'\i as (IGFAE)\\
E-15782 Santiago de Compostela, Spain
\vspace{0.3in}

${}^{\,\dagger}$Joseph Henry Laboratories, Princeton University\\
Princeton NJ 08544, U.S.A.\\

\end{center}
\vskip.5truein

\begin{center}
\bf ABSTRACT
\end{center}

In this paper we study  the
holographic dual, in several spacetime dimensions, of the Higgs branch of
gauge theories with fundamental matter. These theories  contain defects
of  various codimensionalities, where the matter fields are located.  In the
holographic description the matter is added by considering flavor brane
probes in the supergravity backgrounds generated by color branes, while the
Higgs branch is obtained when the color and flavor branes recombine with
each other. We show that, generically, the holographic dual of the Higgs
phase is realized by means of the addition of extra flux on the flavor
branes and by choosing their appropriate embedding in the background
geometry. This suggests a dielectric interpretation in terms of the color
branes, whose vacuum solutions precisely match the F- and  D-flatness
conditions obtained on the field theory side. We further compute the meson
mass spectra in several cases and show that when the defect added has
codimension greater than zero it becomes continuous and gapless.

\vskip2.5truecm
\leftline{US-FT-1/07}
\leftline{PUTP-2225}
\leftline{hep-th/0703094 \hfill March  2007}
\smallskip
\end{titlepage}
\setcounter{footnote}{0}

\tableofcontents

\setcounter{equation}{0}
\section{Introduction}
\medskip
The gauge/gravity correspondence \cite{jm, MAGOO}  
has been a breakthrough in our understanding of both gravity (and string
theory) and gauge field theories. However, a major issue remains to be
the inclusion of open string degrees of freedom, which would correspond
to matter in the fundamental representation in the gauge field theory
side of the correspondence. 

A first step was taken in \cite{KR, KKW}, where it was suggested that 
one can add dynamical open string degrees of freedom by adding $N_f$
intersecting Dq-branes to the original Dp-branes. In the limit in
which the number of Dq-branes is much smaller than the number of
Dp-branes, we can treat the system effectively as $N_f$ probe branes in
the background generated by  the $N_c$ Dp-branes, which, once we take the
decoupling limit, will reduce to the corresponding near horizon geometry.
Generically, the two types  of branes overlap partially, which implies
that the additional Dq-branes create a defect on the worldvolume theory 
of the Dp-branes. In the dual gauge theory description the extra branes
give rise to additional matter, confined to live inside the defect, 
which comes from the Dp-Dq strings. When $q>p$, the decoupling limit
forces the $SU(N_f)$ gauge symmetry on the  Dq-brane  to decouple. It then
appears as a global flavor symmetry for the extra matter, which is in the
fundamental representation of the flavor group. In this context, the
fluctuations of the flavor branes should  correspond to the mesons in the
dual gauge theory. The study of these mesons was started in \cite{KMMW}
for the D7-brane in the $AdS_5\times S^5$ geometry, and it was further
extended to other flavor branes in several backgrounds (\cite{Sonnen}-
\cite{Apreda:2006bu})
(for a review see \cite{R}).

The
dual theories to these brane intersections are in the Coulomb phase.
However one could have more involved situations, such as Higgs phases. On
the field theory side the Higgs phase corresponds to having a non-zero
VEV of the quark fields. As it is well-known (see e.g.
ref. \cite{Giveon:1998sr}) the Higgs branch of gauge theories can be
realized in string theory by recombining color and flavor branes. This
recombination can be described in two different and complementary ways.
From the point of view of the flavor brane (the so-called macroscopic
picture) the recombination is achieved by a non-trivial embedding of the
brane probe in the background geometry and/or by a non-trivial  flux of
the worldvolume gauge field. On the other hand, 
the description of the recombination from the point of view of the color
brane defines  the microscopic picture. In most of the cases this
microscopic picture can be regarded as a dielectric effect \cite{M}, in
which a set of color branes gets polarized into a higher-dimensional
flavor brane. Interestingly, the microscopic description of the Higgs
branch allows a direct relation with the field theory analysis and the
micro-macro matching is essential to understand how gauge theory
quantities are encoded into the configuration of the flavor brane.

In ref. \cite{EGG} the Higgs phase of the D3-D7 intersection was studied
(see also ref. \cite{Guralnik:2004ve}). It was proposed in \cite{EGG,
Guralnik:2004ve} that, from the point of view of the D7-brane, one can
realize  a (mixed Coulomb-)Higgs phase of the D3-D7 system by switching on
an instanton configuration of the worldvolume gauge field of the D7-brane.
This instantonic gauge field lives on the directions of the D7-brane
worldvolume that are orthogonal to the gauge theory directions. The size of
the instanton has been identified in
\cite{EGG, Guralnik:2004ve} with the VEV of the quark fields. The meson 
spectra depends on this size and  was shown in \cite{EGG} to display, in the
limit of infinite instanton size, an spectral flow phenomenon. 

The defect conformal field theory associated to the D3-D3
intersection was studied in ref. \cite{CEGK}, where    the
corresponding fluctuation/operator dictionary was established. The meson
mass spectra of this system when the two sets of D3-branes are separated
was computed analytically in ref. \cite{AR}.  In \cite{CEGK}  the Higgs
branch of the D3-D3 system was identified as a particular holomorphic
embedding of the probe D3-brane in the $AdS_5\times S^5$ geometry, which
was shown to correspond to the vanishing of the F- and D-terms in the dual
superconformal field theory (see also refs.
\cite{Erdmenger:2003kn,Kirsch:2004km}).

The Higgs phase of the
dual to the D3-D5 intersection was discussed in ref. \cite{ARR}. On the
field theory side \cite{WFO} the D3-D5 system  describes the dynamics of a
$2+1$ dimensional defect containing fundamental hypermultiplets living
inside the $3+1$ dimensional
${\cal{N}}=4$ $SYM$. The meson spectra on the Coulomb branch
has been obtained in \cite{AR}.
In \cite{ARR} it was found that, in the
supergravity dual, the Higgs phase corresponds to adding magnetic
worldvolume flux inside the flavor D5-brane  transverse to the D3-branes.
This worldvolume gauge field has the non-trivial effect of inducing
D3-brane charge in the D5-brane worldvolume, which in turn suggests an
alternative microscopical description in terms of D3-branes expanded to a
D5-brane due to dielectric effect \cite{M}. Indeed, the vacuum conditions
of the dielectric theory can be mapped to the $F$ and
$D$ flatness constraints of the dual gauge theory, thus justifying the
identification with the Higgs phase. The Higgs vacua of the field theory
involve a non-trivial dependence of the defect fields on the coordinate
transverse to the defect. In the supergravity side this is mapped to a
bending of the flavor brane, which is actually required by supersymmetry
(see \cite{ST}). Moreover, in \cite{ARR} the spectrum of transverse
fluctuations was computed in the Higgs phase, with the result that the
discrete spectrum is lost. The reason is that the IR theory is modified
because of the non-trivial profile of the flavor brane, so that in the
Higgs phase, instead of having an effective $AdS\times S$ worldvolume for
the flavor brane,  one has Minkowski space, thus loosing the KK-scale
which would give rise to a discrete spectrum. 

In this paper we will generalize the results on the Higgs branch 
of refs. \cite{EGG},
\cite{CEGK} and \cite{ARR} to the 
rest of supersymmetric brane intersections. In general, each type of
intersection is dual to a defect hosting a field theory living inside a
bulk gauge theory. Therefore, we can  label each case by the
codimensionality of the defect. We will see that generically all of them
behave in a similar way to the D3-D5 case, in the sense that the Higgs
phase is achieved by adding extra worldvolume  flux to the flavor brane.
However, as we will see, in not all the cases  the discrete meson
spectra is lost. 

We begin in section 2 by analyzing the codimension zero
defect, which corresponds to the Dp-D(p+4) intersection. We first study
the field theory of the D3-D7 system, where we identify a mixed
Coulomb-Higgs branch which is given by the vanishing of both the D- and
F-terms. This branch is characterized by a non-zero commutator of the
adjoint fields of ${\cal N}=4$ SYM which, from the point of view of the
flavor brane, corresponds to having a non-vanishing flux of the
worldvolume gauge field along the directions orthogonal to the color
brane. We will then describe such non-commutative scalars by using the
Myers action for a dielectric D3-brane and we will argue that,
macroscopically, this configuration can be described in terms of a
D7-brane with a self-dual instantonic gauge field. From this matching
between the D3- and D7-brane descriptions we will be able to extract the
relation proposed in ref. \cite{EGG} between the VEV of the quark field
and the size of the instanton. Afterwards we perform the computation of
the meson spectrum of the general Dp-D(p+4) systems, which in this case
remains discrete. We estimate the value of the mesonic mass gap as a
function of the instanton size. For large instantons this gap is
independent of the size, in agreement with the spectral flow found in
ref. \cite{EGG}, while for small instantons the mass gap is proportional
to the size of the instanton and vanishes in the zero-size limit. 

In section 3 we discuss  the codimension one  defects, whose most
prominent  example is the D3-D5  intersection studied in \cite{ARR}. In
this paper we study the general Dp-D(p+2)  case with worldvolume flux on
the D(p+2)-brane, which also admits a dielectric interpretation. We then
analyze  the meson spectrum, which is continuous and gapless as in the
D3-D5 case studied in \cite{ARR}. We will establish this result for the
full set of fluctuations of the D(p+2)-brane probe, which are analyzed in
appendix
\ref{DpDp+2-decoupling} by using  the same techniques as those employed in
refs. \cite{AR,MT} to study the Coulomb branch. We then study the F1-Dp
intersection which, in particular, for $p=3$  corresponds to the S-dual
version of the D1-D3 system.

In section 4 we study a close relative to the Dp-D(p+2) intersection, 
namely the M2-M5 intersection in M-theory. In this case, we see
that we can dissolve M2-branes by turning on  a three-form flux  on the
M5-brane worldvolume and introducing some bending of the M5-brane. The
supersymmetry of this configuration is explicitly confirmed in appendix
\ref{kappa} by looking at the kappa symmetry of the embedding. 
However, in this case a microscopical description is not at hand,
since it would involve an action for coincident M2-branes which is not
known at present. 

Section 5  is devoted to the analysis of  the codimension two defects,
which correspond to the Dp-Dp intersections. This case, as anticipated in 
\cite{CEGK,Erdmenger:2003kn}, is somehow different, since the Higgs
phase is realized by the choice of a particular embedding of the probe
Dp-brane with no need of extra flux. This case is rather particular
since, as we will show, the profile can be an arbitrary holomorphic curve
in suitable coordinates, although only one of them gives the desired
Higgs phase, while the rest remain unidentified. 

We then finish in section 6 with some conclusions.

\setcounter{equation}{0}
\section{The codimension zero defect}

Let us start considering the D3-D7 intersection, where the D3-branes
are fully contained in the D7-branes as shown in the following array:
\beq
\begin{array}{ccccccccccl}
 &1&2&3& 4& 5&6 &7&8&9 & \nonumber \\
D3: & \times &\times &\times &\_ &\_ & \_&\_ &\_ &\_ &     \nonumber \\
D7: &\times&\times&\times&\times&\times&\times&\times&\_&\_ &
\end{array}
\label{D3D7intersection}
\eeq

Clearly, the D3-D7 string
sector gives rise to extra fundamental matter living in the $3+1$ common
directions. It can be seen that the dual gauge theory is a
${\mathcal{N}}=2$ $SYM$ theory in $3+1$ dimensions obtained by adding
$N_f$ ${\mathcal{N}}=2$ fundamental hypermultiplets to the ${\mathcal{N}}=4$
$SYM$ theory. We can further break the classical conformal invariance of the
theory by adding a mass term for the quark hypermultiplets. The lagrangian is
given by (\cite{EGG}):
\begin{eqnarray}
\label{LFT}
&&{\cal L}= \tau \int d^2 \theta d^2 \bar\theta\,\,
         \Big(\tr (\Phi_I^{\dagger}\,e^V \Phi_I e^{-V}) 
          + Q_i^\dagger e^V Q^i + \tilde Q_i e^{-V} \tilde Q^{i\dagger}
\Big)\,\,+  \rc\rc
  && +\, \tau \int d^2 \theta \Big(\tr ({\cal W}^\alpha{\cal W}_\alpha) +
W\Big) \, +
            \tau \int d^2 \bar \theta \Big(\tr (\bar{\cal W}_{\dot \alpha}
\bar{\cal W}^{\dot\alpha}) + \bar W\Big)\,  \ ,
\end{eqnarray}
where the superpotential of the theory is given by: 

\begin{equation}
W=\tilde{Q}_i(m+\Phi_3)Q^i+{1\over 3}\,\epsilon^{IJK}\,{\rm Tr}\,\Big[\,
\Phi_I\Phi_J\Phi_K\,\Big]\,\,.
\end{equation}
In eq. (\ref{LFT}) we are working in ${\cal N}=1$ language, where 
$Q_i$, ($\tilde{Q}_i$) $i=1,\cdots, N_f$ are the chiral (antichiral)
superfields in the hypermultiplet, while $\Phi_I$ are the adjoint scalars
of ${\mathcal{N}}=4$ $SYM$ once complexified as 
$\Phi_1=X^1+iX^{2}\,$, $\Phi_2=X^3+iX^{4}\,$ and $\Phi_3=X^5+iX^{6}\,$
 where  $X^I$ ($I=1,\cdots 6$) is the  scalar which corresponds to
the direction $I+3$ in the array (\ref{D3D7intersection}).
It is worth  mentioning that an identity matrix in color space
is to be understood to multiply the mass parameter of the quarks  $m$.

We are interested in getting the classical SUSY vacua of this theory, which
can be obtained by imposing the corresponding $D$- and $F$-flatness
conditions that follow from the lagrangian (\ref{LFT}). Let us start
by imposing the vanishing of the
$F$-terms corresponding to the quark hypermultiplets, which amounts to set:

\begin{equation}
\tilde{Q}_i(\Phi_3+m)=0\,\,\,,\qquad\qquad
(\Phi_3+m)Q^i=0\,\, .
\label{Phi3}
\end{equation}
These equations can be satisfied by taking $\Phi_3$ as:

\beq
\Phi_3=\left( \begin{array}{cccccc}
\tilde{m}_1& & & & & \\
 &\ddots& & & \\
  & & \tilde{m}_{N-k}& & &\\
   & & &-m& &\\
   & & & &\ddots& \\
   & & & & & -m
   \end{array}
   \right)\,\, ,
\label{Phi3sol}
 \eeq
where the number of $m'$s is $k$ and, thus,  in order to 
have $\Phi_3$ in the Lie algebra of $SU(N)$, 
one must have $\Sigma_{j=1}^{N-k}\tilde{m}_j=km$.  This choice of $\Phi_3$
lead us to take $Q^i$ and $\tilde{Q}_i$ as:
\beq
\begin{tabular}{c c}
$\tilde{Q}_i=\left( 0\cdots 0,\tilde{q}^1_i\cdots,\tilde{q}^{k}_{i}
\right)\ ,$ & \,\,\,$Q^i=\left(\begin{array}{c} 
0\\ \vdots\\0\\ q^i_1\\ \vdots\\ q^i_{k} 
\end{array}\right)\ .$
\end{tabular}
\label{QQtilde-sol}
\eeq
Indeed, it is trivial to check that the values of $\Phi_3$, $\tilde{Q}_i$ and
$Q^i$ displayed in eqs. (\ref{Phi3sol}) and (\ref{QQtilde-sol}) solve eq.
(\ref{Phi3}).  Since the quark VEV in this solution has some components which
are zero and others that are different from zero, this choice of vacuum leads
to a mixed Coulomb-Higgs phase.

The vanishing of the $F$-terms associated to the adjoint 
scalars gives rise to:
   
\begin{equation}
  [\Phi_1,\Phi_3]=[\Phi_2,\Phi_3]=0\ ,
\end{equation}
 together with the equation:  
\begin{equation}
\label{F3}
Q^i\,\tilde{Q}_i+[\Phi_1,\Phi_2]=0\ .
\end{equation}
In (\ref{F3}) $Q^i\,\tilde{Q}_i$ denotes a matrix in color
space of components $Q^i_a\tilde{Q}_i^b$. For a vacuum election as in
eq. (\ref{QQtilde-sol}) we can restrict ourselves to the lower $k\times k$
matrix block, and we can write eq. (\ref{F3}) as:
\begin{equation}
\label{F}
q^i\tilde{q}_i+[\Phi_1,\Phi_2]=0\ ,
\end{equation}
where now, and it what follows,  it is understood that $\Phi_1$ and $\Phi_2$
are $k\times k$ matrices. 

Eq. (\ref{F}) contains an important piece of
information since it shows that a non-vanishing VEV of the quark fields $q$
and $\tilde q$ induces a non-zero commutator of the adjoint fields $\Phi_1$
and $\Phi_2$. Therefore, in the Higgs branch, some scalars transverse to the
D3-brane are necessarily non-commutative. Notice that $\Phi_1$ and $\Phi_2$
correspond precisely to the directions transverse to the D3-brane which lie on
the worldvolume of the D7-brane  (\ie\ they correspond to the directions
$4,\cdots 7$ in the array (\ref{D3D7intersection})). This implies that the
description of this intersection from the point of view of the D7-branes must
involve a non-trivial configuration of the worldvolume gauge field components
of the latter along the directions $4,\cdots 7$. We will argue in the next
subsection that this configuration corresponds to switching on an instantonic
flux along these directions.

In order to match the field theory vacuum with our brane description we
should also be able to reproduce the $D$-flatness condition arising from the
lagrangian (\ref{LFT}). Assuming that the quark fields $\tilde Q$ and $Q$ are
only non-vanishing on the lower $k\times k$ block, we can write this condition
as:
\begin{equation}
\label{D}
|q^i|^2-|\tilde{q}_i|^2+[\Phi_1,\Phi_1^{\dagger}]+[\Phi_2, 
\Phi_2^{\dagger}]=0\ .
\end{equation}
The constraints (\ref{F}) and (\ref{D}), 
together with the condition $[\Phi^I,\Phi^3]=0$,  define the mixed
Coulomb-Higgs phase of the theory.

\subsection{Gravity dual of the mixed Coulomb-Higgs phase}
\label{MacroD3D7}

As it is well-known, there is a one-to-one correspondence between 
the Higgs phase of ${\cal{N}}=2$ gauge theories and the moduli space of
instantons (\cite{MRD1, MRD2, W}). This comes from the fact that the $F$-
and $D$-flatness conditions can be directly mapped into the ADHM
equations (see \cite{Tong} for a review). Because of this map, we can
identify the Higgs phase of the gauge theory with the space of 4d
instantons. In the context of string theory,  a ${\cal{N}}=2$ theory can
be engineered by intersecting Dp with D(p+4) branes over a $p+1$
dimensional space. In particular, if we consider the D3-D7 system, the
low energy effective lagrangian is precisely given by (\ref{LFT}). In
this context, the Higgsing of the theory amounts to adding some units of
instantonic DBI flux in the subspace transverse to the D3 but contained
in the D7, which provides a natural interpretation of the Higgs
phase-ADHM equations map.

Let us analyze this in more detail. Suppose we have $N$ D3-branes
 and $N_f$ D7-branes. In the field theory limit in which we take
$\alpha'$ to zero but keeping fixed the Yang-Mills coupling of the theory
on the D3's, the gauge dynamics on the D7-brane is decoupled. Then, the
$SU(N_f)$ gauge symmetry of the D7-brane is promoted to a global $SU(N_f)$
flavor symmetry on the effective theory describing the system, which is
${\cal N}=4$ SYM plus $N_f$ ${\mathcal{N}}=2$ hypermultiplets arising
from the D3-D7 strings; and whose lagrangian is the one written in
(\ref{LFT}). The gravity dual of this theory would be obtained by replacing
the branes by their backreacted geometry and taking the appropriate low
energy limit. However, in the limit in which $N_f\ll N$ we can consider
the D7-branes as probes in the  near-horizon geometry created by the
D3-branes, namely $AdS_5\times S^5$:
\begin{equation}
ds^2\,=\,\frac{r^2}{R^2}\,\,dx^2_{1,3}+\frac{R^2}{r^2}d\vec{r}^{\,2}\,,
\label{AdS5metric}
\end{equation}
where $\vec{r}$ is the six-dimensional  vector along the directions
orthogonal to the stack of D3-branes and the radius $R$ is given by
$R^4\,=\,4\pi\, g_s\,N\,(\alpha')^2$. In addition,
$dx^2_{1,3}$ is the metric of the $3+1$ dimensional Minkowski space along
which the D3-branes lie. The type IIB supergravity background also includes 
a 4-form RR potential given by:
\begin{equation}
C^{(4)}\,=\,\Bigg(\,\frac{r^2}{R^2}\,\Bigg)^2\ dx^0\wedge\cdots\wedge
dx^3\ .
\label{C4}
\end{equation}

Let us now write the  $AdS_5\times S^5$ background in a system of
coordinates more suitable four our purposes. Let 
$\vec y\,=\,(\,y^1,\cdots, y^4\,)$ be the coordinates along the directions
$4,\cdots,7$ in the array (\ref{D3D7intersection}) and let us denote by
$\rho$ the length of $\vec y$ (\ie\ $\rho^2\,=\,\vec y\cdot\vec y$).
Moreover, we will call $\vec z=(z^1,z^2)$  the coordinates $8,9$ of 
(\ref{D3D7intersection}). Notice that  $\vec z$   is a vector in the
directions which are orthogonal to both stacks of D-branes. Clearly, 
$r^2\,=\,\rho^2\,+\,\vec{z}^{\,2}$ and the metric (\ref{AdS5metric}) can be
written as:
\begin{equation}
\label{bckgr}
ds^2\,=\,
\frac{\rho^2+\vec{z}^{\,2}}{R^2}\,\,dx^2_{1,3}
+\frac{R^2}{\rho^2+\vec{z}^{\,2}}(d\vec{y}^{\,2}+d\vec{z}^{\,2})\,.
\end{equation}

The Dirac-Born-Infeld (DBI) action for a stack of $N_f$ D7-branes is given by:
\beq
S_{DBI}^{D7}\,=\,-T_7\,\int\,d^8\xi\,e^{-\phi}\,\,
{\rm Str}\,\Bigg\{\sqrt{-\det\, (\,g+F\,)}\Bigg\}\,\,,
\label{DBI-D7-general}
\eeq
where $\xi^a$ is a system of worldvolume coordinates, $\phi$ is the dilaton,
$g$ is the induced metric and $F$ is the field strength of the $SU(N_f)$
worldvolume gauge group\footnote{Notice that, with our notations, $F_{ab}$ is
dimensionless and, therefore, the relation between $F_{ab}$ and the gauge
potential $A$ is 
$F_{ab}=\partial_aA_b-\partial_bA_a+{1\over 2\pi\alpha'}\, [A_a,A_b]$,
whereas the gauge covariant derivative is $D_a=\partial_a+{1\over
2\pi\alpha'}\, A_a$.}. Let us assume that we take 
$\xi^a\,=\,(x^{\mu}\,,\,y^i\,)$ as worldvolume coordinates and that we
consider a D7-brane embedding in which $|\vec{z}|=L$, where $L$ represents
the constant transverse separation between the two stacks of D3- and D7-
branes. Notice that this transverse separation will give a mass 
$L/2\pi\alpha'$ to the D3-D7 strings, which corresponds to the quark mass
in the field theory dual.  For
an embedding with $|\,\vec z\,|=L$, the induced metric takes the form:
\beq
g_{x^{\mu} x^{\nu}}\,=\,{\rho^2\,+\,L^2\over R^2}\,\,\eta_{\mu\nu}\,\,,
\qquad\qquad
g_{y^{i} y^{j}}\,=\,{R^2\over \rho^2\,+\,L^2}\,\,\delta_{ij}\,\,.
\label{inducedgD3-D7}
\eeq
Let us now assume that the worldvolume field strength $F$ has non-zero
entries only along the directions of the $y^i$ coordinates and let us denote
$F_{y^iy^j}$ simply by $F_{ij}$. 
Then, after using
eq. (\ref{inducedgD3-D7}) and the fact that the dilaton is trivial for the
$AdS_5\times S^5$ background, the DBI action (\ref{DBI-D7-general}) takes the
form: 
\begin{equation}
S_{DBI}^{D7}\,=-T_{7}\int\,d^4x\,d^4y\,
 {\rm Str}\,\Bigg\{
\sqrt{\,\det\,\Bigg(\,\delta_{ij}+\Bigg(\frac{\rho^2\,+\,L^2}{R^2}\Bigg)
F_{ij}\Bigg)}\,\,\Bigg\}\ .
\label{DBI-D3D7-reduced}
\end{equation}
The matrix appearing on the right-hand side of eq. (\ref{DBI-D3D7-reduced}) is
a $4\times 4$ matrix whose entries are $SU(N_f)$ matrices. However, inside
the symmetrized trace such matrices can be considered as commutative numbers.
Actually, we will evaluate the determinant in (\ref{DBI-D3D7-reduced}) by
means of the following identity. Let $M_{ij}=-M_{ji}$ be a $4\times 4$
antisymmetric matrix. Then, one can check that:
\beq
\det (\,1\,+\,M\,)\,=\,1\,+\,{1\over 2}\,M^2\,+\,{1\over 16}\,
\,(\,{}^*M\,M\,)^2\,\,,
\label{matrix-identity}
\eeq
where $M^2$ and ${}^*M\,M$ are defined as follows:
\beq
M^2\,\equiv\,M_{ij}\,M_{ij}\,\,,
\qquad\qquad
{}^*M\,M\equiv\,{}^*M_{ij}\,M_{ij}\,\,,
\label{MM}
\eeq
and ${}^*M$ is defined as the following matrix: 
\beq
{}^*M_{ij}\,=\,{1\over 2}\,\epsilon_{ijkl}\,M_{kl}\,\,.
\label{*M}
\eeq
When the $M_{ij}$ matrix is self-dual  (\ie\ when ${}^*M=M$), the three terms
on the right-hand side of (\ref{matrix-identity}) build up a perfect square.
Indeed, one can check by inspection that, in this case, one has:
\beq
\det (\,1\,+\,M\,)\Big|_{{\rm self-dual}}\,=\,\Bigg(\,
1\,+\,{1\over 4}\,M^2\,\Bigg)^2\,\,.
\label{detM(s-d)}
\eeq

Let us apply these results to our problem. First of all, 
by using 
(\ref{matrix-identity}) one can rewrite eq. (\ref{DBI-D3D7-reduced}) as:
\begin{equation}
S_{DBI}^{D7}\,=\,-T_{7}\int\, d^4x\,d^4y\,\,
{\rm Str}\,\Bigg\{
\sqrt{\,1+{1\over 2}\,\Bigg(\frac{\rho^2\,+\,L^2}{R^2}\Bigg)^2 F^2
+\frac{1}{16}\Bigg(\frac{\rho^2\,+\,L^2}{R^2}\Bigg)^{4}
\Big(\,{}^*FF\,\Big)^2}\,\,\Bigg\}\,.
\label{DBI-D3D7-explicit}
\end{equation}

Let us now consider the Wess-Zumino(WZ) piece of the worldvolume action. For
a D7-brane in the $AdS_5\times S^5$ background this action reduces to:
\beq
S_{WZ}^{D7}\,=\,{T_{7}\over 2}\,\int\, \,
{\rm Str}\,\Bigg[\,P[\,C^{(4)}\,]\wedge F\wedge F\,\Bigg]\,\,,
\label{WZactionD7}
\eeq
where $P[\,\cdots\,]$ denotes the pullback of the form inside the brackets to
the worldvolume of the D7-brane. By using the same set of coordinates as in
(\ref{DBI-D3D7-reduced}), and the explicit expression of $C^{(4)}$ (see eq.
(\ref{C4})), one can rewrite $S_{WZ}^{D7}$ as:
\begin{equation}
\label{WZ}
S_{WZ}^{D7}\,=\,T_{7}\int\, \, d^4x\,d^4y\,\,
{\rm Str}\, \Bigg\{\frac{1}{4}\Bigg(\,\frac{\rho^2\,+\,L^2}{R^2}\,\Bigg)^{2}
\,\,{}^*FF\,\Bigg\}\ .
\end{equation}

Let us now consider a configuration in which the worldvolume gauge field is
self-dual in the internal $\RR^4$ of the worldvolume spanned by the $y^i$
coordinates which, as one can check, satisfies the equations of motion of
the D7-brane probe.  For such an instantonic gauge configuration
${}^*F=F$, where 
${}^*F$ is defined as in eq. (\ref{*M}). As in eq. (\ref{detM(s-d)}), 
when $F={}^*F$ the DBI action 
(\ref{DBI-D3D7-explicit}) contains the square root of a perfect square and
we can write:
\beq
S_{DBI}^{D7}({\rm self}-{\rm dual})\,=\,-T_7\,
\int\, \, d^4x\,d^4y\,\,
{\rm Str}\, \Bigg\{\,1\,+\,\frac{1}{4}\,
\Bigg(\,\frac{\rho^2\,+\,L^2}{R^2}\,\Bigg)^{2}\,\,
{}^*FF\,\Bigg\}\,\,.
\label{SDBI-sd}
\eeq
Moreover, by comparing eqs. (\ref{WZ}) and (\ref{SDBI-sd}) one readily
realizes that the WZ action cancels against the second term of the right-hand
side of eq. (\ref{SDBI-sd}). To be more explicit, once we
assume the instantonic character of $F$, the full action for a self-dual
configuration is just:
\beq
S^{D7}({\rm self}-{\rm dual})\,=\,-T_7\,\int\, \, d^4x\,d^4y\,\,
{\rm Str}\, \big[1\big]\,=\,-T_7\,N_f\,\int\, \, d^4x\,d^4y\,\,.
\label{totalaction}
\eeq
Notice that in the total action (\ref{totalaction}) the transverse distance
$L$ does not appear.  This ``no-force" condition is an explicit manifestation
of the SUSY of the system. Indeed, the fact that the DBI action is a square
root of  a perfect square is required for supersymmetry, and actually can be
regarded as the saturation of  a BPS bound.

In order to get a proper interpretation of the role of the instantonic gauge
field on the D7-brane probe, let us recall that for self-dual configurations
the integral of the Pontryagin density ${\cal P}(y)$ is quantized for
topological reasons. Actually, with our present normalization of $F$, 
${\cal P}(y)$ is given by:
\beq
{\cal P}(y)\,\equiv\,\frac{1}{16\pi^2}\,{1\over (2\pi\alpha')^2}\,\,
{\rm tr}\,\Big[\,{}^*FF\,\Big]\,\,,
\label{Poyntriagin}
\eeq
and, if $k\in\ZZ$ is the instanton number, one has:
\begin{equation}
\int d^4y\,\,{\cal P}(y)\,
\,= \,k\ .
\label{instanton-number}
\end{equation}
A worldvolume gauge field satisfying
(\ref{instanton-number}) is inducing $k$ units  of D3-brane charge into the
D7-brane worldvolume along the subspace spanned by the Minkowski coordinates
$x^{\mu}$. To verify this fact, let us rewrite the WZ action
(\ref{WZactionD7}) of the D7-brane as:
\begin{equation}
S_{WZ}^{D7}\,=\,{T_{7}\over 4}\,\int\,d^4x\,d^4y\,
C_{x^0x^1x^2x^3}^{(4)}\,\,
{\rm tr}\,\Big[\,{}^*FF\,\Big] 
\,=\,T_3\int
d^4x\,d^4y\,\,C_{x^0x^1x^2x^3}^{(4)}\,\,
{\cal P}(y)\,\,,
\label{D3induced}
\end{equation}
where we have used (\ref{Poyntriagin}) and the relation
$T_3=(2\pi)^4\,(\alpha')^2\,T_7$ between the tensions of the D3- and
D7-branes. If $C_{x^0x^1x^2x^3}^{(4)}$ does not depend on the coordinate $y$,
we can integrate over $y$ by using eq. (\ref{instanton-number}), namely:
\beq
S_{WZ}^{D7}\,=\,k\,T_3\,\int d^4x\,C_{x^0x^1x^2x^3}^{(4)}\,\,.
\label{D3charge}
\eeq
Eq. (\ref{D3charge}) shows that the coupling of the D7-brane with
$k$ instantons in the worldvolume to the RR potential $C^{(4)}$ of the
background is identical to the one corresponding to $k$ D3-branes, as claimed
above.  It is worth to remark here that the existence of these instanton
configurations relies on the fact that we are considering $N_f>1$ flavor $D7$
branes, \ie\ that we have a non-abelian worldvolume gauge theory.

\subsection{A microscopical interpretation of the D3-D7 intersection with
flux}
\label{microD3D7}

The fact that the D7-branes carry $k$ dissolved D3-branes on them opens up
 the possibility of a new perspective on the system, which could be regarded
not just from the point of view of the D7-branes with dissolved D3s, but
also from the point of view of the dissolved D3-branes which expand due to
dielectric effect \cite{M} to a transverse fuzzy $\RR^4$. To see this,
let us assume that we have a stack of $k$ D3-branes
in the background given by (\ref{bckgr}). These D3-branes are extended along
the four Minkowski coordinates $x^{\mu}$, whereas the transverse coordinates  
$\vec{y}$ and $\vec{z}$ must be regarded as the matrix scalar fields $Y^i$ and
$Z^j$,  taking values in the adjoint representation of  $SU(k)$. Actually, we
will assume in what follows that the $Z^j$ scalars are abelian, as it
corresponds to a configuration in which the D3-branes are localized (\ie\
not polarized) in the  space transverse to  the D7-brane.

The dynamics of a stack of coincident D3-branes is determined by the Myers
dielectric action \cite{M}, which is the sum of a Dirac-Born-Infeld and
a Wess-Zumino part:
\beq
S_{D3}\,=\,S_{DBI}^{D3}\,+\,S_{WZ}^{D3}\,\,.
\eeq
For the background we are considering the Born-Infeld action is:
\beq
S_{DBI}^{D3}\,=\, -T_3\,\int\,\,d^4\xi\,\, {\bf Str}\,\Bigg[\,
\sqrt{-\det\bigg[ P[G+G(Q^{-1}-\delta)G]_{ab}\,\bigg]}\,
\sqrt{\det Q}\,\Bigg]\,\,,
\label{dielectricBI}
\eeq
In eq. 
(\ref{dielectricBI})  $G$ is the background metric, 
${\bf Str}(\cdots)$
represents the symmetrized trace over the $SU(k)$ indices 
and $Q$ is a matrix which depends on
the commutator of the transverse scalars (see below). The Wess-Zumino term for
the D3-brane in the $AdS_5\times S^5$ background under consideration is:
\beq
S_{WZ}^{D3}\,=\,
T_{3}\,\int d^4\xi \,\,\,{\bf Str}\,
\bigg[\,P\big[\,C^{(4)}\,\big]\,\bigg]\,\,.
\label{dielectricWZ}
\eeq
As we are assuming that only the $Y$ scalars are non-commutative, 
the  only elements of the matrix $Q$ appearing in (\ref{dielectricBI}) that
differ from those of the  unit matrix are
given by:
\beq
Q_{y^iy^j}\,=\,\delta_{ij}\,+\,{i\over 2\pi\alpha'}\,
[Y^i,Y^k]\,G_{y^ky^j}\,\,.
\eeq
By using the explicit form of the metric elements along the $y$ coordinates 
(see eq. (\ref{bckgr})), one can rewrite $Q_{ij}$ as:
\beq
Q_{y^iy^j}\,=\,\delta_{ij}\,+\,{i\over 2\pi\alpha'}\,
{R^2\over \hat r^{\,2}}\,\,[\,Y^i\,,\,Y^j\,]\,\,,
\eeq
where $\hat r^{\,2}$ is the matrix:
\beq
\hat r^{\,2}\,=\,(\,Y^i\,)^2\,+\,Z^2\,\,.
\eeq
Let us now define the matrix $\theta_{ij}$ as: 
\beq
i\theta_{ij}\,\equiv\,
{1\over 2\pi\alpha'}\,[Y^i,Y^j]\,\,.
\label{thetaij}
\eeq
It follows from this definition that $\theta_{ij}$ is antisymmetric in the
$i,j$ indices and, as  a matrix of $SU(k)$, is hermitian:
\beq
\theta_{ij}\,=\,-\theta_{ji}\,\,,\qquad\qquad
\theta_{ij}^{\dagger}\,=\,\theta_{ij}\,\,.
\eeq
Moreover, in terms of $\theta_{ij}$, the matrix $Q_{ij}$ can be written as:
\beq
Q_{y^iy^j}\,=\,\delta_{ij}\,-\,{R^2\over \hat r^{\,2}}\,\theta_{ij}\,\,.
\eeq
Using these definitions, we can write the DBI action (\ref{dielectricBI})
for the dielectric 
D3-brane in the $AdS_5\times S^5$ background as:
\beq
S_{DBI}^{D3}\,=\, -T_3\,\int\,\,d^4x\,\,{\bf Str}\,
\Bigg[\,\bigg(\,{\hat r^{\,2}\over R^2}\,\bigg)^2\,
\sqrt{\det\bigg(
\,\delta_{ij}\,-\,{R^2\over \hat r^{\,2}}\,\theta_{ij}\,\bigg)}
\,\,\Bigg]\,\,,
\label{DBI-dielectricD3}
\eeq
where we have chosen the Minkowski coordinates $x^{\mu}$ as our set of
worldvolume coordinates for the dielectric D3-brane.  Similarly, the WZ term
can be written as:
\beq
S_{WZ}^{D3}\,=\,
T_{3}\,\int d^4x \,\,\,{\bf Str}\,
\bigg[\,\bigg(\,{\hat r^{\,2}\over R^2}\,\bigg)^2
\,\bigg]\,\,.
\label{SWZ-D3}
\eeq
Let us now assume that the matrices $\theta_{ij}$ are self-dual with respect
to the $ij$ indices, \ie\ that ${}^*\theta\,=\,\theta$. Notice that, in terms
of the original matrices $Y^i$, this is equivalent to the condition:
\begin{equation}
\label{selfduality}
[Y^i,Y^j]=\frac{1}{2}\epsilon_{ijkl}[Y^k,Y^l]\ . 
\end{equation}
Moreover, the self-duality condition implies that there are three independent
$\theta_{ij}$ matrices, namely:
\beq
\theta_{12}\,=\,\theta_{34}\,\,,\qquad\qquad
\theta_{13}\,=\,\theta_{42}\,\,,\qquad\qquad
\theta_{14}\,=\,\theta_{23}\,\,.
\label{sd-explicit}
\eeq

The description of the D3-D7 system from the perspective of the color
D3-branes should match the field theory analysis performed at the beginning
of this section. In particular, the D- and F-flatness conditions of the
adjoint fields in the Coulomb-Higgs phase of the ${\cal N}=2$ SYM with flavor
should be the same as the ones satisfied by the transverse scalars of the
dielectric D3-brane. In order to check this fact,  let
us define the following complex combinations of the $Y^i$ matrices:
\beq
2\pi\alpha'\,\Phi_1\,\equiv\,{Y^1+iY^2\over \sqrt{2}}\,\,,
\qquad\qquad
2\pi\alpha'\,\Phi_2\,\equiv\,{Y^3+iY^4\over \sqrt{2}}\,\,,
\label{Phi12}
\eeq
where we have introduced the factor $2\pi\alpha'$ to take into account the
standard relation between coordinates and scalar fields in string theory. 
We are going to identify $\Phi_1$ and $\Phi_2$ with the adjoint scalars
of the field theory side.  To verify this identification, let us compute the
commutators of these matrices and, as it was done in \cite{ARR}, let us
match them with the ones obtained from the F-flatness conditions of the
field theory analysis.  From the definitions (\ref{thetaij}) and 
(\ref{Phi12}) and the self-duality condition (\ref{sd-explicit}), 
it is straightforward to check that:
\bear
&&[\,\Phi_1\,,\Phi_2\,]\,=\,-{\theta_{23}\over 2\pi\alpha'}
\,+\,i{\theta_{13}\over 2\pi\alpha'}
\,\,,\rc\rc
&&[\,\Phi_1\,,\Phi_1^{\dagger}\,]\,=\,[\,\Phi_2\,,\Phi_2^{\dagger}\,]\,=\,
{\theta_{12}\over 2\pi\alpha'}\,\,.
\eear
By comparing with the results of the field theory analysis
(eqs. (\ref{F}) and (\ref{D})), we get the
following identifications between the $\theta$'s and the vacuum expectation
values of the matter fields:
\beq
q^i\tilde{q}_i\,=\,{\theta_{23}\over 2\pi\alpha'}\,-\,i
{\theta_{13}\over 2\pi\alpha'}\,\,,
\qquad\qquad
|\tilde{q}_i|^2\,-\,|q^i|^2\,=\,{\theta_{12}\over \pi\alpha'}\,\,.
\label{q-theta}
\eeq
Moreover, from the point of view of this dielectric description, the $\Phi_3$
field in  the field theory is proportional to  $Z^1+iZ^2$. Since the stack of
branes is localized in that directions, $Z^1$ and $Z^2$ are abelian and
clearly we have that
$[\Phi_1,\Phi_3]=[\Phi_2,\Phi_3]=0$, thus matching the last $F$-flatness
condition for the adjoint field $\Phi_3$.

It is also interesting to relate the present ``microscopic" description of
the D3-D7 intersection, in terms of a stack of dielectric D3-branes, to the
``macroscopic" description of subsection \ref{MacroD3D7}, in terms of the
flavor D7-branes. With this purpose in mind, let us compare the actions of
the D3- and D7-branes. First of all, we notice that, when the matrix $\theta$
is self-dual, we can use eq. (\ref{detM(s-d)}) and write the DBI action
(\ref{DBI-dielectricD3}) as:
\beq
S_{DBI}^{D3}({\rm self}-{\rm dual})\,=\,-T_3\int\,d^4x
\,\,{\bf Str}\,\bigg[\,\bigg(\,{\hat r^{\,2}\over R^2}\,\bigg)^2\,+\,
{1\over 4}\,\theta^2
\,\bigg]\,\,.
\label{SD3-sd}
\eeq
Moreover, by inspecting eqs. (\ref{SWZ-D3}) and (\ref{SD3-sd}) we discover
that the WZ action cancels against the first term of the right-hand side of
(\ref{SD3-sd}), in complete analogy to what happens to the D7-brane. Thus, one
has:
\begin{equation}
S^{D3}({\rm self}-{\rm dual})\,=\,-\,{T_3\over 4}\,\,
\int d^{4}x\, {\bf Str}\,\big[\,\theta^2\,\big]\,=\,-
\pi^2\,T_7\,(\,2\pi\alpha'\,)^2
\int d^{4}x\, {\bf Str}\,\big[\,\theta^2\,\big]\,\,,
\label{completeactionD3}
\end{equation}
where, in the last step, we have rewritten the result in terms of the tension
of the D7-brane. Moreover, an important piece of information is obtained by
comparing the WZ terms of the D7- and D3-branes (eqs. (\ref{D3induced}) and
(\ref{SWZ-D3})). Actually, from this comparison we can establish a map
between matrices in the D3-brane description and functions of the $y$
coordinates in the D7-brane approach. Indeed, let us suppose that $\hat f$
is a $k\times k$ matrix and let us call $f(y)$ the function to which  $\hat
f$ is mapped. It follows from the identification between the D3- and
D7-brane WZ actions that the mapping rule is:
\begin{equation}
{\bf Str} [\,\hat f\,]\,\,\Rightarrow\,\, \int d^4y\, {\cal P}(y)\,
f(y)\,\,,
\label{micro-macro}
\end{equation}
where the kernel ${\cal P}(y)$ on the right-hand side of (\ref{micro-macro})
is the Pontryagin density defined in eq. (\ref{Poyntriagin}). Actually, the
comparison between both WZ actions tells us that the matrix $\hat r^2$ is
mapped to the function $\vec y^{\,2}+\vec z^{\,2}$. Notice also that, when
$\hat f$ is the unit
$k\times k$ matrix and $f(y)=1$, both sides of (\ref{micro-macro}) are 
equal to the instanton number $k$ (see eq. (\ref{instanton-number})). Another
interesting information comes by comparing the complete actions of the D3-
and D7-branes. It is clear from (\ref{completeactionD3}) and
(\ref{totalaction}) that:
\beq
(\,2\pi\alpha'\,)^2\,{\bf Str} [\,\theta^2\,]\,\,
\Rightarrow\,\, \int\,d^4y\,\,{N_f\over \pi^2}\,\,.
\label{theta-map}
\eeq
By comparing eq. (\ref{theta-map}) with the general relation
(\ref{micro-macro}), one gets the function that corresponds
to the matrix $\theta^2$, namely:
\beq
(\,2\pi\alpha'\,)^2\,\theta^2\,\,
\Rightarrow \,\,{N_f\over \pi^2 \,\,{\cal P}(y)}\,\,.
\label{theta-instanton}
\eeq
Notice that $\theta^2$ is a measure of the non-commutativity of the adjoint
scalars in the dielectric approach, \ie\ is a quantity that characterizes the
fuzziness of the space transverse to the D3-branes. Eq.
(\ref{theta-instanton}) is telling us that this fuzziness is related to the 
(inverse of the) Pontryagin density for the macroscopic D7-branes. Actually,
this identification is reminiscent of the one found in ref. \cite{SW} between
the non-commutative parameter and the NSNS B-field in the string theory
realization of non-commutative geometry. Interestingly, in our case the
commutator matrix $\theta$ is related to the VEV of the matter fields $q$ and
$\tilde q$ through the F- and D-flatness conditions (\ref{F}) and (\ref{D}). 
Notice that eq. (\ref{theta-instanton}) implies that the quark VEV is somehow
related to the instanton density on the flavor brane. In order to make this
correspondence more precise, let us consider the one-instanton configuration
of the $N_f=2$ gauge theory on the D7-brane worldvolume. In the so-called
singular gauge, the $SU(2)$ gauge field is given by:
\begin{equation}
{A_i\over 2\pi\alpha'}\,=\,2i\Lambda^2
\frac{\bar{\sigma}_{ij}\,\, y^j}{\rho^2(\rho^2+\Lambda^2)}\,\,,
\label{instanton-potential}
\end{equation}
where $\rho^2=\vec y\cdot\vec y$,
$\Lambda$ is a constant (the instanton size) and
the matrices $\bar{\sigma}_{ij}$ are defined as:
\beq
\bar{\sigma}_{ij}\,=\,{1\over 4}\,\,\big(\,\bar\sigma_i\,\sigma_j\,-\,
\bar\sigma_j\,\sigma_i\,\big)\,\,,\qquad
\sigma_i\,=\,(i\vec \tau\,,\,1_{2\times 2}\,)\,,\qquad
\bar\sigma_i\,=\,\sigma_i^{\dagger}\,=\,
(-i\vec \tau\,,\,1_{2\times 2}\,)\,\,.
\label{sigmaij}
\eeq
In (\ref{sigmaij}) the $\vec \tau$'s are the Pauli matrices. Notice that we
are using a convention in which the $SU(2)$ generators are hermitian as a
consequence of the relation  
$\bar{\sigma}_{ij}^{\dagger}\,=\,-\bar{\sigma}_{ij}$.  The non-abelian
field strength $F_{ij}$ for the gauge potential  $A_i$ in 
(\ref{instanton-potential}) can be easily computed, with the result:
\beq
{F_{ij}\over 2\pi\alpha'}\,=\,-{4i\Lambda^2\over (\,
\rho^2\,+\,\Lambda^2)^2}\,\,
\bar{\sigma}_{ij}\,-\,
{8i\Lambda^2\over \rho^2\, (\, \rho^2\,+\,\Lambda^2)^2}\,
\big(\,y^i\,\bar{\sigma}_{jk}\,-\,y^j\,\bar{\sigma}_{ik}\,\big)\,y^k\,\,.
\eeq
Using the fact that the matrices $\bar{\sigma}_{ij}$ are anti self-dual one
readily verifies that $F_{ij}$ is self-dual. Moreover, 
one can prove that:
\beq
{F_{ij}\,F_{ij}\over (\,2\pi\alpha'\,)^2}\,=\,{48 \Lambda^4\over (\,
\rho^2\,+\,\Lambda^2)^4}\,\,,
\label{F2-inst}
\eeq
which gives rise to the following instanton density:
\beq
{\cal P}(y)\,=\,{6\over \pi^2}\,\,{\Lambda^4\over
(\,\rho^2\,+\,\Lambda^2\,)^4}\,\,.
\eeq
As a check one can verify that eq. (\ref{instanton-number}) is satisfied with
$k=1$.

Let us now use this result in (\ref{theta-instanton}) to get some qualitative
understanding of the relation between the Higgs mechanism in field theory and
the instanton density in its holographic description. For simplicity we will
assume that all quark VEVs are proportional to some scale $v$, \ie\ that:
\beq
q, \tilde q\,\sim\,v\,\,.
\eeq
Then, it follows from (\ref{q-theta}) that:
\beq
\theta\,\sim\,\alpha'\,v^2\,\,,
\eeq
and, by plugging this result in (\ref{theta-instanton}) one arrives at the
interesting relation:
\begin{equation}
v\sim \frac{\rho^2+\Lambda^2}{\alpha'\Lambda}\,\, .
\label{holoVEV}
\end{equation}
Eq. (\ref{holoVEV}) should be understood in the holographic sense, \ie\
$\rho$ should be regarded as the energy scale of the gauge theory.
Actually, in the far IR ($\rho\approx 0$) the  relation   (\ref{holoVEV})
reduces to:
\begin{equation}
v\,\sim\,\frac{\Lambda}{\alpha'}\,\,,
\label{v-Lambda}
\end{equation}
which, up to numerical factors, is precisely the relation between the quark
VEV and the instanton size that has been  obtained in \cite{EGG}. Let us now 
consider  the full expression  (\ref{holoVEV})  for $v$. 
For any finite non-zero $\rho$ the quark VEV $v$ is non-zero. Indeed, in both
the  large and small instanton limits  $v$ goes to infinity.
However, in the far IR a subtlety arises, since there the quark VEV goes to
zero in the small instanton limit. This region should be clearly singular,
because a zero quark VEV would mean to unhiggs the theory, which would lead
to the appearance of extra light degrees of freedom.

To finish this subsection, let us notice that
the dielectric effect  considered here is not triggered by the 
influence of any external field other than the metric background. In this
sense it is an example of a purely gravitational dielectric effect, as in
\cite{gravitationaldielectriceffect}.

\subsection{Fluctuations in Dp-D(p+4) with flux}

So far we have seen how we can explicitly map the Higgs phase of the field 
theory to the instanton moduli space in the D7-brane picture through the
dielectric description. In this section we will concentrate on the
macroscopical description and we will consider fluctuations around the
instanton configuration. These fluctuations should correspond to mesons in
the dual field theory.

Since we have a similar situation for all the Dp-D(p+4) intersections,
namely a one to one correspondence between the Higgs phase of the
corresponding field theory and the moduli space of instantons in 4
dimensions, in this section we will work with the general  Dp-D(p+4) system.
Both the macroscopic and the microscopic analysis of the previous section
can be extended in a straightforward manner to the general case, so we will
briefly sketch  the macroscopical computation to set notations, and
concentrate on the fluctuations. In general,  the metric corresponding to a
stack of $N$ Dp-branes in string frame is given by:
\begin{equation}
\label{Dpmetrica}
ds^2\,=\,\Bigg(\frac{r^2}{R^2}\Bigg)^{\alpha}\,\,dx_{1,p}^2\,+\,
\Bigg(\frac{R^2}{r^2}
\Bigg)^{\alpha}\,\,d\vec{r}^{\,\,2}\,\,,
\qquad\qquad\alpha\,=\,\frac{7-p}{4}\,\,,
\end{equation}
where $\vec{r}$ is a $(9-p)$-dimensional vector and $R$ is given by:
\beq
R^{7-p}\,=\,2^{5-p}\,\pi^{{5-p\over 2}}\,
\Gamma\Big({7-p\over 2}\Big)\,g_s\,N\,
(\alpha')^{{7-p\over 2}}\,\,.
\label{RDp}
\eeq
In addition, the type II background generated by the Dp-branes is endowed
with a non-zero dilaton given by:
\begin{equation}
\label{Dpdilaton}
e^{-\Phi}\,=\,\Bigg(\frac{R^2}{r^2}\Bigg)^{\gamma}\,\,,
\qquad\qquad\gamma\,=\,\frac{(7-p)(p-3)}{8}\,\,,
\end{equation}
and there is also a  RR $(p+1)$-form potential, whose expression is:
\begin{equation}
\label{DpRR}
C^{(p+1)}\,=\,\Bigg(\frac{r^2}{R^2}\Bigg)^{2\alpha}\ dx^0\wedge\cdots\wedge
dx^p\,\,,
\end{equation}
where $\alpha$ is the same as in eq. (\ref{Dpmetrica}). 
We will separate again the $\vec{r}$ coordinates in two sets,
namely $\vec r=(\vec y, \vec z)$, where $\vec y$ has four components, and we
will denote $\rho^2=\vec y\cdot\vec y$. As $r^2=\rho^2+\vec{z}^{\,2}$, the
metric (\ref{Dpmetrica}) can be written as:
\begin{equation}
\label{bckgrDp}
ds^2\,=\,\Bigg(\frac{\rho^2+\vec{z}^{\,2}}{R^2}\Bigg)^{\alpha}dx_{1,p}^2
\,+\,
\Bigg(\frac{R^2}{\rho^2+\vec{z}^{\,2}}\Bigg)^{\alpha}(d\vec{y}^{\,2}
+d\vec{z}^{\,2})\,.
\end{equation}
In this background we will consider a stack of $N_f$ D(p+4)-branes
extended along $(x^{\mu}, \vec y)$ at fixed distance $L$ in the transverse
space spanned by the $\vec z$ coordinates (\ie\ with $|\,\vec z\,|=L$). If
$\xi^a=(x^{\mu}, \vec y)$ are the worldvolume coordinates, the action of
a probe D(p+4)-brane is:
\bear
&&S^{D(p+4)}\,=\,-T_{p+4}\int\, d^{p+5}\xi\,\,e^{-\phi}\,
{\rm Str}\,\Bigg\{\,
\sqrt{-\det\big(\,g\,+\,F\,\big)}\Bigg\}\,+\,\rc\rc
&&\qquad\qquad\qquad\qquad
+\,\,{T_{p+4}\over 2}\,\,
\int\, {\rm Str}\,\Bigg\{\,P\bigg(\,C^{(p+1)}\,\bigg)\wedge F\wedge F
\,\Bigg\}\,\,,
\label{SD(p+4)}
\eear
where $g$ is the induced metric and $F$ is the $SU(N_f)$ worldvolume gauge
field strength. In order to write $g$  more compactly, let us define the
function $h$ as follows:
\beq
h(\rho)\,\equiv\,
\Bigg(\frac{R^2}{\rho^2+L^{2}}\Bigg)^{\alpha}\,\,.
\label{h}
\eeq
Then, one can write the non-vanishing elements of the induced metric as:
\beq
g_{x^{\mu}\,x^{\nu}}\,=\,{\eta_{\mu\nu}\over h}\,\,,
\qquad\qquad
g_{y^{i}\,y^{j}}\,=\,h\,\,\delta_{ij}\,\,.
\eeq
Let us now assume that the only non-vanishing components of the worldvolume
gauge field $F$ are those along the $y^i$ coordinates. Following the same
steps as in subsection \ref{MacroD3D7}, the action for the D(p+4)-brane probe
can be written as:
\bear
&&S^{D(p+4)}\,=\,-T_{p+4}\int\, d^4x\,d^4y\,\,
{\rm Str}\,\Bigg\{
\sqrt{\,1+{1\over 2}\,\Bigg(\frac{\rho^2\,+\,L^2}{R^2}\Bigg)^{2\alpha} F^2
+\frac{1}{16}\Bigg(\frac{\rho^2\,+\,L^2}{R^2}\Bigg)^{4\alpha}
\Big(\,{}^*FF\,\Big)^2}\,-\,\rc\rc
&&\qquad\qquad\qquad\qquad\qquad\qquad\qquad\qquad
\,\,-\frac{1}{4}\Bigg(\,\frac{\rho^2\,+\,L^2}{R^2}\,\Bigg)^{2\alpha}
\,\,{}^*FF
\Bigg\}\,,
\label{DBI-DpDp+4-explicit}
\eear
where $F^2$ and ${}^*FF$ are defined as in eqs. (\ref{MM}) and (\ref{*M}). If,
in addition, $F_{ij}$ is self-dual, one can check that the equations of
motion of the gauge field are satisfied and, actually, 
there is a cancellation between the DBI
and WZ parts of the action (\ref{DBI-DpDp+4-explicit}) 
generalizing  (\ref{totalaction}), namely:
\begin{equation}
S^{D(p+4)}({\rm self}-{\rm dual})=-T_{p+4}\int\,{\rm
 Str}[1]\,=\,-N_{f}\,T_{p+4}\int\,d^{p+1}x\int\,d^4y\ .
\end{equation}

We turn now to the analysis of the fluctuations around the self-dual
configuration and the computation of the corresponding meson spectrum 
for this
fluxed Dp-D(p+4) intersection.  We will not compute the whole set of
excitations. Instead, we will focus on the fluctuations of the worldvolume
gauge field, for which we will  write: 
\beq
A=A^{inst}+a\,\,,
\eeq
where $A^{inst}$ is the gauge potential corresponding to a self-dual gauge
field strength $F^{inst}$ and 
$a$ is the fluctuation. The total field strength $F$ reads:
\begin{equation}
F_{ab}=F^{inst}_{ab}\,+\,f_{ab}\,\,,
\end{equation}
with $f_{ab}$ being given by:
\beq
f_{ab}\,=\,\partial_{a}a_{b}-\partial_{b}a_{a}+{1\over 2\pi\alpha'}\,
[A^{inst}_{a},a_{b}]+
{1\over 2\pi\alpha'}\,
[a_{a},A^{inst}_{b}]\,+\,{1\over 2\pi\alpha'}\,[a_{a},a_{b}]\,\,,
\eeq
\noindent where the  indices $a$, $b$  run now over all the worldvolume
directions. Next, let us expand the action  (\ref{SD(p+4)}) in powers of the
field $a$ up to second order. With this purpose in mind,  we rewrite the
square root in the DBI action as:
\beq
\sqrt{-\det\big(\,g\,+\,F^{inst}\,+\,f\,\big)}\,=\,
\sqrt{-\det\big(\,g\,+\,F^{inst}\,\big)}\,\,
\sqrt{\det\big(\,1\,+\,X\,\big)}\,\,,
\label{DetX}
\eeq
where $X$ is the matrix:
\beq
X\,\equiv\,\Big(\,g\,+\,F^{inst}\,\Big)^{-1}\,\,f\,\,.
\label{defX}
\eeq
We will expand the right-hand side of (\ref{DetX}) in powers of $X$ by
using the equation\footnote{The trace used in eqs. (\ref{detX-expansion}) and
(\ref{TrX-TrX2}) should not be confused with the trace over the $SU(N_f)$
indices.}:
\beq
\sqrt{\det\,(1+X)}\,=\,1\,+\,{1\over 2}\,{\rm Tr}\, X\,-\,{1\over 4}\,
{\rm Tr}\, X^2\,+\, {1\over 8}\,\big({\rm Tr}\, X\big)^2\,+\,o(X^3)\,\,.
\label{detX-expansion}
\eeq
To apply this expansion to our problem we need to know previously the value
of $X$, which has been defined in eq. (\ref{defX}). Let us denote by ${\cal
G}$ and 
${\cal J}$ to the symmetric and antisymmetric part of the inverse 
of $g+F^{inst}$, \ie:
\beq
\Big(\,g\,+\,F^{inst}\,\Big)^{-1}\,=\,{\cal G}\,+\,{\cal J}\,\,.
\eeq
One can easily compute the matrix elements of ${\cal G}$, with the result:
\beq
{\cal G}^{\mu\nu}\,=\,h\,\,\eta^{\mu\nu}\,\,,
\qquad\qquad
{\cal G}^{ij}\,=\,{h\over H}\,\delta_{ij}\,\,,
\eeq
where $h$ has been defined in (\ref{h}) and
the function $H$ is given by:
\beq
H\,\equiv\, h^2\,+\,{1\over 4}\,\,\Big(\,F^{inst}\,\Big)^2\,\,.
\label{H-def}
\eeq
Moreover, the non-vanishing elements of ${\cal J}$ are:
\beq
{\cal J}^{ij}\,=\,-{F^{inst}_{ij}\over H}\,\,.
\eeq
Using these results one can easily obtain the expression of $X$ and
the traces of its powers appearing on the right-hand side of
(\ref{detX-expansion}), which are given by:
\bear
&&{\rm Tr}\, X\,=\,{1\over H}\,\,F^{inst}_{ij}\,f_{ij}\,\,,\rc\rc
&&{\rm Tr}\, X^2\,=\,-h^2\,f_{\mu\nu}\,f^{\mu\nu}\,-\,{2h^2\over H}\,
f_{i\mu}\,f^{i\mu}\,-\,{h^2\over H^2}\,\,f_{ij}\,f^{ij}\,+\,
{1\over H^2}\, F^{inst}_{ij}\,F^{inst}_{kl}\,f^{jk}\,f^{li}\,\,.
\label{TrX-TrX2}
\eear
By using these results we get, 
after a straightforward computation, the action up to quadratic order in the 
fluctuations, namely:
\begin{eqnarray}
S^{D(p+4)}&=&-T_{p+4}\int\, {\rm Str}\,
\Bigg\{ 1+\frac{H}{4}f_{\mu\nu}
f^{\mu\nu}+\frac{1}{2}\,f_{i\mu}f^{i\mu}+\frac{1}{4H}f_{ij}f^{ij}\,\,+\rc\rc
&+&\frac{1}{8h^2H}(F^{ij}f_{ij})^2-\frac{1}{4h^2H}\,F^{ij}F^{kl}f_{jk}f_{li}
\,-\,\frac{1}{8h^2}\,\,f_{ij}f_{kl}\epsilon^{ijkl}\Bigg\}\,\,,
\end{eqnarray}
where  we are dropping the superscript in
the instanton field strenght.

From now on we will assume again that $N_f=2$ and that the unperturbed
configuration is the one-instanton $SU(2)$ gauge field written in eq. 
(\ref{instanton-potential}). As in ref. \cite{EGG}, we will concentrate on the
subset of fluctuations
for which $a_i=0$, \ie\ on those for which the fluctuation field $a$ has
non-vanishing  components only along the Minkowski directions.  However, we
should impose this ansatz at the level of the equations of motion in order
to ensure the consistency of the truncation. Let us consider  first  the
equation of motion for $a_i$, which 
after imposing $a_i=0$ reduces to:
\beq
D_i\,\partial^{\mu}\,a_{\mu}\,=\,0\,\,.
\label{ai-eom}
\eeq
Moreover,  the equation for $a_{\mu}$ when $a_i=0$ becomes: 
\beq
H\,D^{\mu}\,f_{\mu\nu}\,+D^i\,f_{i\nu}\,=\,0\,\,,
\label{amu-eq}
\eeq
where now $H$ is given in (\ref{H-def}), with $\big(\,F^{inst}\,\big)^2$ as
in (\ref{F2-inst}).  Eq. (\ref{ai-eom}) is solved by requiring:
\beq
\partial^{\mu}\,a_{\mu}\,=\,0\,\,.
\label{transv}
\eeq
Using this result, eq. (\ref{amu-eq}) can be written as:
\beq
H\,\partial^{\mu}\partial_{\mu}\,a_{\nu}\,+\,\partial_i\partial_i\,a_{\nu}\,+
\,\partial^i\,\Big[\,{A_i\over 2\pi\alpha'}\,,\,a_{\nu}\,\Big]\,+\,
\Big[\,{A_i\over 2\pi\alpha'}\,,\partial_ia_{\nu}\,\Big]\,+\,
\Big[\,{A_i\over 2\pi\alpha'}\,,\Big[\,{A_i\over 2\pi\alpha'}\,,a_\nu
\Big]\Big]\,=\,0\,\,.
\label{fluc-eom-DpDp+4}
\eeq
Let us now adopt the following ansatz for $a_{\mu}$:
\beq
a_{\mu}^{(l)}\,=\,\xi_{\mu}(k)\,f(\rho)\,\,e^{ik_{\mu}x^{\mu}}\,\tau^l\,\,,
\label{amu-ansatz}
\eeq
where $\tau^l$ is a Pauli matrix. 
This ansatz solves eq. (\ref{transv}) provided the following transversality
condition is fulfilled:
\beq
k^{\mu}\,\xi_{\mu}\,=\,0\,\,.
\eeq
Moreover, one can  check that, for this ansatz, one has:
\bear
&& \partial^i\,\big[\,A_i\,,\,a_{\nu}^{(l)}\,\big]\,=\,
\big[\,A_i\,,\partial_ia_{\nu}^{(l)}\,\big]\,=\,0\,\,,\rc\rc
&&\Big[\,{A_i\over 2\pi\alpha'}\,,\Big[\,{A_i\over 2\pi\alpha'}\,,a_\nu^{(l)}
\Big]\Big]\,=\,-\,\frac{8\Lambda^4}{\rho^2(\rho^2+\Lambda^2)^2}\,\,
\xi_{\nu}(k)\,f(\rho)\,\,e^{ik_{\mu}x^{\mu}}\,\tau^l\,\,.
\eear
Let us now use these results in eq. (\ref{fluc-eom-DpDp+4}). Denoting 
$M^2=-k^2$ (which will be identified with the mass of the meson in the dual
field theory) and using eq. (\ref{F2-inst}) to compute the function $H$
(see eq.  (\ref{H-def})), one readily reduces (\ref{fluc-eom-DpDp+4}) to the
following second-order differential equation for the function $f(\rho)$
of the ansatz (\ref{amu-ansatz}):
\begin{equation}
\label{e1}
\Bigg[\,\frac{R^{4\alpha}M^2}{(\rho^2+L^2)^{2\alpha}}
\Big(1+\frac{12(2\pi\alpha')^2\Lambda^4}
{R^{4\alpha}}\frac{(\rho^2+L^2)^{2\alpha}}{(\rho^2+\Lambda^2)^4}\Big)
-\frac{8\Lambda^4}{\rho^2(y^2+\Lambda^2)^2}+\frac{1}{\rho^3}
\partial_\rho(\rho^3\partial_\rho)\,\Bigg]f=0\ .
\end{equation}
In order to analyze eq. (\ref{e1}), let us introduce a new radial variable
$\varrho$ and a reduced mass $\bar M$, which are related to $\rho$ and $M$ as:
\beq
\rho=L\varrho\,\,,\qquad\qquad
\bar{M}^2=R^{7-p}L^{p-5}M^2
\,\,.
\eeq
Moreover, it is interesting to rewrite the fluctuation equation in terms of
field theory quantities. Accordingly, let us introduce the 
quark mass $m_q$ and its VEV $v$ as follows:
\beq
m_q\,=\,{L\over 2\pi\alpha'}\,\,,
\qquad
v\,=\,{\Lambda\over 2\pi\alpha'}\,\,.
\eeq
Notice that the relation between $v$ and the instanton size $\Lambda$ is
consistent with our analysis of subsection \ref{microD3D7} (see eq.
(\ref{v-Lambda})) and with the proposal of ref. \cite{EGG}. On the other
hand,  the Yang-Mills coupling $g_{YM}$ is given by:
\beq
g^{2}_{YM}\,=\,(2\pi)^{p-2}\,(\alpha')^{{p-3\over 2}}\,\,g_s\,\,,
\label{gym}
\eeq
and the effective dimensionless coupling $g_{eff}(U)$ at the energy scale $U$
is \cite{IMSY}:
\begin{equation}
g_{eff}^2(U)\,=\,g^{2}_{YM}\, N\,U^{p-3}\,\,.
\label{effcoupling}
\end{equation}
It is now straightforward to use these definitions to rewrite eq. (\ref{e1})
as:
\begin{eqnarray}
\label{fluctuationsDp+4}
&&\Bigg[\,\frac{\bar{M}^2}{(1+\varrho^2)^{2\alpha}}\Bigg(\,1+
c_{p}(v,m_q)\,
\frac{(1+\rho^2)^{2\alpha}}{(\varrho^2+(\frac{v}{m_q})^2)^4}\,\Bigg)-
\Bigg(\frac{v}{m_q}\Bigg)^4\frac{8}{\varrho^2(\varrho^2+(\frac{v}{m_q})^2)^2}
\,\,+\rc\rc
&&\qquad\qquad\qquad\qquad
+\frac{1}{\rho^3}\partial_{\varrho}(\varrho^3\partial_{\varrho})\,
\Bigg]\,f=0\,\,,
\end{eqnarray}
where $c_{p}(v,m_q)$ is defined as:
\beq
c_{p}(v,m_q)\,\equiv\,
\frac{12\cdot 2^{p-2}
\pi^{\frac{p+1}{2}}}{\Gamma(\frac{7-p}{2})}\,\,
{v^4\over g_{eff}^2(m_q)\,\,m_q^4}\,\,.
\label{cp}
\eeq
Notice that everything conspires to absorb the powers of $\alpha'$ and give
rise to  the effective coupling at the 
quark mass in $c_{p}(v,m_q)$.

The equation (\ref{fluctuationsDp+4}) differs in the $\bar{M}$ term  from the
one obtained in
 \cite{EGG}, where  the term proportional to
$c_{p}(v,m_q)$ is absent. We would like to point out that in order to arrive
to (\ref{fluctuationsDp+4}) we expanded up to quadratic order in the
fluctuations and we have kept all orders in the instanton field. The extra
factor compared to that in (\cite{EGG}) comes from the fact that, for a
self-dual worldvolume gauge field, the unperturbed DBI action actually
contains the square root of a perfect square, which can be evaluated exactly
and shows up in the lagrangian of the fluctuations. 
This extra term is proportional to the inverse of
the effective Yang-Mills coupling. In order to trust the supergravity
approximation the effective Yang-Mills coupling should be large, which would
suggest that the effect of this term is indeed negligible. We will see
however that in the region of small $\frac{v}{m_q}$ the full term is actually
dominating in the IR region and determines the meson spectrum. 
In addition, in order to ensure the validity of
the DBI approximation, we should have slowly varying gauge fields, which
further imposes that $F\wedge F$ should be much smaller than $\alpha'$. 

\begin{figure}
\centerline{\hskip -.1in \epsffile{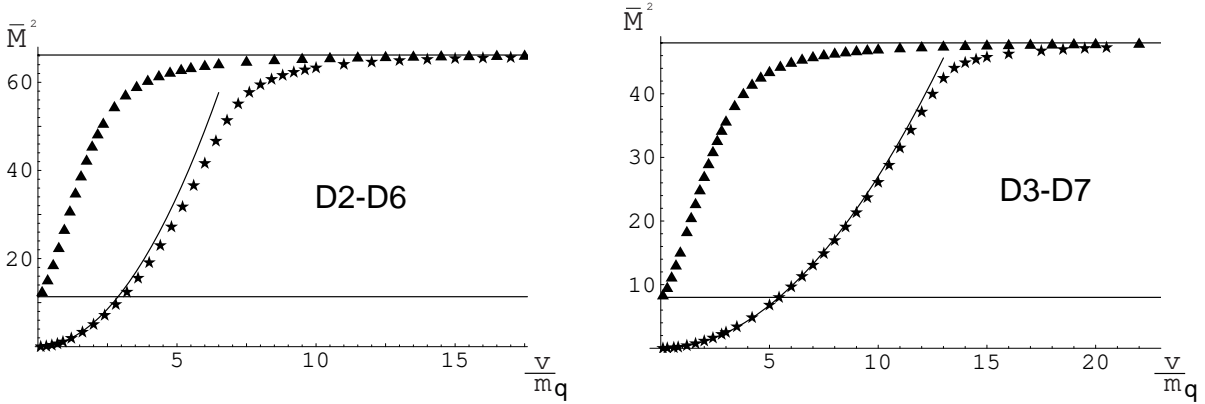}}
\caption{In this figure we  plot the numerical masses for the
first level as a function of 
the instanton size for both the full equation (with stars) and for the
equation obtained in \cite{EGG} (with solid triangles). The quark mass
$m_q$ is such that $g_{eff}(m_q)=1$. The solid line
corresponds to the WKB prediction (\ref{WKB-smallv}) for small $v$. The plot
on the left (right) corresponds to the D2-D6 (D3-D7) intersection.}
\label{massinstanton}
\end{figure}

In order to study the fluctuation equation (\ref{fluctuationsDp+4}) it is
interesting to notice that, after a change of variable,  
(\ref{fluctuationsDp+4}) can be converted into a
Schr\"odinger equation. Indeed, let us change from $\varrho$ and $f$ to the
new variables $z$ and $\psi$, defined as:
\beq
e^z\,=\,\varrho\,\,,\qquad\qquad
\psi\,=\,\varrho\,f\,\,.
\eeq
Notice that $\varrho\to\infty$ corresponds to $z\to +\infty$, while
$\varrho=0$ is mapped to $z=-\infty$. Moreover, one can readily prove that,
in terms of $z$ and $\psi$, eq. (\ref{fluctuationsDp+4}) can be recast as:
\beq
\partial^2_z\,\psi\,-\,V(z)\,\psi\,=\,0\,\,,
\label{Sch-eq}
\eeq
where the potential $V(z)$ is given by:
\bear
&&V(z)\,=\,1\,+\,
\Bigg(\frac{v}{m_q}\Bigg)^4\frac{8}
{\Big(\,e^{2z}+
\big(\frac{v}{m_q}\big)^2\Big)^2}\,-\,\rc\rc
&&\qquad\qquad\qquad-\,\,
\bar M^2\,
{e^{2z}\over \Big(\,e^{2z}+1\,\Big)^{{7-p\over 2}}}\,\,
\Bigg[\,1\,+\,c_{p}(v,m_q)\,
{\big(\,e^{2z}+1\,\Big)^{{7-p\over 2}}\over 
\Big(\,e^{2z}+(\frac{v}{m_q})^2\Big)^4}\,\,\Bigg]\,\,.
\label{Sch-pot}
\eear
Notice that the reduced mass $\bar M$ is just a parameter in $V(z)$.
Actually, in these new variables the problem of finding the mass spectrum can
be rephrased as that of finding the values of $\bar M$  that allow
a zero-energy level for the potential (\ref{Sch-pot}). By using the
standard techniques in quantum mechanics one can convince oneself that such 
solutions exist. Indeed, the potential (\ref{Sch-pot}) is strictly positive
for 
$z\to\pm\infty$ and has some minima for finite values of $z$. The actual
calculation of the mass spectra must be done by means of numerical
techniques. A key ingredient in this approach is the knowledge of the
asymptotic behaviour of the solution when $\varrho\to 0$ and
$\varrho\to\infty$. This behaviour can be easily obtained from the form of
the potential $V(z)$ in (\ref{Sch-pot}). Indeed, for $\varrho\to\infty$, or
equivalently for $z\to +\infty$, the potential $V(z)\to 1$, and the
solutions of (\ref{Sch-eq}) behave as $\psi\sim e^{\pm z}$ which, in terms of
the original variables, corresponds to $f={\rm constant},\,\,\varrho^{-2}$. 
Similarly for $\varrho\to0$ (or $z\to-\infty$) one gets that 
$f=\varrho^{2},\,\,\varrho^{-4}$. Thus, the IR and UV
behaviours of the fluctuation are:
\bear
&&f(\varrho)\,\approx\,a_1\,\varrho^2\,+\,a_2\,\varrho^{-4}\,\,,
\qquad\qquad (\varrho\to 0)\,\,,\rc\rc
&&f(\varrho)\,\approx\,b_1\,\varrho^{-2}\,+\,b_2\,\,,
\qquad\qquad (\varrho\to \infty)\,\,.
\label{UVIRbehaviour}
\eear
The normalizable solutions are those that are regular at $\varrho\approx 0$
and decrease at $\varrho\approx \infty$. Thus they correspond to having
$a_2=b_2=0$ in (\ref{UVIRbehaviour}).
Upon applying a shooting technique, we can determine the values of $\bar M$
for which such normalizable solutions exist. Notice that $\bar M$ depends
parametrically on the quark mass $m_q$ and on its VEV $v$. In general, for
given values of $m_q$ and $v$, one gets a tower of discrete values of  
$\bar M$. In figure \ref{massinstanton} we have plotted the values of the
reduced mass for the first level, as a function of the quark VEV. For
illustrative purposes we have included the values obtained with the
fluctuation equation of ref. \cite{EGG}. As anticipated above, both results
differ significantly in the region of small $v$ and coincide when
$v\to\infty$. Actually, when $v$ is very large we recover the spectral flow
phenomenon described in \cite{EGG}, \ie\ $\bar M$ becomes independent of the
instanton size and equals the mass corresponding to a higher Kaluza-Klein
mode on the worldvolume sphere. However, we see that when $\frac{v}{m_q}$ goes to
zero, the masses of the associated fluctuations also go to zero. Actually,
this limit is pretty singular. Indeed, it corresponds to the small instanton
limit, where it is  expected that the moduli space of instantons becomes
effectively non-compact and  that extra massless degrees of
freedom show up in the spectrum.

It turns out that the mass levels for small $v$ are nicely represented
analytically by means of the WKB approximation for the Schr\"odinger problem
(\ref{Sch-eq}). The WKB method has been very successful \cite{MInahan,RS} in
the calculation of the glueball mass spectra in the gauge/gravity
correspondence and also provides rather reliable predictions for the mass
levels of the mesons \cite{AR}. The WKB quantization rule is:
\beq
(n+{1\over 2})\pi\,=\,\int_{z_1}^{z_2}\,dz\,\sqrt{-V(z)}\,\,,
\,\,\,\,\,\,\,\,\,\,\,\,\,\,
n\ge 0\,\,,
\label{WKBquantization}
\eeq
where $n\in\ZZ$ and $z_1$ and $z_2$ are the turning points of the potential
($V(z_1)=V(z_2)=0$). Following straightforwardly the steps of refs. \cite{RS,
AR}, we obtain the following expression for the WKB values of $\bar M$:
\beq
\bar M_{WKB}^2\,=\,{\pi^2\over \zeta^2}
\,(n+1)\,\bigg(n\,+\,3\,+
{2\over 5-p}\bigg)\,\,,
\label{MWKB}
\eeq
where $\zeta$ is the following integral:
\beq
\zeta\,=\,\int_{0}^{+\infty}\,\,
d\varrho\,\,\sqrt{\,
{1\over (1+\varrho^2)^{{7-p\over 2}}}\,+\,
{c_{p}(v,m_q)\over \Big[\,\big({v\over m_q}\big)^2\,+\,
\varrho^2\,\Big]^4}}\,\,.
\label{zetaWKB}
\eeq
Let us evaluate analytically $\zeta$ when $v$ is small. First of all, as can
be easily checked, we notice that, when $v$ is small,  the second term under
the square root in (\ref{zetaWKB}) behaves as:
\beq
{1\over \Big[\,\big({v\over m_q}\big)^2\,+\,
\varrho^2\,\Big]^2}\,\approx\,{\pi\over 2}\,\,
\bigg({m_q\over v}\bigg)^3\,\,\delta(\varrho)\,\,,
\qquad\qquad
{\rm as}\,\, v\to 0\,\,.
\label{deltaVEV}
\eeq
Then, one can see that this term dominates the integral defining $\zeta$
around $\varrho\approx 0$ and, for small $v$, one can approximate $\zeta$
as:
\beq
\zeta\,\approx\, {\sqrt{c_{p}(v,m_q)}\over 2}\,\,
\int_{-\epsilon}^{\epsilon}\,
{d\varrho\over \Big[\,\big({v\over m_q}\big)^2\,+\,
\varrho^2\,\Big]^2}\,+\,\int_0^{+\infty}\,
{d\varrho\over (1+\varrho^2)^{{7-p\over 4}}}\,\,,
\eeq
where $\epsilon$ is a small positive number and we have used the fact that
the function in (\ref{zetaWKB}) is an even function of $\varrho$. 
Using (\ref{deltaVEV}), one can
evaluate $\zeta$ as:
\beq
\zeta\,\approx\,{\pi\over 4}\,\,\bigg({m_q\over v}\bigg)^3\,
\sqrt{c_{p}(v,m_q)}\,+\,{\sqrt{\pi}\over 2}\,\,
{\Gamma\Big({5-p\over 4}\Big)\over \Gamma\Big({7-p\over 4}\Big)}\,\,.
\label{zeta-approx}
\eeq
Clearly, for $v\to 0$, we can neglect the last term in (\ref{zeta-approx}).
Using the expression of $c_{p}(v,m_q)$ (eq. (\ref{cp})), we arrive at:
\beq
\zeta\,\approx\,{\sqrt{3}\,\cdot 2^{{p-4\over 2}}\,\pi^{{p+5\over 4}}\over
\sqrt{\Gamma\Big({7-p\over 2}\Big)}}\,\,
{m_q\over g_{eff}(m_q)\, v}\,\,,
\eeq
and plugging this result in (\ref{MWKB}), we get the WKB mass of the ground
state ($n=0$) for small $v$:
\beq
\bar M_{WKB}^2\,\approx\,{(17-3p)\,\Gamma\Big({5-p\over 2}\Big)\over
3\cdot 2^{p-3}\,\pi^{{p+1\over 2}}}\,\,\,\,
\Bigg(\,{g_{eff}(m_q)\, v\over m_q}\,\Bigg)^2\,\,.
\label{WKB-smallv}
\eeq
Thus, we predict that $\bar M^2$ is a quadratic function of $v/m_q$ with the
particular coefficient given on the right-hand side of (\ref{WKB-smallv}). 
In figure \ref{massinstanton} we have represented by a solid line the value
of $\bar{M}$ obtained from eq. (\ref{WKB-smallv}). We notice that, for small
$v$, this equation nicely fits the values obtained by the numerical
calculation.

Let us now study the dependence of the mass gap as a function of the quark
mass $m_q$ and the quark VEV $v$. First of all, we notice that the relation
between the reduced mass $\bar M$ and the mass $M$ can be rewritten in terms
of the quark mass $m_q$ and the dimensionless coupling constant
$g_{eff}(m_q)$ as:
\beq
M\,\propto {m_q\over g_{eff}(m_q)}\,\,\bar M\,\,.
\label{M-barM}
\eeq
For large $v$ the reduced mass $\bar M$ tends to a value independent of both
$m_q$ and $v$. Thus, the meson mass $M$ depends only on $m_q$ in a
holographic way, namely:
\beq
M\,\sim {m_q\over g_{eff}(m_q)}\,\,,
\qquad\qquad (v\to\infty)\,\,.
\eeq
Notice that this dependence on $m_q$ and $v$ is exactly the same as in the
unbroken symmetry case, although the numerical coefficient is
different from that found in \cite{AR,MT}. 
On the contrary, for small $v$, after combining eq. (\ref{M-barM}) with  the
WKB result (\ref{WKB-smallv}), we get that the mass gap depends linearly on
$v$ and is independent on the quark mass $m_q$:
\beq
M\sim v\,\,,\qquad\qquad (v\to 0)\,\,,
\eeq
and, in particular, the mass gap disappears in the limit $v\to 0$, which
corresponds to having a zero size instanton.

\setcounter{equation}{0}
\section{The codimension one defect}
\label{DpDp+2section}
In this section we will consider the intersection of Dp- and D(p+2)-branes
according to the array:
\beq
\begin{array}{ccccccccccl}
 &1&\cdots&p-1& p& p+1&p+2 &p+3&\cdots&9 & \nonumber \\
Dp: & \times &\cdots &\times &\times &\_ & \_&\_ &\cdots &\_ &     \nonumber
\\ D(p+2): &\times&\cdots&\times&\_&\times&\times&\times&\cdots&\_ &
\end{array}
\label{DpDp+2intersection}
\eeq
It is easy to verify, by using the standard intersection rules of the type II
theories, that this  Dp-D(p+2) intersection is supersymmetric. Moreover, it is
clear from  (\ref{DpDp+2intersection}) that the D(p+2)-brane is an object of
codimension  one along the gauge theory directions of the Dp-brane
worldvolume. Indeed, for $p=3$ the configuration (\ref{DpDp+2intersection})
was studied in \cite{KR} and shown to be dual to a defect theory 
in which ${\cal N}=4$, $d=4$ super
Yang-Mills theory in the bulk is coupled to ${\cal N}=4$, $d=3$
fundamental hypermultiplets localized at the defect \cite{WFO, EGK}, which is
located at a fixed value of the coordinate $p$ in (\ref{DpDp+2intersection}).
These hypermultiplets are generated by open strings connecting the two types
of D-branes. If we allow a non-zero distance in the $p+4,\cdots, 9$
directions of the two stacks in (\ref{DpDp+2intersection}), the
hypermultiplets become massive and a mass gap is introduced in the theory.
The corresponding meson  spectrum was computed in the probe approximation in
ref.  \cite{AR}.

The  analysis of the Higgs phase of the codimension one defect associated to
the array (\ref{DpDp+2intersection}) has
been worked out recently in \cite{ARR} for the particular case of a 2+1
dimensional defect living in a bulk ${\mathcal{N}}=4$, $d=4$ theory, which
corresponds to the intersection displayed in (\ref{DpDp+2intersection}) for
$p=3$. In that
reference it is argued that the gravity dual description of the Higgs phase of
the theory with ${\cal N}=2$ fundamental hypermultiplets confined to a
codimension one defect is in terms of probe D5-branes in the near horizon of
the D3-brane geometry, once we switch on appropriately a  magnetic
worldvolume gauge field. As in the case
of the codimension zero defect, this worldvolume gauge field has the effect of
introducing extra D3-brane charge in the D5-brane worldvolume, which in turn
can be seen as the macroscopical description of dielectrically expanded
D3-branes. However, the addition of the magnetic field requires a non-trivial
bending of the D5-branes, which now recombine with the D3's rather than 
intersecting them. This bending takes place along the direction 3 in 
(\ref{DpDp+2intersection})  for $p=3$. As in the previous section, the $F$-
and $D$-flatness conditions arise naturally as the vacuum conditions for the
dielectric branes, thus providing a map between the Higgs phase of these
theories and the monopoles in the sphere to which the branes expand. The
required bending then appears naturally as the solution to these $F$- and
$D$-flatness conditions.

We will refer to \cite{ARR} for the field theory analysis, which can be 
extended in a straightforward manner to any dimension. Instead, in this paper
we will focus on the gravity dual for the general case, which will be in
terms of a probe D(p+2)-brane in the Dp-brane background given by
(\ref{Dpmetrica}), (\ref{Dpdilaton}) and (\ref{DpRR}). Let us go to a new
coordinate system, in which we write  the transverse space to the Dp-brane
spanned by $\vec{r}$ in a more suitable manner,  such that the metric 
(\ref{Dpmetrica}) takes the form:
\begin{equation}
ds^2\,=\,\Bigg(\frac{r^2}{R^2}\Bigg)^{\alpha}dx_{1,p}^2+
\Bigg(\frac{R^2}{r^2}\Bigg)^{\alpha}\Big(\,d\rho^2+\rho^2d\Omega_2^2+
d\vec{z}^{\,2}\,\Big)\,\,,
\label{metricDpDp+2}
\end{equation}
where $d\Omega_2^2\,=\,d\theta^2+\sin^2\theta d\varphi^2$ is the line element
of a  unit two-sphere and the coordinates $(\rho,\theta,\varphi)$ parametrize
the directions $p+1$, $p+2$ and $p+3$ in (\ref{DpDp+2intersection}).  
The exponent $\alpha$ in (\ref{metricDpDp+2}) is the same as  in
(\ref{Dpmetrica}) and  $r^2=\rho^2+\vec{z}^{\,2}$.

We shall now consider  a
D(p+2)-brane probe in this background. Its action is:
\beq
S^{D(p+2)}\,=\,-\,T_{p+2}\,\int d^{p+3}\xi\,e^{-\phi}\,\,
\sqrt{-\det (g+F)}\,+\,
T_{p+2}\,\int \,\,\,P\big[\,C^{(p+1)}\,\big]\wedge F\,\,.
\label{DBI-D5}
\eeq
In what follows we will take 
$\xi^a=(x^0,x^1\cdots,x^{p-1},\rho,\theta,\varphi)$ as worldvolume
coordinates. Moreover, we will assume that there exists  
a constant separation on the transverse space,  $\vec{z}^{\,2}=L^2$, which
gives  mass to the quarks,  and we will switch on
a  magnetic worldvolume field on the internal $S^2$ given by: 
\begin{equation}
F\,=\,q\,{\rm Vol}\,(S^2)\,\equiv\,
{\cal F}\,\,,
\label{wvflux}
\end{equation}
where $q$ is a constant and ${\rm Vol}\,(S^2)=\sin\theta d\theta
d\varphi$. As anticipated above, in order to solve the equations of motion
of the probe, we have to consider a non-trivial transverse
$x^p$ field $x^p\,=\,x(\rho)$. Moreover, since nothing depends on the
internal $S^2$, upon integration over this compact manifold, it is
straightforward to see that the action reads:
\newpage
\bear
\label{SDp+2}
S^{D(p+2)}&=&-4\pi T_{p+2}\int\,d^{p} x\,d\rho\,
\Bigg\{\rho^2\,\sqrt{1+\Bigg(\frac{\rho^2+L^2}{R^2}
\Bigg)^{2\alpha}x'^2}\,\sqrt{1+\Bigg(\frac{\rho^2+L^2}{R^2}\Bigg)^{2\alpha}
\frac{q^2}{\rho^4}}\,-\rc\rc
&&\qquad\qquad\qquad\qquad\qquad\qquad\qquad\qquad
-\,q\,\Bigg(\frac{\rho^2+L^2}{R^2}\Bigg)^{2\alpha}x'\Bigg\}\,\,.
\eear
One can check that the Euler-Lagrange equation for $x(\rho)$ derived from
(\ref{SDp+2}) is solved if one requires that:
\beq
x'(\rho)\,=\,{q\over \rho^2}\,\,,
\label{Dp+2-first-order}
\eeq
which can be immediately integrated, giving rise to the following profile of
the transverse scalar:
\begin{equation}
x(\rho)\,=\,x_0\,-\,\frac{q}{\rho}\,\equiv\,{\mathcal{X}}(\rho)\,\,.
\label{bending}
\end{equation}
For this configuration, the two square roots in (\ref{SDp+2}) become equal
and there is a cancellation between the WZ and (part of) the DBI term. 
Then, the energy for such a brane,
which is nothing but minus the lagrangian since our configuration is static,
reduces to:
\begin{equation}
\label{SDp+2onshell}
E\,=\,4\pi
T_{p+2}\int \rho^2\ ,
\end{equation}
where,  as in the Dp-D(p+4) case, the distance  $L$ does not explicitly
appear,  displaying the supersymmetry properties of the configuration.
Indeed, one can verify as in \cite{ST} that the condition 
(\ref{Dp+2-first-order}) is a BPS equation that can be derived from
kappa symmetry of the probe and that the energy (\ref{SDp+2onshell}) saturates
a BPS bound. 

Let us remind the reader that the existence of the bending (\ref{bending}) 
was a key  ingredient in the analysis of \cite{ARR}, where it was shown, for
the particular case of $p=3$,  that it has the effect of spreading
the defect over the whole bulk  which, in turn, led to the loss of the
discrete spectrum. We shall see that, indeed,  the same situation occurs in
the more general case considered here.

\subsection{Microscopical description of the Dp-D(p+2) intersection with
flux}

The flux  of the worldvolume gauge field $F$ of  eq. (\ref{wvflux}) on  the
internal $S^2$ has the non-trivial effect of inducing
Dp-brane  charge in the D(p+2)-brane worldvolume. To verify this fact, let us
point out that $F$ is constrained by a flux
quantization condition \cite{Flux} which, with our notations, reads:
\beq
\int_{S^2}\,F\,=\,{2\pi k\over T_f}\,\,,\qquad
k\in \ZZ\,\,,\qquad T_f\,=\,{1\over 2\pi\alpha'}\,\,.
\label{fluxquantization}
\eeq
By plugging the expression of $F$ given in (\ref{wvflux}) on the quantization
condition (\ref{fluxquantization}), one immediately concludes that the
constant $q$ is restricted to be of the form:
\beq
q\,=\,k\pi\alpha'\,\,,
\label{q-k}
\eeq
where $k$ is an integer. In order to interpret the meaning of $k$, let us
notice that in the WZ piece of
the action for the D(p+2)-brane we have the coupling:
\begin{equation}
T_{p+2}\int\, P[C^{p+1}]\wedge F\,\,.
\end{equation}
Upon using the explicit form of $F$  to integrate it over the two-sphere
and the relation $T_{p+2}\,(2\pi )^2\,\alpha'=T_p$
we have that this coupling reads:
\begin{equation}
k\, T_p\int\, P[C^{p+1}]\,\,,
\end{equation}
where now the integration is over $p+1$ variables. 
Thus, we see that $F$ is inducing $k$ units of 
Dp-brane charge in the
worldvolume of the D(p+2)-brane. This charge is located along the
$\{x^0,\cdots, x^{p-1},\rho\}$ directions. This suggests an alternative
interpretation of the system in terms of dielectric Dp-branes that polarize
to a  D(p+2)-brane,  as anticipated in
$\cite{ARR}$. To be more explicit, let us consider a stack of $k$ coincident
Dp-branes in the background (\ref{metricDpDp+2}). The dynamics of such a
stack is governed by the Myers action \cite{M}, which is given by the
straightforward generalization of eqs. (\ref{dielectricBI}) and
(\ref{dielectricWZ}) to a Dp-brane. We will choose $(x^0,\cdots,
x^{p-1},\rho)$ as worldvolume coordinates and we shall consider the other
coordinates in the metric (\ref{metricDpDp+2}) as scalar fields which, in
general, are non-commutative. Moreover, we shall introduce new
coordinates  $Y^I(I=1,2,3)$ for the two-sphere of the metric
(\ref{metricDpDp+2}). These new coordinates satisfy
$\sum_I\,Y^I\,Y^I\,=\,1$ and the line element $d\Omega_2^2$ is given by:
\beq
d\Omega_2^2\,=\,\sum_I\,dY^I\,dY^I\,\,,
\qquad\qquad
\sum_I\,Y^I\,Y^I\,=\,1\,\,.
\eeq
We will assume that the $Y$'s are the only non-commutative scalars and that
they are represented by $k\times k$ matrices. Furthermore, 
we shall adopt the ansatz in which they  are given by:
\begin{equation}
Y^I=\frac{J^I}{\sqrt{C_2(k)}}\,\, ,
\end{equation}
where the $k\times k$ matrices $J^I$ correspond to the $k$-dimensional
irreducible representation of the $SU(2)$ algebra:
\beq
[J^I,J^J]\,=\,2i\epsilon_{IJK}\,J^K\,\,,
\label{Jcommutator}
\eeq
and $C_2(k)=k^2-1$ is the corresponding quadratic Casimir. Clearly the $Y$'s
parametrize a fuzzy two-sphere. Let us, in addition, assume that we consider
embeddings of the Dp-brane in which the scalars $\vec z$ and $x^p$ are
commutative and such that $|\vec z\,|=L$ and $x^p=x(\rho)$. With these
assumptions it is easy to evaluate the dielectric action for the Dp-brane in
the large $k$ limit, following the same steps as those followed in ref.
\cite{ARR} for the D3-D5 system. The final result exactly coincides with the
macroscopical action (\ref{SDp+2}), once $q$ is related to the integer
$k$ as in the quantization condition (\ref{q-k}). This matching is a
confirmation of our interpretation of the D(p+2)-brane configuration with
flux as a bound state of a stack of coincident Dp-branes. 
Once again we see that the expansion to the dielectric configuration is not 
caused by any other field apart from the metric background, thus constituting
another example of purely gravitational dielectric effect
(\cite{gravitationaldielectriceffect}).

\subsection{Fluctuations in Dp-D(p+2) with flux}
\label{Dp-D(p+2)fluctuations}

Let us  now study the
fluctuations around the Dp-D(p+2) intersection with flux described above.
Without loss of generality we can take the unperturbed configuration as
$z^1=L$, $z^m=0$ $(m>1)$.  Next, let us consider a fluctuation of the type:
\bear
&&z^1\,=\,L\,+\,\chi^1\,\,,
\qquad z^m\,=\,\chi^m\,,\qquad\qquad\qquad (m=2,\cdots, 6-p)\,\,,\rc\rc
&&x^p\,=\,{\cal X}\,+\,x\,\,,\qquad\qquad\qquad 
F\,=\,{\cal F}\,+\,f\,\,,
\eear
where the bending ${\cal X}$ and the worldvolume gauge field ${\cal F}$ are
given by eqs. (\ref{bending}) and  (\ref{wvflux}) respectively and we assume
that $\chi^m$, $x$  and $f$ are small. The induced metric on the D(p+2)-brane
worldvolume can be written as:
\beq
g\,=\,{\cal G}\,+\,g^{(f)}\,\,,
\eeq
with ${\cal G}$ being the induced metric of the unperturbed configuration:
\beq
{\cal G}_{ab}\,d\xi^a\,d\xi^b\,=\,h^{-1}\,dx_{1,p-1}^2\,+\,
h\,\Bigg[\,\Bigg(\,1\,+\,{q^2\over \rho^4 h^2}\,\Bigg)\,d\rho^2\,+\,
\rho^2\,d\Omega_2^2\,\Bigg]\,\,,
\eeq
where $h=h(\rho)$ is the function defined in (\ref{h}). Moreover, $g^{(f)}$
is the part of $g$ that depends on the derivatives of the fluctuations,
namely:
\beq
g^{(f)}_{ab}\,=\,{q\over \rho^2
h}\,\,\Big(\,\delta_{a\rho}\,\partial_{b}\,x\,+\,
\delta_{b\rho}\,\partial_{a}\,x\,\Big)\,+\,
{1\over h}\,\partial_{a}\,x\,\partial_{b}\,x\,+\,h\,
\partial_{a}\,\chi^m\,\partial_{b}\,\chi^m\,\,.
\eeq
Let us next rewrite the Born-Infeld determinant as:
\beq
\sqrt{-\det(g+ F)}\,=\,\sqrt{-\det \big(\,{\cal G}\,+\,{\cal
F}\,\big)}\,
\sqrt{\det\,(1+X)}\,\,,
\label{detX}
\eeq
where the matrix $X$ is given by:
\beq
X\,\equiv\,\bigg(\,{\cal G}\,+\,{\cal  F}\,\bigg)^{-1}\,\,
\,\bigg(\,g^{(f)}\,+\,f\,\bigg)
\,\,.
\label{matrixX}
\eeq
We shall evaluate the right-hand side of (\ref{detX}) by expanding it  in
powers of $X$ by means of eq. (\ref{detX-expansion}). In order to evaluate
more easily the trace of the powers of $X$ appearing on the right-hand side
of this equation, let us separate the symmetric and antisymmetric part in the
inverse of the matrix ${\cal G}\,+\,{\cal  F}$:
\beq
\bigg(\,{\cal G}\,+\,{\cal  F}\,\bigg)^{-1}\,=\,
\hat{\cal G}^{-1}\,+\,{\cal J}\,\,,
\eeq
where:
\beq
\hat{\cal G}^{-1}\,\equiv\,{1\over ({\cal G}+{\cal  F})_S}\,\,,\qquad
{\cal J}\,\equiv\,{1\over ({\cal G}+{\cal  F})_A}\,\,.
\eeq
Notice that $\hat{\cal G}$ is just  the open string metric which, for the
case at hand, is given by:
\beq
\hat{\cal G}_{ab}\,d\xi^a\,d\xi^b\,=\,h^{-1}\,dx_{1,p-1}^2\,+\,h\,
\Bigg(\,1\,+\,{q^2\over \rho^4 h^2}\,\Bigg)\,
\Bigg(\,d\rho^2\,+\,
\rho^2\,d\Omega_2^2\,
\Bigg)\,\,. \label{openstrmetric}
\eeq
Moreover,  the antisymmetric matrix ${\cal J}$ takes the form:
\beq
{\cal J}^{\theta\varphi}\,=\,-{\cal J}^{\varphi\theta}\,=\,
-{1\over \sqrt{\tilde g}}\,\,{q\over q^2\,+\,\rho^4\,h^2}\,\,,
\eeq
where $\theta, \varphi$ are the standard polar coordinates on $S^2$ and 
$\tilde g=\sin^2\theta$ is the determinant of its round metric. It is now
straightforward to show that:
\beq
\tr \,X\,=\,h\,\,\hat{\cal  G}^{ab}\,\partial_a\chi^m\,\partial_b\chi^m\,+\,
{1\over h}\,\hat{\cal G}^{ab}\,\partial_a x\,\partial_b x\,+\,
{q\over q^2\,+\,\rho^4\,h^2}\,\Big[\,
2\rho^2\,\partial_\rho x\,+\,{\epsilon^{ij}f_{ij}\over 
\sqrt{\tilde  g}}\,\Big]\,\,,
\eeq
while, up to quadratic terms in the fluctuations,  $\tr\, X^2$ is given by:
\bear
&&\tr\, X^2\,=\,-f_{ab}f^{ab}\,+\,{2\over h}\,\,
{q^2\over q^2\,+\,\rho^4h^2}\,\Big[\,
\hat{\cal G}^{ab}\,\partial_a x\,\partial_b x\,+\,
\hat{\cal G}^{\rho\rho}\,(\partial_\rho x)^2\,\Big]\,+\,\rc\rc
&&\qquad
+{q^2\over \bigg(q^2\,+\,\rho^4h^2\bigg)^2}\,\Bigg[\,
{1\over 2\tilde g}\,\big(\epsilon^{ij}f_{ij}\big)^2\,-\,4\, \rho^2 \,\,
{\epsilon^{ij}\partial_i x\,f_{j\rho}\over
\sqrt{\tilde g}}\,\Bigg]\,\,,
\eear
where the indices $i,j$ refer to the directions along the $S^2$ and
$\epsilon^{ij}=\pm 1$.  Using these results one can readily compute the DBI
term of the lagrangian density. Dropping constant global factors that do not
affect the equations of motion, one gets:
\bear
&&{\cal L}_{DBI}\,=\,-\rho^2\,\sqrt{\tilde g}\,\Bigg[\,1\,+\,
{q^2\over \rho^4h^2}\,+\,
{h\over 2}\,\bigg(\,1\,+\,{q^2\over \rho^4 h^2}\,\bigg)\,
\hat{\cal G}^{ab}\,\partial_a\chi^m\,\partial_b\chi^m\,+\,\rc\rc
&&\qquad\qquad\qquad\qquad
+{1\over 2 h}\,\hat{\cal G}^{ab}\,\partial_a x\,\partial_b
x\,+\, {1\over 4}\,\bigg(\,1\,+\,{q^2\over \rho^4 h^2}\,
\bigg)\,f_{ab}f^{ab}\,
\Bigg]\,+\,\rc\rc
&&\qquad\qquad\qquad\qquad
+{A(\rho)\over 2}\,\, x\,\epsilon^{ij}\,f_{ij}\,-\,
{q\sqrt{\tilde g}\over h^2}
\,\,\partial_{\rho}x\,-\,{q\over 2\rho^2 h^2}\,\,\epsilon^{ij}\,f_{ij}\,\,,
\eear
where the indices $a,b$ are raised with the open string metric $\hat{\cal G}$
and
$A(\rho)$ is the following function:
\beq
A(\rho)\,\equiv\,{d\over d\rho}\,\Bigg[\,
{q^2\over h^2\Big(\,q^2\,+\,\rho^4h^2\Big)}\,\Bigg]\,\,.
\eeq
To get the above expression of ${\cal L}_{DBI}$ we have integrated by 
parts and made use of
the Bianchi identity for the gauge field fluctuation:
\beq
\epsilon^{ij}\,\partial_i f_{j\rho}\,+\,{\epsilon^{ij}\over 
2}\,\partial_\rho f_{ij}\,=\,0\,\,.
\eeq
Similarly, the WZ term can be written as:
\beq
{\cal L}_{WZ}\,=\,\sqrt{\tilde g}\,\,{q^2\over \rho^2
h^2}\,+\,\sqrt{\tilde g}\, \,{q\over h^2}\,\partial_{\rho}x\,
+\,{q\over 2\rho^2 h^2}\,\epsilon^{ij}\,f_{ij}\,+\, 
{\partial_{\rho}h\over h^3}\,\,x
\epsilon^{ij}\,f_{ij}\,\,.
\eeq
By combining ${\cal L}_{DBI}$ and ${\cal L}_{WZ}$ and droping the term
independent of the fluctuations,  we get that the  total lagrangian density
is given by:
\bear
&&{\cal L}\,=\,-\rho^2\,\sqrt{\tilde g}\,\Bigg[\,{h\over 2}\,
\bigg(\,1\,+\,{q^2\over \rho^4 h^2}\,\bigg)\,
\hat{\cal G}^{ab}\,\partial_a\chi^m\,\partial_b\chi^m\,+\,
{1\over 2 h}\,\hat{\cal G}^{ab}\,\partial_a x\,\partial_b x\,+\,\rc\rc
&&\qquad\qquad
+{1\over 4}\,\bigg(\,1\,+\,{q^2\over \rho^4 h^2}\,\bigg)\,
f_{ab}f^{ab}\,\Bigg]\,-\,{C(\rho)\over 2}\,
x\,\epsilon^{ij}f_{ij}\,\,.
\label{fluct-lag}
\eear
In eq. (\ref{fluct-lag}), and in what follows, the function
$C(\rho)$ is given by:
\beq
C(\rho)\,\equiv\,
{d\over d\rho}\,\Bigg[\,
{\rho^4\over q^2\,+\,\rho^4h^2}\,\Bigg]\,\,.
\label{C}
\eeq
As it is manifest from (\ref{fluct-lag}), the transverse scalars $\chi$ do
not couple  to other fields, while the scalar $x$ is coupled to the
fluctuations $f_{ij}$ of the gauge field strength along the two-sphere. For
the fluxless case $q=0$ these equations were solved in ref. \cite{AR}, where 
it was shown that they give rise to a discrete meson mass spectrum, which can
be  computed numerically and, in the case of the D3-D5 intersection,
analytically. Let us examine here the situation for $q\not=0$. 
The equation of motion of the transverse scalars $\chi$ that follow from 
(\ref{fluct-lag}) is:
\beq
\partial_a\Bigg[\sqrt{\tilde g}\,\rho^2 h\,
\bigg(\,1\,+\,{q^2\over \rho^4 h^2}\,\bigg)
\hat{\cal G}^{ab}\,\partial_b\chi\,\bigg)
\,\Bigg]\,=\,0\,\,.
\label{chi-fluct}
\eeq
By using the explicit form of the open string  metric $\hat{\cal G}^{ab}$ (eq.
(\ref{openstrmetric})), we can rewrite (\ref{chi-fluct}) as:
\beq
\partial_{\rho}\,\Big(\rho^2\partial_\rho 
\chi\Big)\,+\,
\bigg[ \rho^2h^2\,+\,{q^2\over \rho^2}\,\bigg]\,\,
\partial^{\mu}\partial_{\mu}\,\chi\,
+\,\nabla^i\nabla_i\,\chi\,=\,0\,\,.
\eeq
Let us separate variables and
write the scalars in terms of the eigenfunctions of the laplacian in the
Minkowski and sphere parts of the brane geometry as:
\begin{equation}
\chi=e^{ikx}\, Y^l(S^2)\,\xi(\rho)\,\, ,
\end{equation}
where the product $kx$ is performed with the Minkowski metric and $l$ is the
angular momentum on the $S^2$. 
The fluctuation equation for the function $\xi$  is:
\begin{equation}
\partial_{\rho}\,(\,\rho^2\,\partial_{\rho}\xi\,)\,+\,
\Bigg\{\Bigg[\,
R^{4\alpha}\,\frac{\rho^2}{(\rho^2\,+\,L^2)^{2\alpha}}\,+\,
\frac{q^2}{\rho^2}\Bigg]\,M^2\,
-\,l(l+1)\Bigg\}\,
\xi\,=\,0\,\,, 
\label{radial-chi-fluct}
\end{equation}
where $M^2=-k^2$ is the mass of the meson. When the distance $L\not=0$ and 
$q=0$ eq.  (\ref{radial-chi-fluct}) gives rise to a set of normalizable
solutions that occur for a discrete set of values of $M$ \cite{AR}. As argued
in ref. \cite{ARR} for the D3-D5 system, the situation changes drastically
when  the flux is switched on. Indeed, let us  consider  the equation 
(\ref{radial-chi-fluct}) when $L, q\not=0$ in the IR, \ie
when $\rho$ is close to zero. In this case, for small values of $\rho$, eq. 
(\ref{radial-chi-fluct})  reduces to:
\beq
\partial_{\rho}\,\Big(\rho^2\partial_{\rho}\xi\Big)\,+\,
\Bigg[\,{q^2\,M^2\over\rho^2}\,-\,l(l+1)\,\Bigg]\xi\,=0\,\,,
\qquad\qquad(\rho\approx 0)\,\,.
\label{IRfluc}
\eeq
Eq.  (\ref{IRfluc}) can be solved in terms of Bessel functions,
namely:
\beq
\xi\,=\,{1\over \sqrt{\rho}}\,\,
J_{\pm (l+{1\over 2})}\,\Bigg({qM\over \rho}\Bigg),\,
\qquad\qquad(\rho\approx 0)\,\,.
\label{Bessel}
\eeq
Near $\rho\approx 0$ the Bessel function (\ref{Bessel}) oscillates infinitely
as:
\beq
\xi\,\approx\,e^{\pm i{qM\over \rho}}\,\,,
\qquad\qquad(\rho\approx 0)\,\,.
\label{IRfluct-behaviour}
\eeq
The behaviour (\ref{IRfluct-behaviour}) implies that the spectrum of $M$ is 
continuous and gapless. Actually, one can
understand this result by rewriting the function (\ref{Bessel}) in terms
of the coordinate $x^p$ by using (\ref{bending}). Indeed, $\rho\approx 0$
corresponds to large $|x^p|$ and $\xi(x^p)$ can be written in this limit as 
a simple plane wave:
\beq
\xi\,\approx\,e^{\pm iMx^p}\,\,,\qquad\qquad(\,|x^p|\to\infty\,)\,\,.
\label{planewave}
\eeq
Thus, the fluctuation spreads out of the defect locus at fixed $x^p$,
reflecting the fact that the bending has the effect of recombining, rather
than intersecting, the Dp-branes with the D(p+2)-branes. As in ref.
\cite{ARR} we can understand this result by looking at the IR form of the
open string metric (\ref{openstrmetric}). One gets:
\beq
\hat{\cal G}_{ab}d\xi^a\,d\xi^b\,\approx\,{L^{2\alpha}\over R^{2\alpha}}\,\,
\Bigg[\,dx_{1,p-1}^2\,+\,q^2\Big(\,
{d\rho^2\over \rho^4}\,+\, {1\over \rho^2}\,d\Omega_2^2\,\Big)
\,\Bigg]\,\,,
\qquad\qquad(\rho\approx 0)\,\,.
\label{IRmetric}
\eeq
By changing variables from $\rho$ 
to  $u=q/\rho$, this metric can be written as:
\beq
{L^{2\alpha}\over R^{2\alpha}}\,\,\Big[\,dx_{1,p-1}^2\,+\, du^2\,+\,u^2\,
d\Omega_2^2\,\Big]\,\,,
\label{IRMinkowski}
\eeq
which is nothing but the (p+3)-dimensional Minkowski space and, thus, one
naturally expects to get plane waves  as in (\ref{planewave}) as solutions of 
the fluctuation equations. This fact is generic for all the fluctuations of
this system. Recall that the other fields in the Lagrangian (\ref{fluct-lag})
are coupled. However, in appendix A we show that they can be decoupled by
generalizing the results of ref. \cite{WFO,AR}. The decoupled fluctuation
equations can actually be mapped \cite{MT} to that satisfied by the scalars
$\chi$. Thus, we conclude that the full mesonic mass spectrum is continuous
and gapless, as a consequence of the recombination of the color and flavor
branes induced by the worldvolume flux.

\subsection{An S-dual picture: the F1-Dp  intersection}

Let us now have a  look to the S-dual configurations for the IIB cases 
in this section, which will  give us information about the weak 't Hooft
coupling regime of the dual theory. For $p=3$ the S-dual background will be
once again $AdS_5\times S^5$. In this case, the D5-brane gets mapped to a
NS5 brane. However, since the dilaton is zero in this background, at least
formally this situation will be identical to the D3-D5 case already studied
above. In particular we will lose again the discrete spectrum. In turn, we
can look at the $p=1$ case, whose S-dual version is the F1-D3 intersection.
Actually, we will analyze the more general system corresponding to the F1-Dp
intersection, according to the array:
\beq
\begin{array}{ccccccccccl}
 &1&2&\cdots& p+1&p+2 &\cdots&9 & &&\nonumber \\
F1: & \times &\_ &\cdots &\_  &\_  & \cdots &\_ & &    
\nonumber
\\ Dp: &\_&\times&\cdots&\times&\_&\cdots&\_&&&
\end{array}
\label{F1Dpintersection}
\eeq
As in previous cases, we will consider a stack of F1 
strings, which we will treat as a background of 
 type II theory. The corresponding metric is given by:
\begin{equation}
ds^2=H^{-1}dx_{1,1}^2+d\vec{r}^{\, 2}\ ,
\label{F1metric}
\end{equation}
where, in the near-horizon limit, $H=R^6/r^6$, 
with $R^6\,=\,32\pi^2 (\alpha')^3\,g_s^2\,N$. The F1 background is also
endowed with a NSNS $B$ field and a non-trivial dilaton, given by:
\begin{equation}
B=H^{-1}dx_0\wedge dx_1\ ,\,\,\qquad\qquad
e^{-\Phi}=H^{\frac{1}{2}}\,\,.
\end{equation}
Let us now rewrite this solution in terms of a new coordinate system 
more suitable for our probe analysis. First of all, we split the coordinates
transverse to the F1 as
$\vec r=(\vec y, \vec z)$, where the $\vec y$ vector 
corresponds to the directions 
$2\cdots p+1$ and $\vec z$ refers to the coordinates
 transverse to both the F1 and Dp-brane.  Moreover, let us 
assume that $p>1$ and  use spherical
coordinates to parametrize the subspace spanned by the $y$'s, \ie\ $d\vec
y^{\,2}\,=\,d\rho^2\,+\,\rho^2\,d\Omega^2_{p-1}$. Then, the metric
(\ref{F1metric}) can be rewritten as:
\begin{equation}
ds^2=H^{-1}dx_{1,1}^2+d\rho^2+\rho^2d\Omega_{p-1}^2+d\vec{z}^{\,2}\ .
\label{F1metric-polar}
\end{equation}
The dynamics of the Dp-brane probe is determined by the DBI lagrangian, 
which in this case takes the form:
\beq
{\cal L}\,=\,-T_p\,e^{-\phi}\,\,\sqrt{-\det(\,g+{\cal F}\,)}\,\,,
\label{DBIactionF1Dp}
\eeq
where ${\cal F}$ is the following combination of the worldvolume gauge 
field strength $F$ and the pullback $P[B]$ of the NSNS $B$ field:
\beq
{\cal F}\,=\,F\,-\,P[B]\,\,.
\eeq
Let us choose $x^0, \rho$ and the $p-1$ angles parametrizing the 
$S^{p-1}$ sphere as our set of worldvolume coordinates. We will consider
embeddings of the type:
\beq
x^1\,=\,x(\rho)\,\,,\qquad\qquad
|\,\vec z\,|\,=\,L\,\,.
\label{F1Dp-embed-ansatz}
\eeq
Moreover, we will switch on an electric field $F_{0\rho}\equiv F$ in the
worldvolume, such that the only non-vanishing component of ${\cal F}$ is:
\beq
{\cal F}_{0\rho}\,=\,F\,-\,H^{-1}\,x'\,\,,
\label{F1Dp-curlyF}
\eeq
where, from now on, $H$ should be understood as the following function of $\rho$:
\beq
H=H(\rho)\,=\,\Bigg[\,{R^2\over \rho^2+L^2}\,\Bigg]^3\,\,.
\eeq
The form of the lagrangian density (\ref{DBIactionF1Dp}) for 
this ansatz can be straightforwardly computed, with the result:
\beq
{\cal L}\,=\,-T_p\,\rho^{p-1}\,\sqrt{\tilde g}\,\,\,
\sqrt{1+2Fx'-HF^2}\,\,,
\label{F1Dp-L}
\eeq
and the equation of motion for the electric field $F$ is:
\beq
{\partial\over \partial\rho}\,
\Bigg[\,{\partial {\cal L}\over \partial F}\,\Bigg]\,=\,0\,\,.
\eeq
This equation can be immediately integrated, namely:
\beq
{\rho^{p-1}\,\Big(\,HF\,-\,x'\,\Big)\over 
\sqrt{1+2Fx'-HF^2}}\,=\,c\,\,,
\label{electricFeq}
\eeq
where $c$ is a constant. Moreover, from (\ref{electricFeq}) 
we can obtain $F$ as a function of $x'$ and $\rho$:
\beq
F\,=\,H^{-1}\,\Bigg[\,x'\,+\,c\,
{\sqrt{H+(x')^2}\over \sqrt{c^2+\rho^{2(p-1)}\,H}}
\,\,\Bigg]\,\,.
\label{F1Dp-F}
\eeq
Actually, F can be eliminated in a systematic way by means of 
a Legendre transformation. Indeed, let us define the Routhian density ${\cal
R}$ as follows:
\beq
{\cal R}\,=\,F\,{\partial {\cal L}\over \partial F}\,-\,{\cal L}\,\,.
\eeq
By computing the derivative in the explicit expression of ${\cal L}$ in 
(\ref{F1Dp-L}), and by using (\ref{F1Dp-F}), one can readily show that
${\cal R}$ can be written as:
\beq
{\cal R}\,=\,T_p\,\sqrt{\tilde g}\,H^{-1}\,\,\Bigg[\,
\sqrt{c^2+\rho^{2(p-1)}\,H}\,\,\sqrt{H+(x')^2}\,+\,cx'\,\Bigg]\,\,.
\label{F1DpRouthian}
\eeq
The equation of motion for $x$  derived from ${\cal R}$ is just:
\beq
{\partial\over \partial\rho}\,\Bigg[\,{\partial {\cal R}\over \partial x'}\,
\Bigg]\,=\,0\,\,.
\eeq
A particular solution of this equation can be obtained by 
requiring the vanishing of
$\partial {\cal R}/\partial x'$.  By computing explicitly this derivative 
from the expression of ${\cal R}$ in (\ref{F1DpRouthian}) one easily shows
that the value of $x'$ for this particular solution is simply:
\beq
x'\,=\,-{c\over \rho^{p-1}}\,\,,
\label{F1Dpbending}
\eeq
which, for $p\not=2$ can be integrated as:
\beq
x(\rho)\,=\,{c\over p-2}\,\,{1\over \rho^{p-2}}\,+\,{\rm constant}\,\,,
\qquad
(p\not=2)\,\,,
\label{integratedF1Dpbending}
\eeq
while for $p=2$ the Dp-brane has a logarithmic bending of the 
type $x(\rho)\sim -c\log \rho$. Moreover, after substituting
(\ref{F1Dpbending}) on the right-hand side of (\ref{F1Dp-F})  one easily
realizes that the worldvolume gauge field $F$ for this configuration
vanishes, \ie:
\beq
F=0\,\,.
\label{F=0}
\eeq
Actually, it is also easy to verify from (\ref{F1Dp-F}) that the 
requirement of having vanishing electric gauge field on the worldvolume  is
equivalent to have a bending given by eq. (\ref{F1Dpbending}). Notice also
that the on-shell lagrangian density (\ref{F1Dp-L}) for this configuration
becomes 
${\cal L}=-T_p\,\rho^{p-1}\,\sqrt{\tilde g}$, which is independent of the
distance $L$. This suggests that the configuration is supersymmetric, a
fact that we will verify explicitly in the next subsection by looking at
the kappa symmetry of the embedding. 

Notice that the embedding (\ref{F1Dpbending}) depends on the constant $c$. This constant is constrained by a flux quantization condition which, for electric worldvolume gauge fields, was worked out in \cite{Camino:2001at} and reads:
\beq
\int_{S^{p-1}}\,\,{\partial {\cal L}\over \partial F}\,=\, nT_f\,\,,
\qquad\qquad n\in\ZZ\,\,.
\label{F1Dp-fluxquantization}
\eeq
From (\ref{F1Dp-L}) one easily gets:
\beq
{\partial {\cal L}\over \partial F}\,\Bigg|_{ F=0}\,=\,T_p\,\sqrt{\tilde g}\,c\,\,,
\eeq
which allows one to compute the integral on the left-hand side of (\ref{F1Dp-fluxquantization}). Let us express the result in terms of the Yang-Mills coupling, which was written
in terms of string theory quantities in (\ref{gym}). Taking into account that the Dp-brane tension $T_p$ is related to $g_{YM}$ as 
$T_p=T_f^2/g^2_{YM}$, one easily arrives at the following expression of
$c$ in terms of the integer $n$:
\beq
c\,={\alpha' g^2_{YM}\over \Omega_{p-1}}\,\,2\pi n\,\,,
\label{c-quantization}
\eeq
where $\Omega_{p-1}$ is the volume of a unit $S^{p-1}$, namely
$\Omega_{p-1}=2\pi^{{p\over 2}}/\Gamma({p\over 2})$. 
Physically, the integer $n$ represents the number of fundamental strings
 that are reconnected to the Dp-brane. Notice that for $p=3$ eq.
(\ref{c-quantization})  reduces to $c=n\pi\alpha' g_s$, to be compared
with the S-dual relation (\ref{q-k}).

\subsubsection{Supersymmetry}
The supersymmetric configurations of a D-brane probe in a given 
background are those for which
the following condition:
\beq
\Gamma_{\kappa}\,\epsilon\,=\,\epsilon\,\,,
\label{Gammakappa}
\eeq
is satisfied \cite{bbs}. In eq. (\ref{Gammakappa}), $\Gamma_{\kappa}$ is a 
matrix whose explicit
expression depends on the embedding of the probe (see below) and 
$\epsilon$ is a Killing spinor
of the background. For simplicity we will restrict ourselves to study the
kappa symmetry condition (\ref{Gammakappa}) in the type IIB theory.
First of all, let us define the induced
worldvolume gamma matrices as:
\beq
\gamma_{m}\,=\,\partial_{m}\,X^M\,E^{\bar N}_{M}\,\Gamma_{\bar N}\,\,,
\label{inducedgamma}
\eeq
where $\Gamma_{\bar N}$ are constant ten-dimensional Dirac matrices 
and $E^{\bar N}_{M}$ is the
vielbein for the ten-dimensional metric. Then, if $\gamma_{m_1m_2\cdots}$ 
denotes the antisymmetrized
product of the induced gamma matrices (\ref{inducedgamma}), the kappa 
symmetry matrix for a
Dp-brane in the type IIB theory is \cite{swedes}:
\bear
\Gamma_{\kappa}\,=\,{1\over \sqrt{-\det(g+{\cal F})}}
&&\sum_{n=0}^{\infty}\,{(-1)^n\over 2^n n!}\,\gamma^{m_1n_1\,
\cdots  m_n n_n}\,\,
{\cal F}_{m_1 n_1}\,\cdots\,{\cal F}_{m_n n_n}\,\,
\times\rc\rc
&&\times(\sigma_3)^{{p-3\over 2}\,-\,n}\,\, (i\sigma_2)\,\,\Gamma_{(0)}\,\,,
\label{generalgammak}
\eear
where $\Gamma_{(0)}$ denotes:
\beq
\Gamma_{(0)}\,=\,{1\over (p+1)!}\,\,\epsilon^{m_1\cdots m_{p+1}}\,\,
\gamma_{m_1\cdots m_{p+1}}\,\,.
\label{Gammazero}
\eeq
In eq. (\ref{generalgammak}) $\sigma_2$ and $\sigma_3$ are Pauli 
matrices that act on the two Majorana-Weyl components (arranged as a two-dimensional vector) of  the type IIB spinors.

Let us consider a Dp-brane embedded in the geometry (\ref{F1metric-polar}) according to the ansatz  (\ref{F1Dp-embed-ansatz}). Let us assume that 
we parametrize the $S^{p-1}$ sphere by means of the angles $\alpha^1,\cdots, \alpha^{p-1}$. The induced gamma matrices are:
\bear
&&\gamma_{x^0}\,=\,H^{-{1\over 2}}\,\Gamma_{x^0}\,\,,\rc\rc
&&\gamma_{\rho}\,=\,\Gamma_{\rho}\,+\,H^{-{1\over 2}}\,x'\,\Gamma_{x^1}\,\,,\rc\rc
&&\gamma_{\alpha^1\,\cdots\,\alpha^{p-1}}\,=\,\rho^{p-1}\,\,\sqrt{\tilde g}\,\,
\Gamma_{\Omega_{p-1}}\,\,,
\eear
where $\Gamma_{\Omega_{p-1}}\equiv \Gamma_{\alpha^1}\,\cdots\,
\Gamma_{\alpha^{p-1}}$. Using these matrices we can write the kappa symmetry matrix 
$\Gamma_{\kappa}$ in (\ref{generalgammak}) as:
\beq
\Gamma_{\kappa}\,=\,
{\rho^{p-1}\,\sqrt{\tilde g}\over  \sqrt{-\det(g+{\cal F})}}\,\,
(\sigma_3)^{{p-3\over 2}}\,\, (i\sigma_2)\,\Bigg[\,
H^{-{1\over 2}}\,\Gamma_{x^0\rho}\,+\,H^{-1 }\,x'\,\Gamma_{x^0 x^1}\,+\,
{\cal F}\,\sigma_3\,\Bigg]\,\Gamma_{\Omega_{p-1}}\,\,.
\eeq
Let us now study the action of $\Gamma_{\kappa}$ on the Killing spinor $\epsilon$. We shall  impose to $\epsilon$ the projections corresponding to the Dp-brane and the F1-string, namely:
\bear
&&(\sigma_3)^{{p-3\over 2}}\,\, (i\sigma_2)\,\Gamma_{x^0\rho}\,
\Gamma_{\Omega_{p-1}}\,\epsilon\,=\,\epsilon\,\,,\rc\rc
&&\sigma_3\,\Gamma_{x^0 x^1}\,\epsilon\,=\,\epsilon\,\,.
\eear
It is now straightforward to verify that 
\beq
\Gamma_{\kappa}\,\epsilon\,=\,
{\rho^{p-1}\,\sqrt{\tilde g}\over  \sqrt{-\det(g+{\cal F})}}\,\,
\Bigg[\,H^{-{1\over 2}}\,+\,
\Big(\,{\cal F}\,+\,H^{-1 }\,x'\,\Big)
\,(\sigma_3)^{{p-1\over 2}}\,\, (i\sigma_2)\,
\Gamma_{\Omega_{p-1}}\,\Bigg]\,\epsilon\,\,.
\label{F1DpGe}
\eeq
We want to impose that the right-hand side of (\ref{F1DpGe}) be $\epsilon$. It is clear that, if we do not want to impose any further projection to the spinor, we should require that the term that is  not proportional to the unit matrix cancels, which happens when:
\beq
{\cal F}\,+\,H^{-1 }\,x'\,=\,0\,\,.
\eeq
Notice that this condition is equivalent to require the vanishing of $F$ (see eq. (\ref{F1Dp-curlyF})), as claimed. Moreover, by computing the DBI determinant on the denominator of 
(\ref{F1DpGe}) one readily proves that, indeed, eq. (\ref{Gammakappa}) is satisfied by our configuration.

\subsubsection{Fluctuations}
Now we will study the fluctuations around the configuration described by eqs.
 (\ref{F1Dp-embed-ansatz}) and  (\ref{F=0}). We will only analyze the fluctuations on the transverse 
$\vec z$ space, which we  will denote by $\chi$.
 After a straightforward computation, we get that, up to quadratic
 order, the lagrangian density of these fluctuations is:
\begin{equation}
{\cal L}\,=\,-\rho^{p-1}\, \sqrt{\tilde{g}}\,\,\Bigg(\,1+\frac{c^2}
{\rho^{2(p-1)}\,H}\,\Bigg)\,{\cal{G}}^{\mu\nu}\partial_{\mu}\chi\partial_{\nu}\chi\,\,,
\label{F1Dp-fluc-lag}
\end{equation}
where the effective metric ${\cal{G}}_{\mu\nu}$ is given by:
\begin{equation}
{\cal{G}}_{\mu\nu}\,\,
dx^{\mu}dx^{\nu}\,=\,-H^{-1}(dx^0)^2+\Bigg(\,1+
\frac{c^2}{\rho^{2(p-1)}\,H}\,\Bigg)(\,d\rho^2+\rho^2d\Omega_{p-1}^2\,)\,\,.
\label{F1Dp-fluc-metric}
\end{equation}
As a check, one can verify that the equation derived from 
(\ref{F1Dp-fluc-lag}) for $p=3$  (\ie\ for the F1-D3 intersection) matches
precisely that of the transverse scalar fluctuations of the D1-D3 system
(\ie\ eq. (\ref{chi-fluct}) with $p=1$), once the constants $c$ and $q$ are
identified.   This is, of course, expected from S-duality and  implies that
the F1-D3 spectrum is continuous and gapless. For $p>3$ the meson spectrum
displays the same characteristics as in the F1-D3 intersection. However, the
F1-D2 system behaves differently. Indeed, for $p=2$ the profile function
$x(\rho)$ is logarithmic (see eqs. (\ref{F1Dpbending}) and
(\ref{integratedF1Dpbending})). Moreover, one can check that in this case
the effective metric (\ref{F1Dp-fluc-metric}) in the IR region $\rho\sim 0$
corresponds to an space of the type ${\rm Min}_{1,1}\times S^1$. Actually,
by studying the fluctuation equation derived from (\ref{F1Dp-fluc-lag}) for
$p=2$ and  $\rho\sim 0$, one can verify that non-oscillatory solutions can
exist if the KK momentum in the $S^1$ is non-zero. As one can check
by solving numerically the fluctuation equation, in this case the mass
spectrum starts with a finite number of discrete states, followed by a
continuum.

\setcounter{equation}{0}
\section{M2-M5 intersection and codimension one defects in M-theory}
\label{M2M5section}

We will consider now a close relative in M-theory of the 
Dp-D(p+2) intersections, namely the M2-M5 intersection along one common
spatial dimension. The corresponding array is:
\beq
\begin{array}{cccccccccccl}
 &1&2&3& 4& 5&6 &7&8&9 &10 \nonumber \\
M2: & \times &\times &\_ &\_ &\_ & \_&\_ &\_ &\_ &\_    
\nonumber
\\ M5: &\times&\_&\times&\times&\times&\times&\_&\_&\_ &\_
\end{array}
\label{M2M5intersection}
\eeq
Since this configuration can be somehow thought as the
uplift of the D2-D4 intersection  to eleven dimensions, 
we  expect a behaviour similar to the one studied in section
\ref{DpDp+2section}. Indeed, notice that the M5-brane induces a codimension
one defect in the M2-brane worldvolume. As in the previous examples we will
treat the highest dimensional brane (\ie\ the M5-brane) as a probe in the
background created by the lower dimensional object, which in this case is the
M2-brane. The near-horizon metric of the M2-brane background of
eleven-dimensional supergravity is:
\begin{equation}
ds^2=\frac{r^4}{R^4}dx_{1,2}^2+\frac{R^2}{r^2}d\vec{r}^{\,2}\,\, ,
\label{M2metric}
\end{equation}
where $R$ is constant, $dx_{1,2}^2$ represents the Minkowski metric in the
directions $x^0, x^1, x^2$ of the M2-brane worldvolume and $\vec r$ is an
eight-dimensional vector transverse to the M2-brane. The metric
(\ref{M2metric}) is the one of the $AdS_4\times S^7$ space, where the radius
of the $AdS_4$ ($S^7$) factor is $R/2$ ($R$). The actual value of $R$ for a
stack of $N$ coincident M2-branes is:
\beq
R^6=32\pi^2 l_p^6 N\,\,,
\label{M2-radius}
\eeq
where $l_p$ is the Planck length in eleven dimensions. This background is
also endowed with a three-form potential $C^{(3)}$, whose explicit
expression is:
\beq
C^{(3)}\,=\,{r^6\over R^6}\,\,dx^0\wedge dx^1\wedge dx^2\,\,.
\label{C3M2}
\eeq

The dynamics of the M5-brane probe is governed by the so-called PST action
\cite{PST}. In the PST formalism the worldvolume fields are a three-form
field strength $F$ and an auxiliary scalar $a$. This action is given by
\cite{PST}:
\bear
S\,&=&\,T_{M5}\,\int\,d^6\xi\,
\Bigg[\,-\sqrt{-{\rm det}(g_{ij}\,+\,\tilde H_{ij})}\,+\,
{\sqrt{-{\rm det}g}\over 4\partial a\cdot\partial a}\,
\partial_i a\,(\star H)^{ijk}\,H_{jkl}\partial^l a\,
\Bigg]\,+\rc\rc
&&+\,T_{M5}\int\Bigg[\,P[C^{(6)}]\,
+\,{1\over 2}\,F\wedge\,P[C^{(3)}]\,\Bigg]\,\,,
\label{stnueve}
\eear
where $T_{M5}=1/(2\pi)^5\,l_p^6$ is the tension of the M5-brane, $g$ is the
induced metric and $H$ is the following combination of the worldvolume
gauge field $F$ and the pullback of the three-form $C^{(3)}$:
\beq
H\,=\,F\,-\,P[C^{(3)}]\,\,.
\label{stsiete}
\eeq
Moreover, the field ${\tilde H}$ is defined as follows:
\beq
{\tilde H}^{ij}\,=\,{1\over 3!\,\sqrt{-{\rm det}\,g}}\,
{1\over \sqrt{-(\partial a)^2}}\,
\epsilon^{ijklmn}\,\partial_k\,a\,H_{lmn}\,\,,
\label{stocho}
\eeq
and the worldvolume indices in (\ref{stnueve}) are lowered with the induced
metric $g_{ij}$. 

In order to study the embedding of the M5-brane in the M2-brane background,
let us introduce a more convenient set of coordinates. Let us split the
vector $\vec r$ as $\vec r=\,(\vec y,\vec z)$, where $\vec y=(y^1,\cdots,
y^4)$ is the position vector along the directions $3456$  in the array
(\ref{M2M5intersection}) and $\vec z=(z^1,\cdots, z^4)$ corresponds to the
directions $7,8,9$ and $10$. Obviously, if $\rho^2=\vec y\cdot \vec y$, one
has that 
$\vec r^{\,2}=\rho^2+\vec z^{\, 2}$ and 
$d\vec r^{\,2}\,=\,d\rho^2\,+\,\rho^2\,d\Omega_3^2\,+\,d\vec{z}^{\,2}$, where 
$d\Omega_3^2$ is the line element of a three-sphere. Thus, the metric
(\ref{M2metric}) becomes:
\begin{equation}
ds^2=\frac{\big(\,\rho^2+\vec z^{\,2}\,\big)^2}{R^4}\,\,dx_{1,2}^2
+\frac{R^2}{\rho^2+\vec z^{\,2}}\,\,\big(\,d\rho^2+\rho^2d\Omega_3^2
+d\vec{z}^{\,2}\,\big)\,\,.
\end{equation}
We will now choose $x^0$, $x^1$, $\rho$ and the three angular coordinates
that parametrize $d\Omega_3^2$ as our worldvolume coordinates $\xi^i$.
Moreover, we will assume that the vector $\vec z$ is constant and we will
denote its modulus by $L$, namely:
\beq
|\,\vec z\,|\,=\,L\,\,.
\eeq
To specify completely the embedding of the M5-brane we must give  the form
of the remaining scalar $x^2$ as a function of the worldvolume coordinates.
For simplicity we will assume that $x^2$ only depends on the radial
coordinate $\rho$, \ie\ that:
\beq
x^2\,=\,x(\rho)\,\,.
\eeq
Moreover, we will switch on a magnetic field $F$ along the three-sphere of
the M5-brane worldvolume, in the form:
\begin{equation}
F=q\,{\rm Vol}\,(\,S^3\,)\,\, ,
\label{M5flux}
\end{equation}
where $q$ is a constant and ${\rm Vol}\,(\,S^3\,)$ is the volume form of the
worldvolume three-sphere. Notice that the induced metric for this
configuration is given by:
\beq
g_{ij}d\xi^i d\xi^j\,=\,{\big(\,\rho^2+L^2\,\big)^2\over R^4}\,\,dx^{2}_{1,1}
\,+\,{R^2\over \rho^2+L^2}\,\,\Bigg\{\,
\Bigg(\,
1\,+\,{\big(\,\rho^2+L^2\,\big)^3\over
R^6}\,(x')^2\,
\Bigg)\,d\rho^2\,+\,\rho^2\,d\Omega_3^2\,\Bigg\}\,.
\label{inducedmetricM5}
\eeq

In order to write the PST action for our ansatz we must specify the value of
the PST scalar $a$. As pointed out in ref. \cite{PST} the field $a$ can be
eliminated by gauge fixing, at the expense of losing manifest covariance.
Here we will choose a gauge such that the auxiliary PST scalar is:
\beq
a\,=\,x_1\,\,.
\eeq
It is now straightforward to prove that the only non-vanishing component of
the field $\tilde H$ is:
\beq
\tilde H_{x^0\rho}\,=\,-{i\over R^4}\,\,
{\big(\,\rho^2+L^2\,\big)^2\over \rho^3}\,\,
\Bigg(\,
1\,+\,{\big(\,\rho^2+L^2\,\big)^3\over R^6}\,(x')^2\,
\Bigg)^{{1\over 2}}\,q\,\,.
\label{tildeH}
\eeq
Using these results we can write the PST action (\ref{stnueve}) as:
\bear
&&S\,=\,-2\pi^2\,T_{M5}\,\,\int\,d^2x\,d\rho\,\,\Bigg[\,\rho^3\,
\sqrt{1\,+\,{\big(\,\rho^2+L^2\,\big)^3\over R^6}\,(x')^2}\,\,
\sqrt{1\,+\,{\big(\,\rho^2+L^2\,\big)^3\over R^6}\,
{q^2\over \rho^6}}\,+\,\rc\rc
&&\qquad\qquad\qquad\qquad\qquad\qquad
\qquad\qquad\qquad+\,
{\big(\,\rho^2+L^2\,\big)^3\over R^6}\,q\,x'\,\,\Bigg]\,\,.
\label{PSTansatz}
\eear
Let ${\cal L}$ be the lagrangian density for the PST action, which we can
take as given by the expression inside the brackets in (\ref{PSTansatz}). 
Since $x$ does not appear explicitly in the action, 
one can immediately write a first integral of the equation of motion of
$x(\rho)$, namely:
\beq
{\partial {\cal L}\over \partial x'}\,=\,{\rm constant}\,\,.
\label{PSTcyclic}
\eeq
By setting the constant on the right-hand side of (\ref{PSTcyclic}) equal
to zero, this equation reduces to a simple first-order equation for 
$x(\rho)$,
\ie:
\begin{equation}
x'\,=\,-\frac{q}{\rho^3}\,\,,
\label{BPSx}
\end{equation}
which can be immediately integrated to give:
\begin{equation}
x(\rho)\,=\,\bar x\,+\,\frac{q}{2\rho^2}\,\,,
\label{x-explicit}
\end{equation}
where $\bar x$ is a constant. Notice that the flux parametrized by $q$ 
induces a bending of the M5-brane, which is characterized by the non-trivial
dependence of $x$ on the holographic coordinate $\rho$. Actually, when the
first-order eq. (\ref{BPSx}) holds, the two square roots in (\ref{PSTansatz})
are equal and there is a cancellation with the last term in (\ref{PSTansatz}).
Indeed, the  on-shell action takes the form:
\begin{equation}
S=-2\pi^2T_5\int d^2x\,d\rho \,\rho^3\,\, ,
\end{equation}
which is  independent of the M2-M5 distance $L$. This is usually a signal of
supersymmetry and, indeed, we will verify in appendix \ref{kappa} that the
embeddings in which the flux and the bending are related as in  (\ref{BPSx})
are kappa symmetric. Thus, eq. (\ref{BPSx}) can be regarded as the
first-order BPS equation required by supersymmetry. Notice also that the
three-form flux (\ref{M5flux}) induces M2-brane charge in the M5-brane
worldvolume, as it is manifest from the form of the PST action
(\ref{stnueve}). In complete analogy with the Dp-D(p+2) system, we can
interpret the present M-theory configuration in terms of M2-branes that
recombine with the M5-brane. Moreover, in order to gain further insight on
the effect of the bending, let us rewrite the induced metric
(\ref{inducedmetricM5}) when the explicit form of $x(\rho)$ written in eq.
(\ref{x-explicit}) is taken into account. One gets:
\beq
{\big(\,\rho^2+L^2\,\big)^2\over R^4}\,\,dx^{2}_{1,1}
\,+\,{R^2\over \rho^2+L^2}\,\,\Bigg\{\,\Bigg(\,
1\,+\,{q^2\over R^6}\,\,
{\big(\,\rho^2+L^2\,\big)^3\over
\rho^6}\,
\Bigg)\,d\rho^2\,+\,\rho^2\,d\Omega_3^2\,\Bigg\}\,\,.
\label{AdS3-M5metric}
\eeq
From (\ref{AdS3-M5metric}) one readily notices that the UV induced metric at 
$\rho\to\infty$ (or, equivalently when the M2-M5 distance L is zero) takes
the form $AdS_3 (R_{eff}/2)\times S^3 (R)$, where the $AdS_3$ radius 
$R_{eff}$ depends on the flux as:
\beq
R_{eff}\,=\,\Bigg(\,1\,+\,
{q^2\over R^6}\,\Bigg)^{1\over 2}\,\,R\,\,.
\label{Reff}
\eeq
Therefore, our M5-brane is wrapping an  $AdS_3$ submanifold of the  $AdS_4$
background. Actually, there are infinite ways of embedding an  $AdS_3$ within
an $AdS_4$ space and the bending of the probe induced by the flux is
selecting one particular case of these embeddings. In order to shed light on
this, let us suppose that we have an $AdS_4$ metric of the form:
\beq
ds^2_{AdS_4}\,=\,{\rho^4\over R^4}\,\,dx^2_{1,2}\,+\,
{R^2\over \rho^2}\,\,d\rho^2\,\,.
\label{AdS4metric}
\eeq
Let us now change variables from $(x^{0,1}, x^2, \rho)$ to 
$(\hat x^{0,1}, \varrho, \eta)$, as follows:
\beq
x^{0,1}\,=\,2\,\hat x^{0,1}\,\,,\qquad\qquad
x^2\,=\,\bar x\,+\,{2\over \varrho}\,\tanh\eta\,,\qquad\qquad
\rho^2\,=\,{R^3\over 4}\,\,\varrho\cosh\eta\,\,,
\label{changevariables}
\eeq
where $\bar x$ is a constant. In these new variables the $AdS_4$ metric
(\ref{AdS4metric}) can be written as a foliation by $AdS_3$ slices, namely:
\beq
ds^2_{AdS_4}\,=\,{R^2\over 4}\,
(\,\cosh^2\eta\,ds^2_{AdS_3}\,+\,d\eta^2\,)\,\,,
\label{foliation}
\eeq
where $ds^2_{AdS_3}$ is given by:
\beq
ds^2_{AdS_3}\,=\,\varrho^2\,
\big(\,-(d\hat x^{0})^2\,+\,(d\hat x^{1})^2\,\big)\,+\,
{d\varrho ^2\over \varrho^2}\,\,.
\eeq
Clearly the $AdS_3$ slices in (\ref{foliation}) can be obtained by taking
$\eta={\rm constant}$. The radius of such $AdS_3$ slice is $R_{eff}/2$, with:
\beq
R_{eff}\,=\,R\,\,\cosh\eta\,\,.
\label{sliceradius}
\eeq
Moreover, one can verify easily by using the change of variables
(\ref{changevariables})  that our embedding
(\ref{x-explicit}) corresponds to one of such 
$AdS_3$ slices with:
\beq
\eta\,=\,\eta_q\,=\,\sinh^{-1}\,\Big({q\over R^3}\Big)\,\,.
\eeq
Furthermore, one can check that the $AdS_3$ radius $R_{eff}$ of eq. 
(\ref{sliceradius}) reduces to (\ref{Reff}) when $\eta=\eta_q$.

\subsection{Fluctuations}

Let us now study the fluctuations of the M2-M5 intersection. For simplicity 
we will focus on the fluctuations of 
the transverse scalars which, without loss of generality,  
we will parametrize as:
\begin{equation}
z^1=L+\chi^1\ ,\,\qquad\qquad z^m=\chi^m\,\,,\qquad\qquad (m=2,\cdots,
4)\,\,.
\end{equation}
Let us substitute this ansatz in the PST action and keep 
up to second order terms in the fluctuation $\chi$. As the calculation is
very similar to the one performed in subsection \ref{Dp-D(p+2)fluctuations},
we skip the details and give the final result for the effective lagrangian of
the fluctuations, namely:
\beq
{\cal L}\,=\,-\rho^3\,\sqrt{\tilde g}\,\,
{R^2\over \rho^2\,+\,L^2}\,\,\Bigg[\,1\,+\,
{q^2\over R^6}\,\,
{\big(\,\rho^2+L^2\,\big)^3\over
\rho^6}\,\Bigg]\,\hat{\cal G}^{ij}\,\partial_i\chi\,\partial_j\chi\,\,,
\label{lag-fluc-M2M5}
\eeq
where $\tilde g$ is the determinant of the round metric of the $S^3$ and
$\hat{\cal G}_{ij}$ is the following effective metric on the M5-brane
worldvolume:
\beq
\hat{\cal G}_{ij}\,d\xi^i\,d\xi^j\,=\,
{\big(\,\rho^2+L^2\,\big)^2\over R^4}\,\,dx^{2}_{1,1}
\,+\,{R^2\over \rho^2+L^2}\,\,\Bigg(\,
1\,+\,
\,{q^2\over R^6}\,\,
{\big(\,\rho^2+L^2\,\big)^3\over
\rho^6}
\Bigg)\,\Big(\,d\rho^2\,+\,\rho^2\,d\Omega_3^2\,\Big)\,\,.
\label{effect-metricM2M5}
\eeq
Notice the close analogy with the Dp-D(p+2) system studied in subsection
\ref{Dp-D(p+2)fluctuations}. Actually (\ref{effect-metricM2M5}) is the
analogue of the open string metric in this case. The equation of motion for
the scalars can be derived straightforwardly from the lagrangian density 
(\ref{lag-fluc-M2M5}). For $q=0$ this equation was integrated in ref.
\cite{AR}, where the meson mass spectra was also computed. This fluxless
spectra is discrete and displays a mass gap. As happened with the codimension
one defects in type II theory studied in section \ref{DpDp+2section}, the
situation changes drastically when
$q\not=0$. To verify this fact let us study the form of the effective metric
(\ref{effect-metricM2M5}) in the UV ($\rho\to\infty$) and in the IR
($\rho\to 0$). After studying this metric when $\rho\to\infty$, one easily
concludes that the UV is of the form $AdS_3 (R_{eff}/2)\times S^3 (R_{eff})$,
where $R_{eff}$ is just the effective radius with flux written in
(\ref{Reff}). Thus, the effect of the flux in the UV is just a redefinition
of the $AdS_3$ and $S^3$ radii of the  metric governing the
fluctuations. On the contrary, for $q\not=0$ the behaviour of this metric in
the  IR  changes drastically with respect to the fluxless
case. Indeed, for $\rho\approx 0$ the metric (\ref{effect-metricM2M5}) takes
the form:
\beq
{L^{4}\over R^{4}}\,\,
\Bigg[\,dx_{1,1}^2\,+\,q^2\Big(\,
{d\rho^2\over \rho^6}\,+\, {1\over \rho^4}\,d\Omega_2^2\,\Big)
\,\Bigg]\,\,,
\qquad\qquad(\rho\approx 0)\,\,.
\label{IR-M2M5metric}
\eeq
Notice the analogy of (\ref{IR-M2M5metric}) with the IR metric
(\ref{IRmetric}) of the Dp-D(p+2) defects. Actually, the IR limit of the
equation of motion of the fluctuation can be integrated, as in
(\ref{Bessel}),  in terms of Bessel functions, which for
$\rho\approx 0$ behave as plane waves of the form $e^{\pm i Mx}$, where $x$
is the function (\ref{x-explicit}). Notice that $\rho\approx 0$ corresponds
to large  $x$ in  (\ref{x-explicit}). Thus, the fluctuations spread out of
the defect and oscillate infinitely at the IR and, as a consequence, the mass
spectrum is continuous and gapless. In complete analogy with the Dp-D(p+2)
with flux, this is a consequence of the recombination of the M2- and
M5-branes and should be understood microscopically in terms of dielectric
multiple M2-branes polarized into a M5-brane, once such an action is
constructed.

\setcounter{equation}{0}
\section{The codimension two defect} 

We now analyze the codimension two defect, which can be engineered in type
II string  theory as a Dp-Dp intersection over $p-2$ spatial dimensions.
We will consider a single Dp$'$-brane intersecting a stack of $N$
Dp-branes, according to the array:
\beq
\begin{array}{ccccccccccl}
 &1&\cdots&p-2& p-1& p&p+1 &p+2&\cdots&9 & \nonumber \\
Dp: & \times &\cdots &\times &\times &\times & \_&\_ &\cdots &\_ &    
\nonumber
\\ Dp\,': &\times&\cdots&\times&\_&\_&\times&\times&\cdots&\_ &
\end{array}
\label{DpDpintersection}
\eeq
In the
limit of large $N$ we can think of the system as a probe Dp$'$-brane in
the near horizon geometry of the Dp-branes given by (\ref{Dpmetrica}),
(\ref{Dpdilaton}) and (\ref{DpRR}). It is clear from the array 
(\ref{DpDpintersection}) that the Dp$'$-brane produces a defect of
codimension two  in the field theory dual to the stack of Dp-branes. The
defect field theory dual to the D3-D3 intersection was studied in detail in
ref. \cite{CEGK} (see also ref. \cite{Kirsch:2004km}).  Notice also that
this same D3-D3 intersection was considered in \cite{Gukov:2006jk} in
connection with the surface operators of ${\cal N}=4$ super Yang-Mills
theory, in the context of the geometric Langlands program. 

In order to describe the dynamics of the Dp$'$-brane probe, let us relabel the 
$x^{p-1}$ and $x^{p}$ coordinates appearing in the metric (\ref{Dpmetrica}) as:
\beq
\lambda^1\,\equiv\,x^{p-1}\,\,,
\qquad\qquad
\lambda^2\,\equiv\,x^{p}\,\,.
\eeq
Moreover, we will split the coordinates $\vec r$ transverse to the Dp-branes as
$\vec r\,=\,(\vec y\,,\vec z)$, where $\vec y\,=\,(y^1,y^2)$ corresponds to the p+1 and
p+2 directions in (\ref{DpDpintersection}) and $\vec z\,=\,(z^1,\cdots, z^{7-p})$ 
to the remaining transverse coordinates. With this split of coordinates the
background metric reads:
\beq
ds^2\,=\,\Bigg[\,{\vec y^{\,2}\,+\,\vec z^{\,2}\over
R^2}\,\Bigg]^{\alpha}\,\big(\,dx_{1,p-2}^2\,+\,d \vec\lambda^{\,2}\,\big)
\,+\, \Bigg[\,{R^2\over
\vec y^{\,2}\,+\,\vec z^{\,2}}\,\Bigg]^{\alpha}
\,\big(\,d\vec y^{\,2}\,+\,d \vec z^{\,2}\,\big)\,\,,
\eeq
where $dx_{1,p-2}^2$ is the Minkowski metric in the coordinates $x^0,\cdots
x^{p-2}$ and $\alpha$ has been defined in (\ref{Dpmetrica}).

\subsection{Supersymmetric embeddings}
To study the embeddings of the Dp$'$-brane probe in the background  
(\ref{Dpmetrica})-(\ref{DpRR}) let us consider 
$\xi^m\,=\,(x^0,\cdots,x^{p-2},y^1,y^2)$ as worldvolume coordinates. In this
approach $\vec \lambda$ and $\vec z$ are scalar fields that characterize the
embedding. Actually, we will restrict ourselves to the case in which $\vec
\lambda$ depends only on the $\vec y$ coordinates (\ie\ 
$\vec\lambda\,=\,\vec\lambda(\,\vec y\,)$) and the transverse separation 
$|\,\vec z\,|$ is constant, \ie\ $|\,\vec z\,|=L$. 

In order to characterize the embeddings of the probe 
that preserve supersymmetry, let us try to implement the kappa symmetry
condition (\ref{Gammakappa}). The induced gamma matrices 
$\gamma_{x^\mu}$ ($\mu=0,\cdots, p-2$) and $\gamma_{y^i}$ ($i=1,2$) 
can be computed from eq. (\ref{inducedgamma}), with the result:
\bear
&&\gamma_{x^\mu}=\Bigg[\,{\rho^2+L^2\over R^2}\,\Bigg]^{{\alpha\over 2}}\,\,
\Gamma_{x^\mu}\,\,,\rc\rc
&&\gamma_{y^i}=
\Bigg[\,{R^2\over \rho^2+L^2}\,\Bigg]^{{\alpha\over 2}}\,\,\Gamma_{y^i}\,+\,
\Bigg[\,{\rho^2+L^2\over R^2}\,\Bigg]^{{\alpha\over 2}}\,\,
\Big[\,\partial_i\lambda^1\,\Gamma_{\lambda^1}\,+\,\partial_{i}\lambda^2
\,\Gamma_{\lambda^2}\,\Big]\,\,,
\label{DpDpinducedgammas}
\eear
where $\partial_i\equiv \partial_{y^i}$ and, as before, we have 
defined $\rho^2=\vec y\cdot\vec y$.   To simplify matters, let us assume
that $p$ is odd and, thus, we are working on the type IIB theory. The
general expression of the kappa symmetry matrix $\Gamma_{\kappa}$  has been
written in eq. (\ref{generalgammak}). For the present case this matrix reads:
\beq
\Gamma_{\kappa}\,=\,{1\over \sqrt{-\det(g)}}\,\,
\Bigg[\,{\rho^2+L^2\over R^2}\,\Bigg]^{{(p-1)\alpha\over 2}}\,\,
(\sigma_3)^{{p-3\over 2}}\,(i\sigma_2)\,
\Gamma_{x^0\cdots x^{p-2}}\,\,\gamma_{y^1y^2}\,\,.
\label{GammakappaDpDp}
\eeq
The antisymmetrized product $\gamma_{y^1y^2}$ can be straightforwardly 
computed from the expression of the $\gamma_{y^i}$ matrices in
(\ref{DpDpinducedgammas}). One gets:
\bear
&&\Bigg[\,{\rho^2+L^2\over R^2}\,\Bigg]^{\alpha}\,\,
\gamma_{y^1y^2}\,=\,\Gamma_{y^1y^2}\,+\,
\Bigg[\,{\rho^2+L^2\over R^2}\,\Bigg]^{2\alpha}\,\,
\Big(\,\partial_1\lambda^1\,\partial_2\lambda^2\,-\,
\partial_1\lambda^2\,\partial_2\lambda^1\,\Big)\,
\Gamma_{\lambda^1\lambda^2}\,+\,\rc\rc
&&\qquad
+\,\Bigg[\,{\rho^2+L^2\over R^2}\,\Bigg]^{\alpha}\,\,
\Big[\,\partial_2\lambda^1\,\Gamma_{y^1\lambda^1}\,+\,
\partial_2\lambda^2\,\Gamma_{y^1\lambda^2}\,-\,
\partial_1\lambda^1\,\Gamma_{y^2\lambda^1}\,-\,
\partial_1\lambda^2\,\Gamma_{y^2\lambda^2}\,\,\Big]\,\,.
\label{gammay1y2}
\eear
Let us now use this expression to fulfill the condition $\Gamma_{\kappa}\epsilon\,=\,\epsilon$, where $\epsilon$ is a Killing spinor of the Dp-brane background. For a generic value of $p$ these Dp-brane spinors satisfy the projection condition:
\beq
(\sigma_3)^{{p-3\over 2}}\,\, (i\sigma_2)\,
\Gamma_{x^0\cdots x^{p-2}}\,\Gamma_{\lambda^1\lambda^2}\,
\epsilon\,=\,\epsilon\,\,.
\label{Dp-projection}
\eeq
Moreover we will also impose the projection corresponding to the Dp$' $- brane probe, namely:
\beq
(\sigma_3)^{{p-3\over 2}}\,\, (i\sigma_2)\,
\Gamma_{x^0\cdots x^{p-2}}\,\Gamma_{y^1y^2}\,\epsilon\,=\,\epsilon\,\,.
\label{Dp'projection}
\eeq
Notice that (\ref{Dp-projection}) and (\ref{Dp'projection}) are compatible, 
as it should for a supersymmetric intersection. Moreover, they can be
combined to give:
\beq
\Gamma_{y^1y^2}\,\epsilon\,=\,\Gamma_{\lambda^1\lambda^2}\,\epsilon\,\,,
\label{DpDpcondition}
\eeq
which implies that:
\bear
&&\Bigg[\,{\rho^2+L^2\over R^2}\,\Bigg]^{\alpha}\,\,
\gamma_{y^1y^2}\,\epsilon\,=\,\Bigg[\,1\,+\,
\Bigg[\,{\rho^2+L^2\over R^2}\,\Bigg]^{2\alpha}\,\,
\Big(\,\partial_1\lambda^1\,\partial_2\lambda^2\,-\,
\partial_1\lambda^2\,\partial_2\lambda^1\,\Big)\,\Bigg]
\Gamma_{\lambda^1\lambda^2}\,\epsilon\,+\,\rc\rc
&&\qquad
+\,\Bigg[\,{\rho^2+L^2\over R^2}\,\Bigg]^{\alpha}\,\,
\Big[\,(\partial_2\lambda^1\,+\,\partial_1\lambda^2)
\Gamma_{y^1\lambda^1}\,+\,
(\partial_2\lambda^2\,-\,\partial_1\lambda^1)
\Gamma_{y^1\lambda^2}\,\Big]\,\epsilon\,\,.
\label{secondgammay1y2}
\eear
We can now use this result to compute $\Gamma_{\kappa}\,\epsilon$, 
where $\Gamma_{\kappa}$ is given in (\ref{GammakappaDpDp}). By using the
condition (\ref{DpDpcondition}) one easily gets that the terms of the first
line of the right-hand side of (\ref{secondgammay1y2}) give contributions
proportional to the identity matrix, while those on the second line of
(\ref{secondgammay1y2}) give rise to terms that contain matrices that do not
act on $\epsilon$ as the identity unless we impose some extra projections
which would reduce the amount of preserved supersymmetry. Since we do not
want this to happen, we require that the coefficients of
$\Gamma_{y^1\lambda^1}$ and $\Gamma_{y^1\lambda^2}$ in
(\ref{secondgammay1y2}) vanish, \ie:
\beq
\partial_1\lambda^1\,=\,\partial_2\lambda^2\,\,,\qquad\qquad
\partial_2\lambda^1\,=-\,\partial_1\lambda^2\,\,.
\label{CR}
\eeq
Notice that eq. (\ref{CR}) is nothing but the Cauchy-Riemann equations. 
Indeed, let us define the following complex combinations of worldvolume
coordinates and scalars
\footnote{The complex worldvolume coordinate $Z$ should not be confused with the real transverse scalars $\vec z$. Notice also that $\rho^2=|Z|^2$. }:
\beq
Z\,=\,y^1\,+\,iy^2\,\,,\qquad\qquad
W\,=\,\lambda^1\,+\,i\lambda^2\,\,.
\eeq
In addition, if we define the holomorphic and antiholomorphic derivatives as:
\beq
\partial\,=\,{1\over 2}\,(\partial_1\,-\,i\partial_2)\,\,,\qquad\qquad
\bar\partial\,=\,{1\over 2}\,(\partial_1\,+\,i\partial_2)\,\,,
\eeq
then  (\ref{CR}) can be written as:
\beq
\bar\partial\,W\,=\,0\,\,,
\eeq
whose general solution is an arbitrary holomorphic function of $Z$, namely:
\beq
W\,=\,W(Z)\,\,.
\eeq
It is also straightforward to check that for these holomorphic embeddings the induced metric 
takes the form:
\beq
\Bigg[\,{\rho^2+L^2\over R^2}\,\Bigg]^{\alpha}\,\,dx^2_{1,p-2}\,+\,
\Bigg[\,{R^2\over \rho^2+L^2}\,\Bigg]^{\alpha}\,\,
\Bigg[\,1\,+\,\Bigg[\,{\rho^2+L^2\over R^2}\,\Bigg]^{2\alpha}\,\,
\partial W\bar\partial \bar W\,\Bigg]\,dZ\,d\bar Z\,\,,
\label{DpDp-ind-metric-holo}
\eeq
whose determinant is:
\beq
\sqrt{-\det (g)}\,=\,\Bigg[\,{\rho^2+L^2
\over R^2}\,\Bigg]^{{(p-3)\alpha\over 2}}\,\,
\Bigg[\,1\,+\,\Bigg[\,{\rho^2+L^2\over R^2}\,\Bigg]^{2\alpha}\,\,
\partial W\bar\partial \bar W\,\Bigg]\,\,.
\eeq
Using this result one can easily verify that the condition $\Gamma_{\kappa}\epsilon=\epsilon$ is indeed satisfied. Moreover, 
for  these holomorphic embeddings the DBI lagrangian density takes the
form:
\beq
{\cal L}_{DBI}\,=\,-T_p\,e^{-\phi}\,\,\sqrt{-\det (g)}\,=\,-T_p\,
\Bigg[\,1\,+\,\Bigg[\,{\rho^2+L^2\over R^2}\,\Bigg]^{2\alpha}\,\,
\partial W\bar\partial \bar W\,\Bigg]\,\,,
\label{DpDpLDBI}
\eeq
where we have used the value of $e^{-\phi}$ for the Dp-brane background 
displayed in eq. (\ref{Dpdilaton}). On the other hand, from the form of
the RR potential 
$C^{(p+1)}$ written in (\ref{DpRR}) one can readily check that, for these holomorphic embeddings,  the WZ piece of the lagrangian  can be written as:
\beq
{\cal L}_{WZ}\,=\,T_p\,\Bigg[\,{\rho^2+L^2\over R^2}\,\Bigg]^{2\alpha}\,
\partial W\bar\partial \bar W\,\,.
\eeq
Notice that, for these holomorphic embeddings, the WZ lagrangian 
${\cal L}_{WZ}$ cancels against the second term of ${\cal L}_{DBI}$ (see
eq. (\ref{DpDpLDBI})). Thus, once again, the on-shell action is
independent of the distance $L$, a result which is a consequence of
supersymmetry and holomorphicity.

Notice that, from the point of view of supersymmetry, any holomorphic
 curve $W(Z)$ is allowed. Obviously, we could have $W=$constant. In this
case the probe sits at a particular constant point of its transverse
space and does not recombine with branes of the background. If, on the
contrary, $W(Z)$  is not constant, Liouville theorem ensures us that it
cannot be bounded in  the whole complex plane. The points at which $|W|$
diverge are spikes of the probe profile,  and one can interpret them  as
the points where the probe and background branes merge. Notice that, as
opposed to the other cases studied in this paper, the non-trivial profile
of the embedding is not induced by the addition of any worldvolume field.
Thus,  we are not dissolving any further charge in the probe brane and a
dielectric interpretation is not possible now. 

The field theory dual  for the $p=3$ system has been worked out 
in refs.  \cite{CEGK} and \cite{Kirsch:2004km}. The dual gauge theory 
for this D3-D3 intersection was shown to correspond to two ${\cal N}=4$
four-dimensional theories coupled to each other through a two-dimensional
defect that hosts a bifundamental hypermultiplet. The Coulomb branch
corresponds to taking the embedding $W=$ constant.  Moreover, one can
seek for a Higgs branch arising from the corresponding $D$ and $F$
flatness conditions of the supersymmetric defect theory.  Actually, it
was shown in \cite{CEGK, Kirsch:2004km} that this Higgs branch
corresponds to the embedding $W=c/Z$, where $c$ is a constant.
Interestingly, only for these embeddings the induced UV metric is of the
form $AdS_3\times S^1$. Indeed, one can check that the metric
(\ref{DpDp-ind-metric-holo}) for $p=3$ (and $\alpha=1$) and for the
profile $W=c/Z$ reduces in the UV to that of the $AdS_3\times S^1$ space,
where the two factors have the same radii 
$R_{eff}=\sqrt{1+{c^2\over R^4}}\,\,R$. Thus, as in the M2-M5 intersection of section \ref{M2M5section}, the constant $c$ parametrizes the particular $AdS_3\times S^1$ slice of the 
$AdS_5\times S^5$ space that is occupied by our D3-brane probe.

\subsection{Fluctuations of the Dp-Dp intersection}

Let us now study the fluctuations around the previous configurations.  
We will concentrate on the fluctuations of the scalars transverse to both
types of branes, \ie\ those along the $\vec z$ directions. Let $\chi$ be
one of such fields. Expanding the action up to quadratic order in the
fluctuations it is easy to see that the lagrangian density for  $\chi$ is:
\beq
{\cal L}\,=\,-\Bigg[\,{R^2\over \rho^2+L^2}\,\Bigg]^{\alpha}\,\,
\Bigg[\,1\,+\,\Bigg[\,{\rho^2+L^2\over R^2}\,\Bigg]^{2\alpha}\,
\partial W\bar\partial \bar W\,\Bigg]\,
{\cal G}^{mn}\,\partial_m\chi\,\partial_n\chi\,\,,
\eeq
where ${\cal G}_{mn}$ is the induced metric (\ref{DpDp-ind-metric-holo}).
Let us parametrize the complex variable $Z$  in terms of polar
coordinates  as $Z=\rho\, e^{i\theta}$ and let us separate variables  in
the fluctuation equation as
\beq
\chi\,=\,e^{ikx}\,e^{il\theta}\,\,\xi(\rho)\,\,,
\eeq
where the product $kx$ is performed with the Minkowski metric of the defect. 
If $M^2=-k^2$, the equation of motion for the radial
 function $\xi(\rho)$ takes the form:
\beq
\Bigg[\,\,\Bigg[\,{R^2\over \rho^2+L^2}\,\Bigg]^{2\alpha}\,\,\Bigg[\,1\,+\,
\Bigg[\,{\rho^2+L^2\over R^2}\,\Bigg]^{2\alpha}\,
\partial W\bar\partial \bar W\,\Bigg]\,M^2\,-\,{l^2\over \rho^2}\,+\,
{1\over \rho}\,\,
\partial_{\rho}\big(\,\rho\,\partial_\rho\big)\,\,
\Bigg]\,\xi(\rho)\,=\,0\,\,.
\label{DpDpflucteq}
\eeq
For $W=$constant, eq. (\ref{DpDpflucteq}) was solved in ref. \cite{AR}, 
where it was shown to give rise to a mass gap and a discrete spectrum of
$M$. As in the case of the codimension one defects, this conclusion
changes completely when we go to the Higgs branch. Indeed, let us
consider the embeddings with $W\sim 1/Z$. One can readily prove that for
$\rho\to\infty$ the function $\xi(\rho)$ behaves as 
$\xi(\rho)\sim c_1\rho^{l}+c_2\rho^{-l}$, 
which is exactly the same behaviour as in the 
$W=$constant case.  However, in the opposite limit $\rho\to 0$ 
the fluctuation equation can be solved in terms of Bessel functions which
oscillate infinitely as $\rho\to0$. Notice that, for our Higgs branch
embeddings, $\rho\to 0$ means $W\to\infty$ and, therefore, the
fluctuations are no longer localized at the defect, as it happened in the
case of the Dp-D(p+2) and M2-M5 intersections. Thus we conclude that,
also in this case, the mass gap is lost and the spectrum is continuous.

\section{Conclusions}
 
In this paper we have studied the holographic description of the Higgs
branch of a large class of theories with fundamental matter.  These 
theories are embedded in string theory as supersymmetric systems of
intersecting branes. The strings joining both kind of  branes 
give rise to bifundamental matter confined to the
intersection, which once the suitable field theory limit is taken,
becomes fundamental matter with a flavor symmetry.

The general picture that emerges from our results is that the Higgs phase
is realized by recombining both types of intersecting branes. From the
point of view of the higher dimensional flavor brane the recombination
takes place when a suitable embedding is chosen and/or some flux of the
worldvolume gauge field is switched on. This  flux is
dissolving color brane charge in the flavor branes and, thus,  it is
tempting to search for a microscopical description from the point of view
of those dissolved branes. Indeed, we have seen that the vacuum
conditions of the dielectric description (when this description is
available) match exactly the F- and D-flatness
constraints that give  rise to the Higgs phase on the field theory side,
which gives support to  our holographic description of the Higgs branch.

The first case studied was the Dp-D(p+4) intersections, where the flavor 
D(p+4)-branes fill completely the worldvolume directions of the color 
Dp-brane. Following \cite{EGG}, we argued that the holographic description
of the Higgs branch of this system corresponds to having a self-dual
gauge field along the directions of the worldvolume of the D(p+4)-brane
that are orthogonal to the Dp-brane. To confirm this statement we have
worked out in detail the microscopic description of this system and we
have computed the meson mass spectra as a function of the quark VEV.

We also analyzed other intersections that are dual to gauge theories
containing defects of non-vanishing codimension. The paradigmatic example
of these theories is the Dp-D(p+2) system, where a detailed microscopic
description can be found. Other cases include the M2-M5 intersection in
M-theory as well as the Dp-Dp system, which gives rise to a codimension
two defect. In this latter case the field theory limit does
not decouple the flavor symmetry, so we actually have a $SU(N)\times SU(M)$
theory. In addition, the profile of the intersection is only constrained to
be holomorphic in certain coordinates, but is otherwise unspecified. In
any case, it turns out that conformal invariance in the UV is preserved only
for two particular curves, which can be shown to correspond to the Coulomb
and Higgs phases (see \cite{CEGK}). In all these non-zero codimension defect
theories we studied the meson spectrum and we have shown that it is
continuous and  that the mass gap is lost. The reason behind this result is
the fact that, due to the recombination of color and flavor branes in the
Higgs branch, the defect can spread over the whole bulk, which leads to an
effective Minkowski worldvolume metric in the IR for the flavor brane.
This implies the loss of a KK scale coming from a compact manifold and,
therefore, the disappearance  of the discrete spectrum. Notice that the 
case of the Dp-D(p+4) system  is different,
since in this case  the defect fills the whole color brane and there is no
room for spreading on the Higgs branch.

Also the Dp-Dp case deserves special attention, since it behaves in a
completely different manner to all the other intersections. As we
already mentioned  the intersection profile is not uniquely
fixed by supersymmetry. However,  just for two of all the possible embeddings
we recover conformal invariance in the UV. While one of them corresponds to
the Coulomb phase, the other corresponds to the Higgs phase. It should be
stressed that in this case there is no need for extra flux to get the Higgs
phase, which in this sense is purely geometrical. The other important
difference is that in this case the field
theory limit does not decouple any of the gauge symmetries. Then, our fields
will be bifundamentals under the gauge group on each Dp-brane. Taking into
account the relation with the surface operators in gauge theories
\cite{Gukov:2006jk},  it would be interesting to gain more understanding of
this system.

Let us now discuss some of the possible extensions of our work. Notice that
our analysis has been performed in the probe approximation, in which we
neglect the backreaction of the flavor branes on the geometry. This
approximation is valid when the number of flavor branes is small as compared
to the number of color branes. The analysis of the backreacted geometry
corresponding to the Higgs branch is of obvious interest. In particular it
would be very exciting to find the way in which the backreacted geometry
encodes some of the phenomena that we have uncovered in the probe
approximation. Actually, the backreacted geometry corresponding to the D3-D5
intersection was found in refs. \cite{Lunin:2006xr,Gomis:2006cu}. Also, it
would be interesting to see if one can apply the smearing procedure proposed
in \cite{Casero:2006pt} (see also \cite{unquenched}) to find a solution of
the equations of motion of the gravity plus branes systems studied in this
paper. 

Another problem of great interest is trying to describe holographically
(even in the probe approximation) the Higgs branch of theories with less
supersymmetry. The most obvious case to look at   would be that of branes
intersecting on the conifold, such as the D3-D7 systems in the
Klebanov-Witten model \cite{KW} and its generalizations. Actually, the
supersymmetric D3-D5 intersections with flux on the conifold and on more
general Sasaki-Einstein cones were obtained in ref.
\cite{conifold,SEdefects}. These configurations are the analogue of the ones
analyzed in section
\ref{DpDp+2section}, and it would be desirable to find its field theory
interpretation.

\section*{Acknowledgments}

We are grateful to F. Benini, 
S. Benvenuti, F. Canoura, J. Erdmenger. J. Grosse, I.
Klebanov, P. Merlatti, C. N\'u\~nez and A. Paredes for discussions at
different stages of this work. The  work of DA and AVR was 
supported in part by MCyT and  FEDER  under  grant
FPA2005-00188,  by Xunta de Galicia (grant   PGIDIT06PXIB206185PR and
Conselleria de Educacion) and by  the EC Commission under  grant
MRTN-CT-2004-005104. The work of DR-G has been partially suported by a
MEC-Fulbright fellowship FU-2006-07040. DR-G is grateful to Universidad
Autonoma de Madrid, Universidad de Granada and Universidade de Santiago de
Compostela for their hospitality and support during various stages of this
work. He would like also to specially thank Antonio Garcia-Garcia for
helping him with the move in, as well as for many invaluable discussions.

\vskip 1cm
\renewcommand{\theequation}{\rm{A}.\arabic{equation}}
\setcounter{equation}{0}
\medskip
\appendix

\setcounter{equation}{0}
\section{Fluctuations of the Dp-D(p+2) intersection}
\label{DpDp+2-decoupling}
\medskip
In this appendix we will complete the analysis of the fluctuations of the
Dp-D(p+2) system of section \ref{DpDp+2section}. Recall that the Lagrangian
that governs these fluctuations has been written in eq. (\ref{fluct-lag}). In
section  \ref{DpDp+2section} we already studied the equation of motion of the
transverse scalars $\chi$ and we concluded that the associated meson mass
spectrum is continuous and gapless. The other fields in (\ref{fluct-lag}) are
the scalar $x$ (which is transverse to the D(p+2)-brane and is directed along
the  ${\rm p}^{th}$ direction of the Dp-brane worldvolume) and the 
gauge field  $f_{ab}$. The equation of motion of $x$  reads:
\begin{equation}
\partial_a\,\Big[\,{\rho^2\over h}\,
\sqrt{\tilde{g}}\,
\,\hat{\mathcal{G}}^{ab}\,\partial_b\,x\Big]\,-\,\frac{C}{2}\,
\epsilon^{ij}\,f_{ij}\,=\,0\,\,,
\label{x-equation}
\end{equation}
while that of  the gauge field is:
\begin{equation}
\partial_a\,\Big[\,\rho^2\sqrt{\tilde{g}}
\Big(1+\frac{q^2}{\rho^4 h^2}\,\Big)\,f^{ab}\,\Big]
\,-\,C\,\epsilon^{bi}\partial_i\,x\,=\,0\,\,,
\label{f-equation}
\end{equation}
where $h$ and $C$ are the functions of $\rho$ defined in eqs. (\ref{h}) and
(\ref{C}) and $\epsilon^{bi}$ is zero unless $b$ is an index along the
two-sphere. Notice that eqs. (\ref{x-equation}) and (\ref{f-equation}) are
coupled. Let us decouple them by using the same method as that applied in 
\cite{AR} for the $q=0$ case. As in ref. \cite{AR} we define the following
two types of vector spherical harmonics for the two-sphere:
\beq
Y_i^l(S^2)\equiv \nabla_i\,Y^l(S^2)\,\,,
\qquad\qquad
\hat Y_i^l(S^2)\,\equiv\,{1\over \sqrt{\tilde g}}\,
\tilde g_{ij}\,\epsilon^{jk}\,
\nabla_k\,Y^l(S^2)\,\,,
\label{S2har}
\eeq
where $Y^l(S^2)$ is a  scalar harmonic  on $S^2$. We now study the
different types of modes following closely the analysis of ref. \cite{AR}. 
\subsection{Type I modes}
The type I modes are the ones that involve the scalar field $x$ and the
components of the gauge field strength along the $S^2$ directions. The ansatz
that we will adopt for $x$ is the following:
\beq
x\,=\,\Lambda(x^\mu,\rho)\,Y^l(S^2)\,\,.
\eeq
Moreover, if $a_a$ denote the components of the gauge field potential for
$f_{ab}$, we will take:
\beq
a_{\mu}=0\,\,,\qquad\qquad
a_{\rho}=0\,\,,\qquad\qquad
a_i\,=\,\phi(x^\mu,\rho)\,\hat Y_i^l(S^2)\,\,.
\eeq
Using this ansatz, the equation for $x$ becomes:
\beq
\rho^2\,\partial^{\mu}\partial_{\mu}\,\Lambda\,+\,
\partial_{\rho}\,\Bigg[\,
{\rho^6\over q^2\,+\,\rho^4\,h^2}\,\,\partial_{\rho}\,\Lambda\,
\Bigg]\,-\,l(l+1)\,{\rho^4\over q^2\,+\,\rho^4\,h^2
}\,\,\Lambda\,-\,C\,l(l+1)\,\phi\,=\,0\,\,.
\eeq
Then, by using the property $\nabla^i\hat Y_i^l=0$, which follows 
directly from the definition (\ref{S2har}), one can check that
(\ref{f-equation}) reduces to:
\beq
\partial^{\mu}\partial_{\mu}\,\phi\,+\,
\partial_{\rho}\,\Bigg[\,
{\rho^4\over q^2\,+\,\rho^4\,h^2}
\,\,\partial_{\rho}\,\phi\,
\Bigg]\,-\,l(l+1)\,
{\rho^2 \over q^2\,+\,\rho^4\,h^2}
\,\phi\,-\,C\,\Lambda\,=\,0\,\,.
\eeq
Let us now define  the function $V(x^\mu,\rho)$ as:
\beq
V\,=\,\rho\,\Lambda\,\,,
\eeq
and the following second-order differential operator ${\cal O}^2$:
\beq
{\cal O}^2\,\psi\,\equiv \,{1\over C}\,
\Bigg[\,\partial^{\mu}\partial_{\mu}\,\psi\,+\,
\partial_\rho\,\Bigg(\,{\rho^4\over q^2\,+\,
\rho^4\,h^2}\,
\partial_\rho\psi\,\Bigg)\,\Bigg]\,\,.
\eeq
Then, the equations of $V$ and $\phi$ can be written as  the following system
of  differential equations:
\bear
&&{\cal O}^2\,\,V\,=\,\Bigg[\,{1\over \rho}\,+\,l(l+1)\,
{\rho^2 \over \Big(\,q^2\,+\,
\rho^4\,h^2\,\Big)C}\,\Bigg]\,V\,+\,
{ l(l+1)\over \rho}\,\phi\,\,,\rc\rc
&&{\cal O}^2\,\,\phi\,=\,
l(l+1)\,
{\rho^2 \over \Big(\,q^2\,+\,
\rho^4\,h^2\,\Big)C}\,\,\phi\,+\,
{1\over \rho}\,V\,\,.
\eear
In order to decouple this system, let us define the functions $Z^\pm$ as:
\beq
Z^+\,=\,V\,+\, l\phi\,\,,\qquad\qquad\qquad
Z^-\,=\,V\,-\,(l+1)\,\phi\,\,.
\eeq
In terms of $Z^\pm$ the equations of the fluctuations are:
\bear
&& {\cal O}^2\,\,Z^+\,=\,\Bigg[\,l(l+1)\,
{\rho^2\over
\Big(\,q^2\,+\,\rho^4\,h^2\,\Big)C}\,+\,
{l+1\over \rho}\,\Bigg]\,Z^+\,\,,\rc\rc
&& {\cal O}^2\,\,Z^-\,=\,\Bigg[\,l(l+1)\,
{\rho^2\over \Big(\,q^2\,+\,
\rho^4\,h^2\,\Big)C}\,-\,
{l\over \rho}\,\Bigg]\,Z^-\,\,.
\label{Z+-}
\eear
Following {\cite{MT}, one can analytically map  eqs. (\ref{Z+-}) to
the one for the transverse scalars (\ref{radial-chi-fluct}). First of all,
let us introduce the reduced variables $\varrho$, $\bar M$ and $\bar q$,
defined as:
\beq
\varrho={\rho\over L}\,,\qquad \bar M^2=-{R^{4\alpha}\over L^{4\alpha-2}}\,
k^2\,,\qquad
\bar q
= \frac{q}{L^{2(1-\alpha)}R^{2\alpha}}\,\,.
\label{redvbles}
\eeq
Then, after substituting $Z^\pm=e^{ikx}\phi^\pm$  in (\ref{Z+-}) and  a
little algebra, the equation for these modes can be written as:
\bear
\varrho^{l+1}\,\partial_{\varrho}\,\left(\varrho^4\,Q\,
\partial_{\varrho}\phi^+\right)+\left[\bar
  M^2\,\varrho^{l+1}-(l+1)\,\partial_{\varrho}\left(\varrho^{4+l}\,Q
   \right)
\right]\phi^+=0\,,
\label{typeIphi+} \\ \rc
\varrho^{-l}\,\partial_{\varrho}\,\left(\varrho^4\,Q
\,\partial_{\varrho}\phi^-
\right)+\left[\bar
  M^2\varrho^{-l}+l\,\partial_{\varrho}\left(\varrho^{3-l}\,Q\right)
\right]\phi^-=0\,,\label{typeIphi-}
\eear
where $Q=Q(\varrho)$ is the following function:
\beq
Q(\varrho)\,\equiv\,{1\over \bar
q^{\,2}+\frac{\varrho^4}{(1+\varrho^2)^{2\alpha}}}\,\,.
\eeq
Moreover, in terms of these reduced variables, 
the equation  for the transverse scalars (\ref{radial-chi-fluct}) reads:
\begin{equation}
\partial_{\varrho}(\varrho^2\partial_{\varrho}\xi)\,+\,
\left[\frac{\bar{M}^2}{\varrho^2\,Q}\,\,-
\,l(l+1)\,\right]\,\xi\,=\,0\,\,.
\label{trans-scalars-reduced}
\end{equation}

In order to relate (\ref{trans-scalars-reduced}) and (\ref{typeIphi+})
let us  rewrite $\xi\,=\,\varrho^{-(l+1)}F^{+}$  and multiply the transverse
scalar equation (\ref{trans-scalars-reduced})  by $\varrho^{l+3}\,Q$. Then, 
one can check that the term with $F^{+}$ has a constant coefficient and, 
therefore, once we differentiate with respect to
$\varrho$,  the function $F^{+}$  appears in
the equation only through its derivatives. Then, upon defining
$\partial_{\varrho}F^{+}\,=\,\varrho^l\, g^{+}$, we conclude that the
resulting equation for $g^{+}$ is simply:
\begin{equation}
\varrho^{l}\,\partial_{\varrho}\,\left(\varrho^4\,Q\,
\partial_{\varrho}g^{+}\right)+\left[\bar
  M^2\,\varrho^{l}-l\,\partial_{\varrho}\left(\varrho^{3+l}\,Q\right)
\right]\,g^{+}\,=\,0\ .
\end{equation}
This equation is exactly the same as the one for the $\phi^+$ mode 
(eq. ({\ref{typeIphi+}))
once we identify $l+1$ in the $\phi^+$ equation with $l$ in the equation for
$g^{+}$.

It is easy to see that an alternative route can be followed relating the
transverse scalar equation (\ref{trans-scalars-reduced})
to the equation (\ref{typeIphi-}) for $\phi^-$, namely by defining
$\xi\,=\,\varrho^l\,F^{-}$. Then, after multiplying the equation by 
$\varrho^{2-l}\,Q$ and taking the $\varrho$ derivative, we see that, again,
$F^{-}$ appears only through its derivatives. Then,  we can define
$\partial_{\varrho}F^{-}\,=\,\varrho^{-(l+1)}\,g^{-}$ and the equation for
$g^{-}$ becomes:
\begin{equation}
\varrho^{-(l+1)}\,\partial_{\varrho}\,\left(\varrho^4\,Q\,
\partial_{\varrho}g \right)+\left[\bar
  M^2\varrho^{-(l+1)}+(l+1)\,\partial_{\varrho}\left(\varrho^{3-(l+1)}\,Q
  \right)
\right]\,g^{-}=0\,\,,
\end{equation}
which, indeed,  is identical to the equation (\ref{typeIphi-})  for the
$\phi^-$ modes once we take into account that $l$ is now to be identified
with $l+1$ in the equation for $g^{-}$.

To sum up, we have that the mapping of \cite{MT}:
\bear
\phi^+_{\,l=L}&=&\varrho^{-L-1}\,\partial_\varrho\left(\varrho^{L+2}\,
\xi_{\,l=L+1}\right)\,\,,\rc\rc
\phi^-_{\,l=L}&=&\varrho^{L}\,\partial_\varrho\left(\varrho^{1-L}\,
\xi_{\,l=L-1}\right)\,,
\eear
also works in the  Dp-D(p+2) intersection with flux studied here. As a
consequence of this result we can conclude that the mass spectrum of the
type I modes  displays the same features of that corresponding to  the
transverse scalars, namely it is  continuous and has no mass gap.

\subsection{Type II modes}

Consider now a configuration with $x=0$ and take the following ansatz for the
gauge field:

\beq
a_{\mu}=\phi_{\mu}(x,\,\rho)\,Y^l(S^2)\,\,,\qquad\qquad
a_{\rho}=0\,\,,\qquad\qquad
a_i\,=0\,\,.
\eeq

\noindent with the extra condition on $\phi$:

\begin{equation}
\partial^{\mu}\,\phi_{\mu}\,=\,0\ .
\end{equation}

\noindent Due to this condition, since $x=0$, the equations of motion for
$x$, $a_{\rho}$ and $a_i$ are trivially satisfied. The only remaining
non-trivial equation is that for $a_{\mu}$, which reads:
\beq
\bigg[ \rho^2\,
h^2\,+\,{q^2\over \rho^2}\,\bigg]\,\,
\partial^{\nu}\partial_{\nu}\,\phi_{\mu}\,+\,
\partial_{\rho}\,\Big(\rho^2\partial_\rho 
\phi_{\mu}\Big)\,-\,l(l+1)\,\phi_{\mu}\,=\,0\,\,.
\eeq
Now, if we write $\phi_{\mu}$ in a plane-wave basis: 
\begin{equation}
\phi_{\mu}\,=\,e^{ikx}\,\xi_{\mu}\,\, ,
\end{equation}
then this equation becomes:t
\beq
\partial_{\rho}\,\Big(\rho^2\partial_\rho 
\xi_{\mu}\Big)\,+\Bigg\{\bigg[\,\rho^2\,
h^2\,+\,{q^2\over
    \rho^2}\,\bigg]\, M^2\,-\,l(l+1)\Bigg\}\,\xi_{\mu}=\,0\,\,.
\eeq
Notice that this equation is the same as that in (\ref{radial-chi-fluct}) for
the transverse scalars.

\subsection{Type III modes}

Let us take now as ansatz for the gauge field:
\beq
a_{\mu}=0\,\,,\qquad\qquad
a_{\rho}=\phi(x,\,\rho)\,Y^l(S^2)\,\,,\qquad\qquad
a_i\,=\tilde{\phi}(x,\,\rho)\,Y^l_i(S^2)\,\,.
\eeq

With this ansatz it is straightforward to check that
$f_{ij}=0$. Therefore the equation of motion for $x$ is directly
satisfied if the take $x=0$. This leads to an equation of motion of $a_i$
which reads:
\begin{equation}
\partial^{\mu}\partial_{\mu}\tilde{\phi}\,+\,\partial_{\rho}
\bigg[\,\frac{\rho^4}{q^2+\,\rho^4\,h^2}\,
\big(\partial_{\rho}\tilde{\phi}-\phi\big)\,\bigg]\,=\,0\,\,.
\end{equation}
Moreover, the equation of motion of $a_{\rho}$ is:
\begin{equation}
\rho^2\partial^{\mu}\partial_{\mu}\phi\,+\,l(l+1)\,
\frac{\rho^4}{q^2+\rho^4\,
h^2}\,\big(\partial_{\rho}\tilde{\phi}
-\phi\big)\,=\,0\,\,,
\end{equation}
while that of  $a_{\mu}$ is:
\begin{equation}
\partial_{\mu}\big(\,l(l+1)\,\tilde{\phi}\,-\,\partial_{\rho}\,
(\rho^2\phi)\,\big)\,=\,0\,\,.
\label{phi-tildephi}
\end{equation}
Clearly,  eq. (\ref{phi-tildephi})  is satisfied if:
\begin{equation}
l(l+1)\,\tilde{\phi}\,=\,c\,+\,\partial_{\rho}(\rho^2\phi)\ ,
\end{equation}
where $c$ is an integration constant. Given this condition, it is
straightforward to see that the remaining two equations are indeed
equivalent. Thus, we arrive to the following equation  for $\phi$:
\begin{equation}
\rho^2\partial^{\mu}\partial_{\mu}\phi\,-\,l(l+1)\,
\frac{\rho^4}{q^2+\rho^4\,h^2}\,
\phi\,+\,\frac{\rho^4}{q^2+\rho^4\,h^2}\,
\partial_{\rho}^2(\rho^2\phi)\,=\,0\,\,.
\end{equation}
Writing $\phi=e^{ikx}\zeta(\rho)$  with $M^2=-k^2$, we get the following
differential equation for $\zeta(\rho)$: 
\begin{equation}
\partial_{\rho}^2(\rho^2\zeta)\,+\,\Bigg[\,\Big(\,h^2\rho^2\,+\,
{q^2\over \rho^2}\,\Big)\,M^2
\,-\,l(l+1)\,\,\Bigg]\,\zeta\,=\,0\,\,.
\label{typeIII-zeta}
\end{equation}
Eq. (\ref{typeIII-zeta})  can also be easily related to the one
corresponding to the transverse scalars. Indeed, it is a simple exercise to
verify that, if one defines $\xi=\rho\zeta$, eq. (\ref{typeIII-zeta}})
becomes exactly (\ref{radial-chi-fluct}). In particular, this fact implies
that the mass spectra of these type III modes is also continous and gapless.

\renewcommand{\theequation}{\rm{B}.\arabic{equation}}
\setcounter{equation}{0}
\medskip

\setcounter{equation}{0}
\section{Supersymmetry of the M2-M5 intersection}
\medskip
\label{kappa}

\medskip
In this appendix we will verify that the M2-M5 intersections with flux
studied in section \ref{M2M5section} are supersymmetric. We will verify this
statement by looking at the kappa symmetry of the M5-brane embedding, which
previously requires the knowledge of the Killing spinors of the background.
In order to write these spinors in a convenient way, 
let us rewrite the $AdS_4\times S^7$ near-horizon metric (\ref{M2metric}) of
the M2-brane background as:
\beq
ds^2\,=\,{r^4\over R^4}\,\,dx_{1,2}^2\,+\,{R^2\over r^2}\,dr^2\,+\,
R^2\,d\Omega_7^2\,\,,
\eeq
where $d\Omega_7^2$ is the line element of a unit seven-sphere and 
$R$ is given in eq. (\ref{M2-radius}). In what follows we shall represent 
the metric of $S^7$  in terms of polar coordinates $\theta^1,\cdots \theta^7$:
\beq
d\Omega_7^2\,=\,(d\theta^1)^2\,+\,\sum_{k=2}^7\,\,
\Bigg(\,\prod_{j=1}^{k-1}\,(\sin\theta^j)^2\,\Bigg)\,(d\theta^k)^2\,\,.
\eeq
Moreover, we shall consider the vielbein:
\bear
&&e^{x^\mu}\,=\,{r^2\over R^2}\,\,dx^\mu\,\,,\qquad\qquad
(\mu=0,1,2)\,\,,\rc\rc
&&e^{r}\,=\,{R\over r}\,\,dr\,\,,\rc\rc
&&e^{\theta^i}\,=\,R\,\,\Bigg(\,\prod_{j=1}^{i-1}\,\sin\theta^j\,\Bigg)\,
d\theta^i\,\,,\qquad\qquad (i=1,\cdots, 7)\,\,,
\eear
where, in the last line,  it is understood that for $i=1$ the product is absent. 

The Killing spinors of this background are obtained by solving the equation 
$\delta\psi_M\,=\,0$, where the supersymmetric variation of the gravitino in 
eleven dimensional supergravity is given by:
\beq
\delta\psi_M\,=\,D_M\,\epsilon\,+\,{1\over 288}\,
\Bigg(\,\Gamma_{M}^{\,\,N_1\cdots N_4}\,-\,8\delta_{M}^{N_1}\,
\Gamma^{\,\,N_2\cdots N_4}\,\Bigg)\,\epsilon\,\,
F^{(4)}_{N_1\cdots N_4}\,\,.
\label{gravitino}
\eeq
In eq. (\ref{gravitino}) $F^{(4)}=dC^{(3)}$, where $C^{(3)}$ has been written 
in eq. (\ref{C3M2}). In order to write  equation (\ref{gravitino}) more
explicitly,  let us define the matrix:
\beq
\Gamma_*\,\equiv\,\Gamma_{x^0x^1x^2}\,\,.
\eeq
Notice that $\Gamma_*^2=1$. From the equations
$\delta\psi_{x^\alpha}\,=\,\delta\psi_{r}\,=0$ we get the value of the derivatives
of $\epsilon$ with respect to the $AdS_4$ coordinates, namely:
\bear
&&\partial_{x^\alpha}\,\epsilon\,=\,-{r^2\over R^3}\,\,
\Gamma_{x^\alpha r}\,\big(\,1\,+\Gamma_*\,)\,\,,\rc\rc
&&\partial_{r}\,\epsilon\,=\,-{1\over r}\,\Gamma_*\,\epsilon\,\,.
\label{AdS4-dependence}
\eear
First of all, 
let us solve the first equation in (\ref{AdS4-dependence}) by taking 
$\epsilon=\epsilon_1$ with $\Gamma_*\epsilon_1\,=\,-\epsilon_1$, where 
$\epsilon_1$ is independent of the Minkowski coordinates $x^\alpha$.  The second
equation in (\ref{AdS4-dependence}) fixes the dependence of $\epsilon$ on 
$r$, which is
\beq
\epsilon_1\,=\,r\eta_1\,\,,\qquad\qquad \Gamma_*\eta_1\,=\,-\eta_1\,\,,
\eeq
where $\eta_1$ only depends on the coordinates of the $S^7$. 

Let us now find a second solution of eq. (\ref{AdS4-dependence}),  given
by the ansatz:
\beq
\epsilon_2\,=\,\big(\,
f(r)\,\Gamma_r\,+\,g(r)\,x^{\alpha}\,\Gamma_{x^\alpha}\,\big)\,\eta_2\,\,,
\eeq
where $f(r)$ and $g(r)$ are functions to be determined and
$\eta_2$  is a spinor independent of $x^\alpha$ and $r$. By plugging 
this ansatz in
(\ref{AdS4-dependence}) we get the conditions:
\bear
&&g(r)\,=\,-{2r^2\over R^3}\,f(r)\,\,,\qquad\qquad 
\Gamma_*\eta_2\,=\,-\eta_2\,\,,\rc\rc
&&f'(r)\,=\,-{f(r)\over r}\,\,,\qquad\qquad\,\,\,\,\,\,
g'(r)\,=\,{g(r)\over r}\,\,,
\eear
which can be immediately integrated, giving rise to the following spinor:
\beq
\epsilon_2\,=\,\Big(\,{1\over r}\,\,\Gamma_r\,-\,{2r\over R^3}\,\,
x^{\alpha}\,\Gamma_{x^{\alpha}}\,\Big)\,\eta_2\,\,,\qquad\qquad
\Gamma_*\eta_2\,=\,-\eta_2\,\,,
\eeq  
where $\eta_2\,=\,\eta_2(\theta)$. Then, a general Killing spinor of $AdS_4\times
S^7$ can be written as $\epsilon_1+\epsilon_2$, namely as:
\beq
\epsilon\,=\,r\,\eta_1(\theta)\,+\,
\Big(\,{1\over r}\,\,\Gamma_r\,-\,{2r\over R^3}\,\,
x^{\alpha}\,\Gamma_{x^{\alpha}}\,\Big)\,\eta_2(\theta)\,\,,\qquad\qquad
\Gamma_*\eta_i(\theta)\,=\,-\eta_i(\theta)\,\,.
\label{asd4S7KS}
\eeq

The dependence of the $\eta_i$'s on the angle $\theta^1$ can be 
determined from the
condition $\delta\psi_{\theta^1}=0$, which reduces to:
\beq
\partial_{\theta^1}\,\epsilon\,=\,
-{1\over 2}\,\,\Gamma_{r\theta^1}\,\Gamma_*\,\epsilon\,\,.
\label{thetaonedep}
\eeq
It can be checked that eq. (\ref{thetaonedep}) gives rise to the following
dependence of the spinor  $\eta_i$ on the angle $\theta^1$:
\beq
\eta_i\,=\,e^{{\theta^1\over 2}\,\Gamma_{r\theta^1}}\,\,\tilde \eta_i\,\,,
\eeq
where $\tilde \eta_i$ does not depend on $\theta^1$. Similarly, one can get the
dependence of the $\eta_i$'s on the other angles of the seven-sphere. The result
can be written as:
\beq
\eta_i(\theta)\,=\,U(\theta)\,\hat\eta_i\,\,,
\label{eta-hateta}
\eeq
with $\hat\eta_i$ being  constant spinors such that 
$\Gamma_*\hat\eta_i\,=\,-\hat\eta_i$
and $U(\theta)$ is the following rotation
matrix:
\beq
U(\theta)\,\,=\,e^{{\theta^1\over 2}\,\Gamma_{r\theta^1}}\,\,
\prod_{j=2}^7\,e^{{\theta^{j}\over 2}\,
\Gamma_{\theta^{j-1}\,\,\theta^{j}}}\,\,.
\label{Utheta}
\eeq
Notice that   $\epsilon$ depends on two arbitrary constant spinors $\hat\eta_1$
and  $\hat\eta_2$ of sixteen components  each one  and, thus, this background
has the maximal number of supersymmetries, namely thirty-two. 

\subsection{Kappa symmetry}
\medskip
The number of supersymmetries preserved by the M5-brane probe is the number
of independent solutions of the equation $\Gamma_{\kappa}\epsilon=\epsilon$,
where 
$\epsilon$ is one of the Killing spinors (\ref{asd4S7KS}) and 
$\Gamma_{\kappa}$ is the kappa symmetry matrix of the PST
formalism \cite{PST, APPS}.
In order to write the expression of this matrix, let
us define the following quantities:
\beq
\nu_p\,\equiv {\partial_p a\over \sqrt{-(\partial a)^2}},
\,\,\,\,\,\,\,\,\,\,\,\,\,\,\,\,\,\,\,\,
\,\,\,\,\,\,\,\,\,\,\,\,\,\,\,\,\,\,\,\,\,\,
t^m\,\equiv\,{1\over 8}\,
\epsilon^{mn_1n_2p_1p_2q}\,\tilde H_{n_1n_2}\,\tilde H_{p_1p_2}\,\nu_q\,\,.
\label{csiete}
\eeq
Then, the  kappa symmetry matrix is:
\beq
\Gamma_{\kappa}=-{\nu_m\gamma^m\over \sqrt{-{\rm det} (g+\tilde H)}}\,\,
\Bigg[\,\gamma_n t^n\,+\,
{\sqrt{-g}\over 2}\,\gamma^{np}\,\tilde H_{np}\,
+\,{1\over 5!}\gamma_{i_1\cdots i_5}\,\epsilon^{i_1\cdots i_5n}\nu_n\,\,
\Bigg]\,\,.
\label{cocho}
\eeq
In eq. (\ref{cocho}) $g$ is the induced metric on the worldvolume, 
$\gamma_{i_1i_2\cdots }$  are antisymmetrized
products of the worldvolume Dirac matrices 
$\gamma_i=\partial_iX^M\,E^{\underline N}_M\,
\Gamma_{\underline N}$ and the indices are raised with the inverse of $g$.

We shall consider here the embedding with $L=0$, which corresponds to
having massless quarks. In the polar coordinates we are using for the $S^7$
this corresponds to taking:
\beq
\theta^1\,=\,\cdots\,=\,\theta^4\,=\,{\pi\over 2}\,\,.
\label{M5embedding}
\eeq
Moreover, we shall denote the three remaining angles of the $S^7$ as 
$\chi^i\equiv \theta^{4+i}$, $(i=1,2,3)$. We will describe the M5-brane embeddings
by means of the following set of worldvolume coordinates:
\beq
\xi^i\,=\,(x^0,x^1,r,\chi^1,\chi^2,\chi^3)\,\,,
\eeq
and we will assume that: 
\beq
x\equiv x^2=x(r)\,\,.
\eeq
The induced metric for such embedding is given by (\ref{inducedmetricM5})
with $L=0$ and
$\rho=r$, namely:
\beq
g_{ij}\,d\xi^i\,d\xi^j\,=\,{r^4\over R^4}\,dx_{1,1}^2\,+\,
{R^2\over r^2}\,\Big(\,1+\,{r^6\over R^6}\,x'^2\,\Big)\,dr^2\,+\,
R^2\,d\Omega_3^2\,\,.
\eeq
The induced Dirac matrices for this embedding  are:
\bear
&&\gamma_{x^{\mu}}\,=\,{r^2\over R^2}\,\,\Gamma_{x^{\mu}}\,\,,
\qquad\qquad (\mu=0,1)\,\,,\rc\rc
&&\gamma_{r}\,=\,{R\over r}\,\,
\big(\Gamma_r\,+\,{r^3\over R^3}\,\,x'\,\,\Gamma_{x^2}\,\big)\,\,,\rc\rc
&&\gamma_{\Omega_3}\,\equiv\gamma_{\chi^1\,\chi^2\,\chi^3}\,=\,
R^3\,\sqrt{\tilde g}\,\Gamma_{\Omega_3}\,\,,
\eear
where $\sqrt{\tilde g}\,=\,\sin^2\chi^1\sin\chi^2$ and 
$\Gamma_{\Omega_3}\equiv\Gamma_{\chi^1\,\chi^2\,\chi^3}$. Notice also that:
\bear
&&\gamma^{x^0}\,=\,-{R^2\over r^2}\,\,\Gamma_{x^0}\,\,,
\qquad\qquad\gamma^{x^1}\,=\,{R^2\over r^2}\,\,\Gamma_{x^1}\,\,,\rc\rc
&&\gamma^{r}\,=\,{r\over R}\,\,\Big(\,1+\,{r^6\over R^6}\,x'^2\,\Big)^{-1}\,\,
\Big(\,\Gamma_r\,+\,{r^3\over R^3}\,\,x'\,\,\Gamma_{x^2}\,\Big)\,\,.
\eear
We will also assume that we have switched on a magnetic worldvolume gauge
field $F$, parametrized as in (\ref{M5flux}) in terms of a flux number $q$. 
Moreover, in the gauge $a=x^1$, the only non-vanishing component of $\nu_p$
is:
\beq
\nu_{x^1}\,=\,-i\,{r^2\over R^2}\,\,,
\eeq
and one can check that the only non-vanishing component of $\tilde H$  is
$\tilde H_{x^0r}$, whose expression is given by (\ref{tildeH}) with $L=0$ and
$\rho=r$. Also, the worldvolume vector $t^m$ defined in (\ref{csiete})
is zero and one can verify that 
the kappa symmetry matrix $\Gamma_{\kappa}$ takes the form:
\beq
\Gamma_{\kappa}\,=\,
{\sqrt{\tilde g}\,r^3\over \sqrt{-{\rm det} (g+\tilde H)}}\,\,\Bigg(\,
{q\over  R^3}\,+\,\Gamma_{\Omega_3}\,\Bigg)\,\,\Bigg(\,
\Gamma_{rx^2}\,+\,{r^3\over R^3}\,x'\,\Bigg)\,\,\Gamma_*\,\,.
\eeq
For the embeddings we are interested in the function $x(r)$ is given by:
\beq
x\,=\,\bar x\,+{q\over 2}\,{1\over r^2}\,\,,
\label{M5profile}
\eeq
where $\bar x$ is a constant (see eq. (\ref{x-explicit})). In order to
express the form of $\Gamma_{\kappa}$ for these embeddings, let us define the
matrix ${\cal P}$ as:
\beq
{\cal P}\,\equiv\,\Gamma_{x^2 r}\,\Gamma_{\Omega_3}\,\,.
\eeq
Notice that ${\cal P}^2\,=\,1$. Moreover, the kappa symmetry matrix can be written
as:
\beq
\Gamma_{\kappa}\,=\,-{1\over 1+{q^2\over R^6}}\,\,
\Bigg(\,{\cal P}\,+\,{q^2\over R^6}\,+\,
{q\over R^3}\,\,\Gamma_{x^2 r}\,\,\big(1-{\cal P}\big)\,\,\Bigg)\,\,
\,\,\Gamma_*\,\,.
\label{Gkappa-embedding}
\eeq
Let us represent the Killing spinors $\epsilon$ on the M5-brane worldvolume 
as is eq. (\ref{asd4S7KS}).  By using the explicit function 
$x(r)$ written  in eq. (\ref{M5profile}), one gets:
\beq
\epsilon\,=\,{1\over r}\,\,
\Big(\, \Gamma_{r}\eta_2\,-\,{q\over  R^3}\,\Gamma_{x^2}\,\eta_2\,\Big)\,+\,
r\,\Big(\,\eta_1\,-\,{2 \bar x\over R^3}\,\Gamma_{x^2}\,\eta_2\,\Big)\,-\,
{2r\over R^3}\,\,x^p\,\Gamma_{x^p}\,\eta_2\,\,,
\label{KS-onthe wv}
\eeq
where the index $p$ can take the values $0,1$ and
we have organized the right-hand side of (\ref{KS-onthe wv}) according to
the different dependences on $r$ and $x^p$. By substituting
(\ref{Gkappa-embedding}) and (\ref{KS-onthe wv}) into the equation 
$\Gamma_{\kappa}\epsilon=\epsilon$ and comparing the terms on the two sides of
this equation that have the same dependence on the coordinates, one gets the
following three equations:
\bear
&&\big(\,{\cal P}\,+\,1\,\big)\,\eta_2\,=\,0\,\,,\rc\rc
&&\bigg[\,1\,-\,{q\over  R^3}\,\Gamma_{x^2r}\,\bigg]\,
\big(\,{\cal P}\,-\,1\,\big)\,
\bigg(\,\eta_1\,-\,{2\bar x\over R^3}\,\Gamma_{x^2}\,\eta_2\,
\bigg)\,=\,0\,\,,\rc\rc
&&\bigg[\,{q\over  R^3}\,\Gamma_{x^2r}\,-\,1\,\bigg]\,
\Gamma_{x^p}\,\big(\,{\cal P}\,+\,1\,\big)\,\eta_2\,=\,0\,\,.
\label{kappasystem}
\eear

In order to solve these equations, let us classify the sixteen spinors
$\eta_1$ according to their ${\cal P}$-eigenvalue as:
\beq
{\cal P}\,\eta_1^{(\pm) }\,=\,\pm \eta_1^{(\pm) }\,\,.
\label{M5conditionone}
\eeq
Notice that ${\cal P}$ and $\Gamma_*$ commute and, then, the condition of
having well-defined ${\cal P}$-eigenvalue is perfectly compatible with having
negative  $\Gamma_*$-chirality. We can now solve the system
(\ref{kappasystem}) by taking
$\eta_2=0$ (which solves the first and third equation) and choosing $\eta_1$ to be
one of the eight spinors $\eta_1^{(+)}$
of positive ${\cal P}$-eigenvalue. Thus, this solution of (\ref{kappasystem}) is:
\beq
\eta_1\,=\,\eta_1^{(+)}\,\,,\qquad\qquad
\eta_2\,=\,0\,\,.
\eeq
Another  set of solutions corresponds to taking spinors $\eta_1^{(-)}$ of negative 
${\cal P}$-eigenvalue and a spinor $\eta_2$ related to $\eta_1^{(-)}$ as follows:
\beq
\eta_2\,=\,{R^3\over 2\bar x}\,\,\Gamma_{x^2}\,\,\eta_1^{(-)}\,\,.
\label{second-kappa-spinor}
\eeq
Notice that the second equation in (\ref{kappasystem}) is automatically satisfied.
Moreover, as $[{\cal P},\Gamma_{x^2}\,]=0$, the spinor $\eta_2$ in
(\ref{second-kappa-spinor}) has negative ${\cal P}$-eigenvalue and, therefore, the
first and third equation in the system (\ref{kappasystem})  are also satisfied. 

In order to complete the proof of the supersymmetry of our M2-M5 configuration we
must verify that the kappa symmetry conditions found above can be fulfilled at all
points of the M5-brane worldvolume. Notice that, when evaluated for the embedding 
(\ref{M5embedding}), the spinors $\eta_{1,2}$ depend on the angles $\chi^i$ that
parametrize the $S^3\subset S^7$. To ensure that the conditions
(\ref{M5conditionone}) and (\ref{second-kappa-spinor}) can be imposed at all
points of the $S^3$ we should be able to translate them into some algebraic
conditions for the constant spinors $\hat\eta_i$. Recall (see eq.
(\ref{eta-hateta})) that the spinors $\eta_i$ and $\hat \eta_i$ are related by
means of the matrix 
$U(\theta)$. Let us denote by $U_{*}(\chi)$ the rotation matrix restricted to the
worldvolume, \ie:
\beq
U_{*}(\chi)\,\equiv\,U(\theta)_{\big|\theta^1=\cdots =\theta^4={\pi\over 2}}\,\,.
\eeq
Moreover, let us define $\hat{\cal P}$ as the result of conjugating the matrix 
${\cal P}$ with the rotation matrix $U_{*}(\chi)$:
\beq
\hat{\cal P}\,\equiv U_{*}(\chi)^{-1}\,\,{\cal P}\,\,U_{*}(\chi)\,\,.
\eeq
A simple calculation by using (\ref{Utheta}) shows that $\hat{\cal P}$ is the
following constant matrix:
\beq
\hat{\cal P}\,=\,\Gamma_{x^2\theta^4}\,\Gamma_{\Omega_3}\,\,.
\eeq
Moreover, from the definition of $\hat{\cal P}$ it follows that:
\beq
{\cal P}\,\eta_1^{(\pm) }\,=\,\pm \eta_1^{(\pm) }\qquad
\Longleftrightarrow\qquad
\hat{\cal P}\,\hat\eta_1^{(\pm) }\,=\,\pm \hat\eta_1^{(\pm) }\,\,.
\eeq
Therefore, the condition (\ref{M5conditionone}) for $\eta_1$ is equivalent to
require that the corresponding constant spinor $\hat\eta_1$ be an eigenstate of
the constant matrix $\hat{\cal P}$.  Finally, as $[U_{*}, \Gamma_{x^2}]=0$, eq. 
(\ref{second-kappa-spinor}) is equivalent to the following condition, to be
satisfied by the constant spinors $\hat \eta_1$ and $\hat\eta_2$:
\beq
\hat\eta_2\,=\,{R^3\over 2\bar x}\,\,\Gamma_{x^2}\,\,\hat\eta_1^{(-)}\,\,.
\label{second-constant-spinor}
\eeq
Taken together, these results prove that the kappa symmetry condition
$\Gamma_{\kappa}\,\epsilon\,=\,\epsilon$ can be imposed at all points of the
worldvolume of our M5-brane embedding and that this configuration is
${1\over 2}$-supersymmetric.

\end{document}